\documentclass[11pt]{article}  
 \pdfoutput=1 
\usepackage{jheppub}
\usepackage{titlesec}
\usepackage{tikz, amsmath}
\usetikzlibrary{calc}
\usepackage{bbm}
\usepackage{xcolor}
\usepackage{dsfont}
\usepackage{subfigure}
\usepackage{hyperref}
\usepackage{microtype}
\usepackage{caption}
\usepackage{amsmath}
\usepackage{float}
\usepackage{graphics}
\usepackage{psfrag}
\usepackage{color}
\usepackage{slashed}
\usepackage{subfigure}
\usepackage{graphicx}
\usepackage{array,multirow,axodraw}
\usepackage{amsfonts}
\usepackage{amssymb}
\usepackage{mathtools}
\usepackage{amsmath}
\usepackage{amsfonts}
\usepackage{mathrsfs}
\usepackage{multirow}
\usepackage{feynmp}
\usepackage{array}
\usepackage{textcomp} 
\usepackage{phaistos} 
\usepackage[utf8]{inputenc}
\usepackage{pifont} 
\usepackage{pgf}

\usepackage{subfigure}
\usepackage{multirow}

\newcommand{\beq}{\begin{eqnarray}}
\newcommand{\eeq}{\end{eqnarray}}

\newcommand{\bmp}{\noindent\begin{minipage}{16cm}}
\newcommand{\emp}{\end{minipage}\vskip 7mm} 

\usepackage{dcolumn}
\usepackage{bm}
\usepackage{slashed} 
\usepackage[normalem]{ulem}

\usepackage{epsfig}

\newcommand{\bea}{\begin{eqnarray}}
\newcommand{\eea}{\end{eqnarray}}

\newcommand{\ba}{\begin{eqnarray}}
\newcommand{\ea}{\end{eqnarray}}

\title{ Composite Higgs and Dark Matter Model in $SU(6)/SO(6)$}
\author[a, b]{Giacomo Cacciapaglia,}
\emailAdd{g.cacciapaglia@ipnl.in2p3.fr}
\affiliation[a]{Institut de Physique des Deux Infinis de Lyon (IP2I), CNRS/IN2P3 UMR5822, 4 rue Enrico Fermi,
69622 Villeurbanne Cedex, France} 

\affiliation[b]{Universit\'e Claude Bernard Lyon 1, Universit\'e de Lyon, 92 rue Pasteur, 69361 Lyon Cedex 07, France}

\author[c]{Haiying Cai,}
\emailAdd{haiying.cai@apctp.org}
\affiliation[c]{Asia Pacific Center for Theoretical Physics, Pohang, Gyeongbuk 790-784, Republic of Korea}

\author[a, b]{Aldo Deandrea,}
\emailAdd{deandrea@ipnl.in2p3.fr}

\author[d]{Ashwani Kushwaha.}
\emailAdd{ashwanik@iisc.ac.in}
\affiliation[d]{Centre for High Energy Physics, Indian Institute of Science, Bangalore-560012, India\\ }

\abstract{We consider a realisation of composite Higgs models in the context of $SU(6)/ SO(6)$ symmetry, which features a custodial bi-triplet, two Higgs doublets and dark matter candidates. This model can arise from an underlying gauge-fermion theory. The general vacuum structure is explored using the top partial compositeness to generate a special vacuum characterised by a single angle aligned with the first Higgs doublet. We present the CP and Dark Matter $\mathbb{Z}_2$ parity in  two different pNGB bases and analyse  the spectra  in the absence of tadpoles and tachyons. For the phenomenology,  we discuss  the  constraints from electroweak precision tests and  from a potentially light CP-odd singlet (other than the Dark Matter) in the model. \\[.1cm]}

\begin{document}
\maketitle

\newpage

\section{Introduction}

The discovery of a scalar resonance at the LHC~\cite{Aad:2012tfa, Chatrchyan:2012xdj, Aad:2015zhl}, compatible with the Standard Model (SM) Higgs Boson~\cite{Higgs:1964pj, Englert:1964et, Guralnik:1964eu}, has been a great experimental achievement and a step forward in our understanding of particle physics. The LHC experiments have  delivered a precise determination of the mass of the new resonance and a fairly good understanding of the properties of the new boson in terms of its couplings~\cite{Khachatryan:2016vau}. These results are a  powerful tool to test theories beyond the SM, which typically predict deviations in the couplings of the Higgs boson while aiming at providing new explanations for phenomena not  explained in the SM. A challenging  example of such phenomena is the interpretation of the Dark Matter  in the universe by the presence of a  new neutral stable particle.

A time-honoured extension of the SM consists of adding a new strong interaction at the electroweak scale, thus explaining the dynamical origin of the electroweak symmetry breaking~\cite{Weinberg:1975gm,Dimopoulos:1979es,Eichten:1979ah} and the lightness of the Higgs boson as arising from a pseudo Nambu Goldstone Boson (pNGB) of a broken global symmetry~\cite{Kaplan:1983fs}. This Composite pseudo-Goldstone Higgs paradigm  postulates the existence of a new strong sector above the TeV scale, which confines and breaks chiral symmetry of the underlying sector. The latter produces a set of pNGBs, which includes the Higgs doublet field as a condensation  of fermions  in the strong sector.  In the following, we will call this composite  dynamics {\it{Hypercolor}} (HC), and the condensing fermions {\it{hyper-fermions}} (HF). The underlying theories are selected in order to allow for a vacuum of the HC dynamics that does not break the SM gauge symmetry, contrary to more traditional Technicolor theories without a Higgs boson~\cite{Weinberg:1975gm}. The coupling of the SM to the strong sector explicitly breaks the flavour symmetry group of the theory to a smaller group, thereby generating an effective potential for the composite pNGBs. For a feasible theory,   the  electroweak symmetry breaking should be triggered along the direction of composite Higgs boson  after minimising the effective potential  (an explicit calculation requires use of the strong dynamics and can not be performed perturbatively). While the original idea dates from the early 80's, it was resurrected in the 2000's following the insurgence of the AdS/CFT correspondence~\cite{Maldacena:1997re}.  The construction is based on relating  strongly coupled conformal field theories (CFT) to weakly coupled 5D theories in AdS~\cite{Gubser:1998bc, Witten:1998qj}, where no Ultra-Violet (UV, meaning a higher energy more fundamental formulation of the theory) underlying setup is constructed and the formulation is only based on the effective field theory approach. The Higgs boson, therefore, arises as a component of the gauge bosons in the AdS bulk~\cite{Contino:2003ve}, in analogy to gauge-Higgs unification models~\cite{Manton:1979kb,Hosotani:1983xw}. Detailed models were proposed based on the minimal coset $SO(5)/SO(4)$~\cite{Agashe:2004rs, Agashe:2005dk, Contino:2006qr}, which are minimal in terms of the effective Lagrangian descriptions but somewhat lack a more microscopic understanding offered by underlying gauge-fermion theories.

The present work is therefore based on composite Higgs models that admit an underlying gauge-fermion theory, where the global symmetry is broken by the formation of a bilinear condensate of some new hyper-fermions
 $\psi$, charged under both HC interactions and the electroweak gauge symmetries. Few patterns of symmetry breaking are thus allowed, according to the real, pseudo-real or complex nature of the representation of the hyper-fermions. The following patterns of symmetry breaking are allowed~\cite{Peskin:1980gc}: $SU(2N)/Sp(2N)$ for fermions in a pseudo-real representation, $SU(N)/SO(N)$ for real and $SU(N)\times SU(N)/SU(N)$ for complex ones. It follows that the minimal model features the coset $SU(4)/Sp(4)$~\cite{Cacciapaglia:2014uja}, which can be obtained by a $\mathcal{G}_{\rm HC} = SU(2)_{\rm HC}$ gauge group with 4 Weyl fermions in the fundamental representation~\cite{Ryttov:2008xe,Galloway:2010bp}.

In this work we are interested in exploring non-minimal models with a larger coset. In fact, a simple and small extension of the fermionic sector allows for more pNGBs and for unbroken global symmetries in the coset. Our main goal is thus to find a Dark Matter candidate within the pNGB spectrum. This kind of analysis has already been presented for complex representations, where $SU(4)\times SU(4)/SU(4)$ is the minimal case~\cite{Ma:2015gra}, and for pseudo-real ones in $SU(6)/Sp(6)$~\cite{Cai:2018tet}, the latter being the next extension of the minimal model. We will focus on real representation, leading to the coset $SU(6)/SO(6)$. This is an extension of the minimal $SU(5)/SO(5)$ model, first studied by Georgi, Kaplan and Dugan in Ref.~\cite{Dugan:1984hq}. The minimal case shares many similar features with the Georgi-Machacek model~\cite{Georgi:1985nv}, which is based on elementary scalars, however it cannot feature a Dark Matter candidate. As we will see, a common feature of all composite Higgs models with a Dark Mater candidate is the presence of two Higgs doublet multiplets. A general analysis of composite two-Higgs-doublet models (2HDM) can be found in Refs~\cite{Mrazek:2011iu,Bertuzzo:2012ya}.
An intrinsic problem for composite Higgs models with a bi-triplet under $SU(2)_L \times SU(2)_R$  is that top interactions  tend to induce a vacuum expectation value for the triplets in a custodial violating way. Therefore, following Ref.~\cite{Agugliaro:2018vsu}, we will work with models that feature fermion partial compositeness with the top spurions in an adjoint representation of the global $SU(6)$ symmetry. We will show that this  non-minimal extension naturally leads to an unbroken $\mathbb{Z}_2$ symmetry, under which the second doublet and a singlet are odd and serving as Dark Matter candidates. The presence of   this  parity imposes some restrictions on the UV completions of the model, requiring  the SM fermions only couple  to one pNGB Higgs doublet.  In particular,  we will assume  a mass degeneracy  in the underlying theory,  so that a global $SO(4) \times SO(2)$ symmetry is preserved.  However this plays no crucial role for the DM parity as the latter is a reflection symmetry of the vacuum.

The paper is organised as follows: In section~\ref{Model}, we describe the underlying gauge theory  and construct the  low-energy effective Lagrangians with SM symmetry naturally embedded. We further illustrate the  discrete  parities  and show that  Wess-Zumino-Witten topological term conserves the Dark Matter $\mathbb{Z}_2$ parity.  Then the contribution to the pNGB potential through gauge loop and  hyper-fermion masses is commented. In section~\ref{PartialComp}, we  focus on  top partial compositeness  to generate the consistent vacuum  in absence of tadpoles, which is followed by a brief description of  possible UV completion.  In section~\ref{pNGBMass},  the mass spectra of the non-Higgs pNGBs are analysed in terms of free parameters with tachyon-free region identified. In section~\ref{pheno}, we discuss the constraint from Electroweak precision Test on the misalignment angle, and interpret interesting ALP signatures observable at the LHC  and future colliders.  Finally we give our conclusions and the model implications in section~\ref{Conclusion}.
The technical part for the calculation indicated in the main text as well as the mechanism to generate a neutrino mass  are presented  in the Appendix.

\section{The model}\label{Model}

The model is based on the coset of pattern $SU(6)/SO(6)$, where $SO(6) \supset SO(4) \sim SU(2)_L \times SU(2)_R$ contains the SM electroweak group in an explicitly custodial invariant way (the hypercharge is the $T_{3R}$ generator of $SU(2)_R$ ). Like in the SM Higgs sector, the global $SO(4)$ custodial symmetry is explicitly broken by the gauge couplings and spontaneously broken down to its diagonal $SU(2)_{D}$ by the Higgs expectation value. Potentially harmful contributions to the $\rho$ parameter are cured by the above construction~\cite{Georgi:1984af}. We will then embed the gauging of the electroweak interactions similarly to Ref.~\cite{Dugan:1984hq}.

We begin by defining the underlying model that has a new strong $\mathcal{G}_{\rm HC}$ gauge group with 6 Weyl fermions transforming as a real representation of the gauge group. Upon condensation, a condensate bilinear in the fermions, $\langle \psi^i \psi^j \rangle$, is expected to form at an infrared scale $\Lambda_{\rm HC}$ close to  $\mathcal{O}(1)$~TeV, where the reality of the HF representation guarantees that the condensate is symmetric in the flavour indices~\cite{Kosower:1984aw}.  In the fermionic underlying theory, the SM quantum numbers of the fermions $\psi^i$ generating the global symmetry, can be assigned as 
\begin{eqnarray}
\Psi = \left(\begin{array}{c} \psi_{1/2} \\ \psi_{-1/2}  \\ \psi_{0} \\ \tilde{\psi_{0}} \end{array} \right)\,,
\end{eqnarray}
where $\psi_{1/2}$ and $\psi_{-1/2}$ are $SU(2)_L$ doublets forming a bi-doublet $(2,2)$ of the custodial symmetry, while $\psi_{0}$ $ \tilde{\psi_{0}}$  are  singlets. Thus, our model expands the one of Georgi {\it et al.}~\cite{Dugan:1984hq} by one Weyl singlet. 
In our notation the charge is $Q$=$T_{3}$+$Y$.
The symmetric condensate $\langle \Psi \Psi^T \rangle$ thus break $SU(6)/SO(6)$, generating 20 pNGBs in the low energy effective theory.
 Note that further massive fermions, in the same representation of $\Psi$ or different, can be added in order to render the theory quasi-conformal above $\Lambda_{\rm HC}$, by attracting the theory towards an infra-red fixed point~\cite{Holdom:1981rm}. 
 This will not affect the properties of the light pNGBs.
  
The Higgs sector of the SM is thus replaced by the following Lagrangian for the HFs:
\begin{eqnarray}
\mathcal{L}= -\frac{1}{4}G_{\mu\nu}G^{\mu\nu} + i\bar{\Psi} \sigma^{\mu}D_{\mu}\Psi - \frac{1}{2} \Psi^T M \Psi + \mbox{h.c.}\,,
\end{eqnarray}
where $G^{\mu\nu}$ is the field strength of new strong group $\mathcal{G}_{\rm HC}$ and the covariant derivative contains both the electroweak gauge interaction and the HC ones:
\begin{eqnarray}
D_{\mu}=\partial_{\mu}-ig_2 W_{\mu}^{i}T_{L}^{i}-ig_1 B_{\mu}T_{R}^{3}-ig_{HC}G_{\mu}^{a}T_{HC}^{a}\,,
\end{eqnarray}
where $g_{HC}$ is the HC gauge coupling, $g_2$ is the electroweak $SU(2)_{L}$ gauge coupling and $g_1$ is the hypercharge coupling. $W_{\mu}$ and $B_{\mu}$ are the electroweak $SU(2)_{L}\times U(1)_{Y}$ gauge bosons respectively.
A gauge invariant mass term can be added, where $M$ is a matrix in flavour space given by
\begin{eqnarray} \label{eq:HFmass}
M  =\left(
\begin{array}{cc|cc}
& i M_{1}\sigma_2 &  \\
-i M_{1} \sigma_2 &  & \\ \hline
& & M_{2}\mathbbm{1}_{2} 
\end{array} \right)\,,
\end{eqnarray}
where $M_{1}$ is a Dirac mass for the charged doublets and $M_{2}$ a Majorana mass for the neutral singlets. 
Note that the mass for the doublets is automatically invariant under the custodial symmetry,  while we have chosen equal masses for the two singlets to preserve a $SO(2)$ symmetry for simplicity.~\footnote{Without loss of generality, we could give different masses to the two singlets, thus explicitly breaking the $SO(2)\sim U(1)$ symmetry, just like gauge interactions breaking the custodial symmetry. This more general set up would not affect the remainder  discussion of this model, but simply add one more free parameter.}.  In fact the above mass term explicitly breaks $SU(6)\rightarrow SO(4) \times SO(2)$, while for $M_{1}=M_{2}$ the unbroken symmetry is enhanced to $SO(6)$. The latter can be identified with the global symmetry left unbroken by the condensation, which preserves the electroweak symmetry. It is thus convenient to define the Higgs (and other pNGBs) around this vacuum, given by the 2-index symmetric matrix
\begin{eqnarray}
\Sigma_{\rm EW}  =\left(
\begin{array}{cc|cc}
& i \sigma_2 &  \\
-i \sigma_2 &  & \\ \hline
& & \mathbbm{1}_{2} 
\end{array} \right)\,,
\end{eqnarray}
with $\sigma_{2}$ being the second Pauli matrix. The 35-dimensional adjoint representation of $SU(6)$ is easily seen to
split into 15 unbroken generators $T_{a}$ forming the adjoint of $SO(6)$ and 20 broken ones $X_{i}$ , which reside in a
2-index symmetric traceless 20 of $SO(6)$. Under the symmetry $SO(4) \times SO(2)$, which includes the custodial symmetry of the Higgs, the 20 pNGBs transform as 
\begin{equation}
(3,3)_{0} \oplus (2,2)_{2} \oplus (2,2)_{-2} \oplus (1,1)_{0} \oplus (1,1)_{-4} \oplus (1,1)_{4}\,.
\end{equation}
These further decompose under the SM gauge group ($SU(2)_L \times U(1)_Y$)
as $3_{\pm 1} + 3_{0} + 2_{\pm 1/2} + 2_{\pm 1/2} + 3 \times 1_{0}$.
We see that compared to the minimal $SU(5)/SO(5)$ model that contains a bi-triplet, a bi-doublet and a singlet, our model contains an additional bi-doublet and two singlets. Furthermore, the Higgs doublets and two singlets are charged under the global $SO(2)$.

\subsection{The vacuum and sigma model}

The relevant degrees of freedom in the low energy dynamics can be described in terms of a non-linearly transforming pion matrix
\begin{eqnarray}
\xi(x)=e^{\left(\frac{2 \sqrt{2} i  }{f} ~ \Pi \right)}, ~~  \text{with} ~~ \Pi = \sum_{i=1}^{20} \pi_{i} X^{i}   
\end{eqnarray}
where $f$ is the decay constant, which sets the scale of the condensation, and $\pi_{i}$ are the Goldstone Bosons associated to the 20 broken generators $X^{i}$.
It is convenient, however, to describe the pNGB dynamics in terms of a linearly transforming matrix, defined as 
\begin{eqnarray}
\Sigma (x) &=& e^{\left(\frac{i \sqrt{2}}{f}\Pi \right)} \cdot \Sigma_{\rm EW} \cdot e^{\left(\frac{i \sqrt{2}}{f}\Pi^{T} \right)}  = e^{\left(\frac{2 \sqrt{2} i}{f}\Pi \right)} \cdot \Sigma_{\rm EW}\,,
\end{eqnarray}
where we defined them around the electroweak preserving vacuum in order to classify their quantum numbers under the SM gauge symmetries.
In this basis, the pNGB matrix $\Pi$ is parameterised as:
\begin{eqnarray}
2 ~\Pi = \left(\begin{array}{cccc}
\varphi + \frac{\eta _1}{\sqrt{3}} \mathbbm{1}_{2} &  \Lambda    &  \sqrt{2} {H}_{1}  &  \sqrt{2} {H}_{2} \\
\Lambda^{\dagger}     & - \varphi + \frac{\eta _1}{\sqrt{3}} \mathbbm{1}_{2} & -\sqrt{2}  \widetilde{{H}_{1}} & - \sqrt{2} \widetilde{{H}_{2}} \\         
\sqrt{2} {H}_{1} ^{\dagger} & - \sqrt{2}  \widetilde{{H}_{1}}^{\dagger} &2 \left( \frac{\eta _3}{\sqrt{2}}-\frac{\eta
	_1}{\sqrt{3}}  \right) & \sqrt{2} \eta _2 \\
\sqrt{2} {H}_{2} ^{\dagger} & - \sqrt{2}  \widetilde{{H}_{2}}^{\dagger}  &  \sqrt{2} \eta _2 & -2 \left( \frac{\eta _3}{\sqrt{2}}+\frac{\eta
	_1}{\sqrt{3}}  \right)\\
\end{array} \right)\,,
\end{eqnarray}
where $H_{1}$ and $H_{2}$ are the two doublets that may play the role of Higgs doublet fields:
\begin{eqnarray}
H_1 = \left( \begin{array}{c} G_+ \\ \frac{h + i G_0 }{\sqrt{2}} \end{array} \right) \,, \qquad       H_2 = \left( \begin{array}{c} H_+ \\ \frac{H_0 + i A_0 }{\sqrt{2}} 
\end{array} \right) \,, \qquad  \widetilde{H}_{1,2} = i \sigma_2 H_{1,2}^*\,,
\end{eqnarray}
The other degrees of freedom correspond to a complex isospin triplet $\Lambda$ with hypercharge one, a real triplet $\varphi$ and 3 singlets $\eta_{1,2,3}$. The 6 components of $\Lambda$ and the 3 components of $\varphi$ transform as a $(3,3)$ under $SU(2)_{L}\times~SU(2)_{R}$:
\begin{eqnarray}
\varphi = \sigma^{a} \varphi^{a}  \equiv 
\begin{pmatrix}
\varphi^{0}         & \sqrt{2}\varphi^{+} \\
\sqrt{2}\varphi^{-} & - \varphi^{0} \\
\end{pmatrix} \,,  \qquad
\Lambda = \sigma^{a} \Lambda^{a} \equiv 
\begin{pmatrix}
\sqrt{2}\Lambda^{+}         & 2 \Lambda^{++} \\
2\Lambda_{0} & - \sqrt{2}\Lambda^{+} \\
\end{pmatrix}
\end{eqnarray}
A more detailed discussion on generators of this model is presented in Appendix~\ref{a:GBosons}.


The breaking of the electroweak symmetry can be achieved by giving a vacuum expectation value to any of the two Higgs doublets. Being Goldstones, the two vacuum expectation values can be re-expressed in terms of two angles, $\alpha_{1} = \frac{v_{1}}{ f}$ 
and $\alpha_{2} = \frac{v_{2}}{ f}$, which rotate the vacuum of the strong sector away from $\Sigma_{\rm EW}$. This can be expressed by the following $SU(6)$ rotation:
\begin{equation}
U (\alpha_{1},\alpha_{2}) = e^{i \sqrt{2} ( \alpha_{1} \, {\bf{X_{10}}}  +   \alpha_{2} \, {\bf{X_{14}}} )}\,. 
\end{equation}
Defining  $\alpha^{2} = (\alpha_{1}^{2}+\alpha_{2}^{2})$, and $\tan(\beta) = \alpha_{2}/\alpha_{1} \equiv v_2/v_1$, the rotation matrix reads
\begin{eqnarray}
\footnotesize
U(\alpha,\beta) = 
\left(
\begin{array}{cccccc}
1 & 0 & 0 & 0 & 0 & 0 \\
0 & \cos ^2\left(\frac{\alpha }{2}\right) & \sin ^2\left(\frac{\alpha }{2}\right) & 0 &
\frac{i \cos (\beta ) \sin (\alpha )}{\sqrt{2}} & \frac{i \sin (\alpha ) \sin (\beta
	)}{\sqrt{2}} \\
0 & \sin ^2\left(\frac{\alpha }{2}\right) & \cos ^2\left(\frac{\alpha }{2}\right) & 0 &
-\frac{i \cos (\beta ) \sin (\alpha )}{\sqrt{2}} & -\frac{i \sin (\alpha ) \sin (\beta
	)}{\sqrt{2}} \\
0 & 0 & 0 & 1 & 0 & 0 \\
0 & \frac{i \cos (\beta ) \sin (\alpha )}{\sqrt{2}} & -\frac{i \cos (\beta ) \sin (\alpha
	)}{\sqrt{2}} & 0 & ~1-2 \sin^2 (\frac{\alpha}{2}) \cos ^2(\beta ) & -\sin
^2\left(\frac{\alpha }{2}\right) \sin (2 \beta ) \\
0 & \frac{i \sin (\alpha ) \sin (\beta )}{\sqrt{2}} & -\frac{i \sin (\alpha ) \sin (\beta
	)}{\sqrt{2}} & 0 & -\sin ^2\left(\frac{\alpha }{2}\right) \sin (2 \beta ) & 1-2 \sin^2 (\frac{\alpha}{2}) \sin^2(\beta ) \\
\end{array}
\right)\,.
\end{eqnarray}
This can be used to define a new vacuum, $\Sigma_{0}(\alpha,\beta)= U(\alpha,\beta) \cdot \Sigma_{\rm EW} \cdot  U(\alpha,\beta)^{T}$, and a new basis of pNGBs
\begin{equation}
\Sigma(x)_{\alpha,\beta} = \xi(x)_{\alpha,\beta}\cdot \Sigma_{0}(\alpha,\beta) = U(\alpha, \beta) \cdot \xi(x) \cdot \Sigma_{\rm EW} \cdot U(\alpha, \beta)^{T}\,,
\end{equation}
where 
\begin{eqnarray}
\xi(x)_{\alpha,\beta} = U(\alpha,\beta) \cdot \xi(x)\cdot  U(\alpha,\beta)^{\dagger}\,.
\end{eqnarray}
Interestingly the vacuum misalignment with 2 angles in the $h$ and $H_0$ directions  can be achieved by a combined operation:
\begin{eqnarray}
U(\alpha,\beta) =  U(\beta) \cdot U(\alpha) \cdot  U(\beta)^{\dagger}\,,
\end{eqnarray}
with $U(\beta)$ defined in terms of a generator unbroken by $\Sigma_{\rm EW}$,
\begin{eqnarray}
\footnotesize
U(\beta)= e^{- i \sqrt{2} \,\beta \, \bf{S_{15}} } =  \left(
\begin{array}{cc|cc}
\mathbbm{1}_{2} & 0 & 0 & 0 \\
0&  \mathbbm{1}_{2} & 0 & 0  \\  \hline
0 & 0 & \cos \beta  & - \sin \beta \\
0 & 0 &   \sin \beta & \cos \beta  
\end{array} \right)\,,
\end{eqnarray}
and
\begin{eqnarray}
U({\alpha} )=
\left(
\begin{array}{cccccc}
1 & 0 & 0 & 0 & 0 & 0 \\
0 & \cos ^2\left(\frac{\alpha }{2}\right) & \sin ^2\left(\frac{\alpha }{2}\right) & 0 &
\frac{i}{\sqrt{2}}  \sin (\alpha ) & 0 \\
0 & \sin ^2\left(\frac{\alpha }{2}\right) & \cos ^2\left(\frac{\alpha }{2}\right) & 0 &
-\frac{i }{\sqrt{2}} \sin (\alpha ) & 0 \\
0 & 0 & 0 & 1 & 0 & 0 \\
0 & \frac{i }{\sqrt{2}} \sin (\alpha ) & -\frac{i }{\sqrt{2}} \sin (\alpha ) & 0 & \cos (\alpha
) & 0 \\
0 & 0 & 0 & 0 & 0 & 1 \\
\end{array}
\right)\,.
\end{eqnarray}
This allows to re-write the pNGB matrix as
\begin{equation}
\Sigma(x)_{\alpha,\beta}  =U_{\beta} \cdot U_{\alpha} \cdot \xi^\prime(x). \Sigma_{EW} \cdot U_{\alpha}^{T} \cdot U_{\beta}^{T}. \label{eq:vacuum}
\end{equation}
with  $\xi^\prime(x)=U(\beta)^\dagger \xi (x) U(\beta)^* $,  where the  $\beta$ dependence can be  removed by the pNGB field  redefinition.  Thus we can further simplify the vacuum structure by  rotating away the outer  $U(\beta)$, at the price of absorbing it into  the spurions that explicitly break this $SU(6)$ rotation. In fact, because both the electroweak gauging and the mass term are invariant under $U(\beta)$,  any Lagrangian term involving them will be independent on $\beta$.~\footnote{Note that the same cannot be said for $U(\alpha)$, as $\alpha$ contains the electroweak breaking vacuum expectation value.}  For the top couplings which do not commute with this rotation,  specific treatment will be discussed in the Yukawa section.
Then we can  study the theory around the simpler vacuum that only depends on $\alpha$:
\begin{eqnarray}
\Sigma(x) =U_{\alpha} \cdot \xi^\prime(x). \Sigma_{EW}\cdot U_{\alpha}^{T}.
\end{eqnarray} 
At the leading order, the chiral Lagrangian is given by the kinetic term for $\Sigma$,
\begin{eqnarray}
\mathcal{L} = \frac{1}{16}f^{2}Tr[(D_{\mu}\Sigma(x))^{\dagger} \cdot D^{\mu}\Sigma(x)]\,,\label{Lag}
\end{eqnarray}  
with
\begin{equation}
D_{\mu}\Sigma(x) = \partial_{\mu}\Sigma(x) -
\bigg[ ig_{2} W^{a}_{\mu}( T^{a}.\Sigma(x)+\Sigma(x).T^{aT} ) + 
ig_{1} B_{\mu} ( Y.\Sigma(x)+\Sigma(x).Y^{T} )\bigg]\,.
\end{equation}
We recall that in our notation the electric charge equals $Q = T^{3}+Y$.
This Lagrangian contains the masses of $W$ and $Z$ gauge bosons
\begin{equation}
m_{W}^{2}=\frac{1}{4}g_2^{2} f^{2}\sin^{2}({\alpha})\,,\;\; m_{Z}^{2} = \frac{m_{W}^{2}}{cos^{2}\theta_{W}}\,, \;\; \mbox{with} \;\; f \sin(\alpha) = v_{\rm SM}\,,
\label{mass} \end{equation}    
the latter fixing the relation between the pNGB decay constant $f$ and the electroweak scale $v_{\rm SM} = 246$~GeV via the vacuum misalignment angle $\alpha$. The linear couplings between the Higgs boson $h$ and the vector bosons are given by
\begin{eqnarray}
g_{hWW}=g_{hWW}^{SM}\cos(\alpha),~~~g_{hZZ}=g_{hZZ}^{SM}\cos(\alpha)\,, \label{HWZ}
\end{eqnarray}
showing explicitly the universal coupling modification for composite Higgs models~\cite{Belyaev:2016ftv,  Vecchi:2013bja, Ferretti:2016upr}.
The couplings of two scalars to gauge bosons are reported in the Appendix \ref{a:GBvertex}.
A more general vacuum where custodial triplet CP even field $\mathcal{I}m\Lambda_{0}$ acquires a vacuum expectation value and leads to custodial violated vacuum~\cite{Agugliaro:2018vsu} is studied in Appendix~\ref{a:GeneralVac}.

\subsection{Discrete symmetries: CP and DM parity} \label{parities}

From our choice of degenerate singlet masses, it follows that our model features a  $SO(2) \sim U(1)$ global symmetry that is preserved by gauge interactions and by the hyper-fermion mass only  for $\alpha =0$, i.e. before the electroweak symmetry breaking. This symmetry actually coincides with the rotation $U(\beta)$ described in the previous section. In the Appendix~\ref{complex basis}, we provide an alternative complex basis ($U(1)$ basis) where  the two Higgs doublets form a complex bi-doublet, $H_1 + i H_2$, and two CP-odd singlets form a complex singlet  $\eta_3 - i \eta_2$,  which are charged under this  symmetry. In the complex basis,  the $U(\beta)$ is a $e^{\pm i \beta}$ mapping in the lower $(2\times2)$ block  compared with the real basis  ($SO(2)$ basis). As a consequence once the electroweak symmetry breaking is induced,  the ratio of  two vacuum expectation values $\tan \beta$ can be rotated  away as a phase, while the trace of $SO(2)$ symmetry breaking  is encoded in the mass splitting for the real and imaginary part of the complex field. However when we examine the gauge boson couplings with 2 pNGB fields, we  can find  no  interaction for a single $\eta_2$ particle at the quadratic order, signalling a protection from an extra symmetry. Indeed, even after the Higgs obtains its VEV, our theory still enjoys a discrete  $\mathbb{Z}_2$ parity  (with determinant $-1$), under which the second doublet and one singlet are odd. This parity  denoted  by
\begin{eqnarray}
\Omega_{\rm DM} = \left( \begin{array}{ccc}
\mathbbm{1}_{2} &   &  \\
& \mathbbm{1}_{2} &   \\
&   & \sigma_{3}  \\
\end{array} \right) \,, \quad  \Omega_{\rm DM} \Sigma_\alpha (H_2, \eta_2)\Omega_{\rm DM} =  \Sigma_\alpha (- H_2, -\eta_2)\,,
\end{eqnarray}
can be used to define a Dark Matter candidate.~\footnote{This  DM symmetry is also preserved for unequal singlet masses, which explicitly break the global $SO(2)$. However in such case, since the HF mass $M$ does not commute with $U(\beta)$ rotation, the $\beta$ angle will appear as a physical parameter in the masses and couplings generated by the hyper-fermion mass term.}  In the complex basis, the dark matter candidate is simply the imaginary part of the complex doublet and singlet.  Although  $\Omega_{\rm DM} $ is  a good symmetry within the $\alpha$ angle misaligned vacuum,  such parity will be  broken by the $U(\beta)$ rotation due to  the property of $ \Omega_{\rm DM} U (\beta) \Omega_{\rm DM}  \neq  U(\beta)^\dagger $.  This indicates that  the SM fermions can not couple to the second Higgs in order to conserve the Dark Matter parity, as we will discuss more in details in section~\ref{potential}.
In fact it is the inert condition  that guarantees  a Dark Matter candidate present in the pNGB spectrum,  regardless  of whether the  $SO(2)$ symmetry is broken or not by the HF mass term.

In our effective theory, the CP parity is well defined from the  operation which can hermitian conjugate the sigma matrix and change the sign of CP-odd particles. In the real  basis, the CP operator is accidentally  the remnant symmetry of $U(\beta)$, (note that $\Omega_{\rm CP}$  is not a subgroup of $U(\beta)$ in the complex basis):
\begin{eqnarray}
\Omega_{\rm CP} = \left( \begin{array}{ccc}
\mathbbm{1}_{2} &   &  \\
& \mathbbm{1}_{2} &   \\
&  & - \mathbbm{1}_{2}  \\
\end{array} \right) \,,       \quad  \Omega_{\rm CP} \Sigma_{(\alpha, \beta)} (\phi_{\rm odd}) \Omega_{\rm CP} = \Sigma_{(\alpha, \beta)}^\dagger (-\phi_{\rm odd})
\end{eqnarray}
with $\phi_{\rm odd}$ denoting the CP-odd pNGBs in this model. The operation of $\Omega_{\rm CP}. \Sigma_{0}(x)^\dagger. \Omega_{\rm CP}$, where $\Sigma_0 (x)$ is the sigma field before EWSB, is to change the sign of odd particles and conjugate them if not neutral. Thus combined with  the operation of $\Omega_{\rm CP}.U(\alpha, \beta).\Omega_{\rm CP}  = U(\alpha, \beta)^*$, a hermitian conjugation effect is  produced for the sigma field after the vacuum misalignment.  Unlike the DM parity discussed above,  the  CP operation  is  compatible with both the  $\alpha$ and $\beta$ angles in our model. As we can see, under this  parity the CP-even fields are $h$, $H_{0, \pm}$ and $\lambda_0 =- \frac{i}{2}(\Lambda_0-\Lambda_0^*)$ and  the other fields are CP-odd.  

\subsection{Wess-Zumino-Witten topological term}\label{sec:WZW}
\begin{table}
\footnotesize
 \centering
\begin{tabular}{|c|c|c|}
 \hline
   $\eta_{1} W_{\mu \nu }^+ \tilde{W}_{\mu \nu }^- $  & $\eta _3 W_{\mu \nu }^+ \tilde{W}_{\mu \nu }^-$ &   $\varphi_0 W_{\mu \nu }^+ \tilde{W}_{\mu \nu }^-$  \\
\hline 
 $\frac{ (7 \cos (2 \alpha )+17) }{4 \sqrt{6} f \sin^2\left(\theta _W\right)}$  & $\frac{ \sin ^2(\alpha ) }{2 f \sin^2\left(\theta _W\right)}$ &  $\frac{ \sin ^2(\alpha ) }{2 \sqrt{2} f \sin^2\left(\theta _W\right)}$
\\ \hline \hline
  $\frac{(\Lambda _{0} +\Lambda_{0}^*) }{\sqrt{2}} W_{\mu \nu }^+ \tilde{W}_{\mu \nu }^- $  & $\Lambda _+ W_{\mu \nu }^- \tilde{Z}_{\mu \nu }$ &   $\varphi_+ W_{\mu \nu }^- \tilde{Z}_{\mu \nu }$  \\
\hline
 $\frac{3 \sin ^2(\alpha ) }{2  f \sin^2\left(\theta
   _W\right)}$  & $-  \frac{\sin ^2\left(\frac{\alpha }{2}\right) \left(1-2 \cos (\alpha )-3 \cos \left(2 \theta_W\right)\right)}{\sqrt{2} f \sin^2\left(\theta _W\right) \cos \left(\theta _W\right) }$ & $- \frac{\cos ^2\left(\frac{\alpha }{2}\right) \left(1+2 \cos (\alpha )-3 \cos \left(2 \theta_W\right)\right)}{\sqrt{2} f \sin^2\left(\theta _W\right) \cos \left(\theta _W\right) }$   \\ \hline   \hline
  $\Lambda _{++}  W_{\mu \nu }^- \tilde{W}_{\mu \nu }^- $ & $\Lambda _+ W_{\mu \nu }^- \tilde{A}_{\mu \nu }$& $\varphi_+ W_{\mu \nu }^- \tilde{A}_{\mu \nu }$   \\ \hline    $ \frac{\sin ^2(\alpha ) }{\sqrt{2} f \sin^2(\theta _W)} $ & $  \frac{3 \sqrt{2} \sin ^2\left(\frac{\alpha }{2}\right)}{f  \sin \left(\theta _W\right)}$ & $  \frac{3 \sqrt{2} \cos ^2\left(\frac{\alpha }{2}\right)}{f  \sin \left(\theta _W\right)}$  \\ \hline
  \hline  $\eta _1 A_{\mu \nu } \tilde{A}_{\mu \nu }$ & $\begin{array}{c} \eta _3 \\ \frac{(\Lambda _0+\Lambda_0^*)}{\sqrt{2}} \end{array} $  $A_{\mu \nu } \tilde{A}_{\mu \nu }$ & $\varphi_0 A_{\mu \nu } \tilde{A}_{\mu \nu }$  \\ \hline 
 $\frac{ \sqrt{6} }{ f }$ & $0$ & $\frac{ 3 \sqrt{2} }{ f }$ \\ \hline
   \hline  $\eta _1 Z_{\mu \nu } \tilde{A}_{\mu \nu }$ & $\begin{array}{c} \eta _3 \\ \frac{(\Lambda _0+\Lambda_0^*)}{\sqrt{2}} \end{array} $  $Z_{\mu \nu } \tilde{A}_{\mu \nu }$ & $\varphi_0 Z_{\mu \nu } \tilde{A}_{\mu \nu }$  \\ \hline 
 $\frac{2 \sqrt{6} }{ f  \tan (2 \theta_W)}$ & $ 0 $ & $\frac{6 \sqrt{2} }{ f  \tan (2 \theta_W)}$  \\ \hline
    \hline  $\eta _1 Z_{\mu \nu } \tilde{Z}_{\mu \nu }$ & $\begin{array}{c} \eta _3 \\ \frac{(\Lambda _0+\Lambda_0^*)}{\sqrt{2}} \end{array} $  $Z_{\mu \nu } \tilde{Z}_{\mu \nu }$ & $\varphi_0 Z_{\mu \nu } \tilde{Z}_{\mu \nu }$  \\ \hline 
 $\frac{\left(2+ 7 \cos^2 ( \alpha )+ 3
   \cos \left(4 \theta _W\right)\right)}{ \sqrt{6} f \sin^2\left(2 \theta _W\right)} $ & $\frac{\sin ^2(\alpha )}{f  \sin^2\left(2 \theta _W\right)}$ & $   \frac{ \left(2-5 \cos^2 (\alpha )+3 \cos \left(4 \theta _W\right)\right)}{ \sqrt{2} f \sin^2\left(2 \theta _W\right)}$  \\ \hline
\end{tabular}
\caption{The coefficients for the WZW interactions in $SU(6)/SO(6)$ CHM, and we need to  times a prefactor $\frac{e^2 d_{\psi}}{48 \pi^2 }$ for each term.  }\label{tab:wzw}
\end{table}

A term in the effective Lagrangian that typically breaks parities in the pNGB sector is the Wess-Zumino-Witten (WZW) topological term~\cite{Wess:1971yu, Witten:1983tw}. It allows CP-odd pNGBs to decay to gauge bosons via the anomaly of the global symmetry.  The coupling of one scalar to two gauge bosons originate from the following action~\cite{Kaymakcalan:1983qq}:
\begin{multline}
 S_{\rm WZW} = C \int Tr[(d\mathcal{A} \mathcal{A} +\mathcal{A} d\mathcal{A})d\Sigma \Sigma^{\dagger}+(d\mathcal{A}^{T}\mathcal{A}^{T}+\mathcal{A}^{T}d\mathcal{A}^{T})\Sigma^{\dagger}d\Sigma] \nonumber \\ + C \int Tr[\mathcal{A} d\Sigma\mathcal{A}^{T} \Sigma^{\dagger}-d\mathcal{A}^{T}d\Sigma^{\dagger}\mathcal{A} \Sigma]\,,
\end{multline}
with the overall coefficient $C =  -i \frac{d_{\psi}}{48 \pi^2 }$, and $d_{\psi}$  stands for  the rep dimension of  $\psi$ in the HC gauge theory.  The differential forms  $\mathcal{A}$ and $d\Sigma$ are defined to be:
\begin{equation}
\mathcal{A}= \mathcal{A}_{\mu}dx^{\mu}\,, \quad d\mathcal{A} = \partial_{\mu} \mathcal{A}_{\nu}dx^{\mu}dx^{\nu}\,, \quad d\Sigma=\partial_{\mu}\Sigma dx^{\mu}\,.
\end{equation}
where $\mathcal{A}_\mu$ is the EW gauge  fields embedded in the global $SU(6)$ symmetry, so that $ \mathcal{A}_\mu = g \left(\sum_{i =1,2,3}  W_\mu^i  T_L^i+  \tan (\text{$\theta $w}) B_\mu T_R^3  \right)$.
After integrating out by parts,  the above action yields anomaly interactions in the form of $\frac{\kappa_{VV'}}{f}V^{\mu \nu} \tilde{V'}_{\mu \nu}$, where $V_{\mu \nu}$ is the tensor field for gauge bosons $W^{\pm}, Z$, $A$ and  $\tilde{V}_{\mu \nu} = \epsilon_{\mu\nu \rho \sigma} V^{\rho \sigma}$. We find out that  two Higgs doublets,  $\eta_{2}$ and  $\lambda_0 = -\frac{i}{\sqrt{2}}(\Lambda_0 - \Lambda_0^*)$ remain anomalous free. This indicates that  the WZW interactions conserve the DM parity so that no  decays are induced for the  two CP-odd fields ($\eta_{2}$  and $A_0$ in the second Higgs doublet) thanks to the  remaining $\mathbb{Z}_{2}$ symmetry  after a global $O(2)$ symmetry is broken.  However we  can obtain the anomaly interactions for other CP-odd scalars  and the explicit expressions for these  coefficients are listed in Table~\ref{tab:wzw}.  We can observe  that  for   the  $SU(6)/SO(6)$ model,  the WZW coefficients for the charged triplet fields $\phi_+$, $\Lambda_+$ and $\Lambda_{++}$ are exactly the same as in the $SU(5)/SO(5)$ scenario due to the similar bi-triplet structure~\cite{Ferretti:2016upr}.  However the major  difference arises from the neutral scalars because of  a more complicated singlet sector. For $\eta_1$, $\eta_3$, $\phi_0$ and $\frac{1}{\sqrt{2}}(\Lambda_0 + \Lambda_0^*)$,  their couplings  with $W, Z$, $A$ gauge bosons can be schematically parameterised  as:
\begin{eqnarray}
\mathcal{L}_{WZW} & \supset &  
   \frac{e^2 d_{\psi}}{48 \pi^2 f } \bigg( \kappa_{WW,1} \eta_1 W_{\mu\nu}^+ \widetilde{W}_{\mu\nu}^-  +  \kappa_{WW,2} \left( \varphi^{0}  + \sqrt{2} \eta_3+ 3 \left(\Lambda_0 + \Lambda_0^*\right) \right) W_{\mu\nu}^+ \widetilde{W}_{\mu\nu}^-  \nonumber \\ &+ & \kappa_{AA} \left(\eta_1+  \sqrt{3} \varphi^{0}  \right) A_{\mu\nu}\widetilde{A}_{\mu\nu} +  \kappa_{ZA}  \left(\eta_1+ \sqrt{3} \varphi^{0} \right) Z_{\mu\nu}\widetilde{A}_{\mu\nu} \nonumber \\ &+ &  \left(\kappa_{ZZ,1} ~\eta_{1}+ \kappa_{ZZ,2}~ \varphi^{0} + \kappa_{ZZ,3} ~(  \eta_{3} + \frac{1}{\sqrt{2}}  \left(\Lambda_{0}+\Lambda_{0}^{*} \right) ) \right)  Z_{\mu\nu}\widetilde{Z}_{\mu\nu} \bigg)
 \end{eqnarray}
where those $\kappa_{VV'}$  can be read off from the full coefficients in Table~\ref{tab:wzw}. In particular we find out
$ \kappa_{AA}= \sqrt{6}$,  $\kappa_{ZA} = 2 \kappa_{AA} / \tan (2 \theta_W) $ $ \kappa_{ZZ, 1} =c_1 + \frac{\cos(4 \theta_W)}{2 \sin^2(2 \theta_W)} \kappa_{AA}$ and $\kappa_{ZZ, 2} = c_2 + \frac{ \sqrt{3} \cos(4 \theta_W)}{2 \sin^2(2 \theta_W)} \kappa_{AA}$. For the latter two expressions, the $c_{1,2}$ come from the piece  orthogonal to  the photon generator. One interesting pattern exists  that only the $\eta_1$ and $\phi_0$ have non-vanishing anomaly couplings to $AA$ (diphoton) and $ZA$.   While for   $\eta_3$ and $\frac{1}{\sqrt{2}} \left(\Lambda_{0}+\Lambda_{0}^{*} \right)$, they share the same  coupling $\kappa_{ZZ,3}$ to a pair of  $Z$ bosons,  but subleading as $ \propto \sin^2 (\alpha)$.  


\subsection{Loop-induced potential for the pNGBs} \label{potential}

We will now briefly discuss the effect of loops of tops and gauge bosons on the pNGB potential. 
The hyper-fermion mass also plays a role.


\subsubsection*{The top and bottom Yukawas: bilinear Operators}

We will assume here that the Yukawa couplings are generated via 4-fermion operators
connecting the elementary quarks (and leptons) to the hyper-fermions, which are bilinear in the elementary
fields. As we are interested in the contribution to the pNGB potential, we focus on the fermions with the largest couplings, i.e. top and bottom quarks.
The  4-fermion interactions, bilinear in the SM fields $q_{L}$,  $t_{R}^c$ and $b_R^c$, are given by:
\begin{eqnarray} \label{eq:bilinears}
\frac{1}{\Lambda^{2}_{t}} \underset{i= 1,2}{\sum}  \left( y_{t_i} (q_{L}t_{R}^c)_{\delta}^{\dagger} \Psi^{T}P_{t_i}^{\delta}\Psi + y_{b_i} (q_{L}b_{R}^c)_{\delta}^{\dagger} \Psi^{T}P_{b_i}^{\delta}\Psi \right)\,,
\end{eqnarray} 
where $\delta$ is a $SU(2)_{L}$ index and $P_{1,2}^{\delta}$ (defined in Appendix.\ref{a:4fermioperator}) are projectors that select the two components of $\langle \Psi \Psi^T \rangle$ transforming as the Higgs doublets. The above equation is a generalisation of the four fermion interactions introduced in Ref.~\cite{Galloway:2010bp, Cacciapaglia:2014uja}.  This  operator  is  required to observe the $SU(2)_L \times U(1)_Y$  symmetry as well,  so that by setting $\alpha \to 0$,  we can find that $t_{L,R}$ couples to $\tilde{H}_{1,2}$ and $b_{L, R}$ couples to $H_{1,2}$. This scenario is inspired by Extended Technicolor models~\cite{Dimopoulos:1979es,Eichten:1979ah}, where they are generated by gauge interactions, and by Bosonic Technicolor model~\cite{Samuel:1990dq}, where they are generated by scalar exchanges (similarly to what happens in partial composite Higgs models~\cite{Chung:2005yz}). 
In order to explain the large hierarchies in the quark masses it is clear that the generation of 4-fermion operators for different flavours cannot happen just at one single scale $\Lambda_{t}$. Also, the scale where the top interactions cannot be too high without suppressing the mass, while generating all masses at the same scale will also induce flavour-violating processes that are highly constrained. All this points towards a multi-scale scenario.
Another mechanism involves linear couplings of the elementary fermions, thus generating fermionic partial compositeness~\cite{Kaplan:1991dc}, which we will consider in the next section.


In the Effective Lagrangian, the 4-fermion interactions are replaced by effective operators in terms of pNGB matrices, with the couplings in Eq.\eqref{eq:bilinears} replaced by effective couplings that contain form factors form the strong sector: $y_{ti} \to \tilde{Y}_{ti}$ and $y_{bi} \to \tilde{Y}_{bi}$.
As we already discussed, it is most convenient to study the theory in the basis where the $\beta$  will be  absorbed into  spurions  and  pNGB redefinition according to Eq. \eqref{eq:vacuum} . This implies that the projectors transform  by $P_{1,2}^{\delta} \to U_{\beta}^{T} \cdot P_{1,2}^{\delta} \cdot U_{\beta}$, which is equivalent to rotate the Yukawa couplings:
\begin{eqnarray}
Y_{t1} &=& \cos \beta \tilde{Y}_{t1} + \sin \beta \tilde{Y}_{t2} \,, \quad
Y_{t2} =- \sin \beta \tilde{Y}_{t1} + \cos \beta \tilde{Y}_{t2}. 
\end{eqnarray}
(and similarly for the bottom) with $Y_{t1,2}$ carrying dependence on $\beta$. Simultaneously the $H_1$, $H_2$,  $\eta_2$ and $\eta_3$ which are  charged under the $U(\beta) \in SO(2)$,  will undergo the rotation  of $U(\beta)^\dagger \xi (x) U(\beta)^*$, such that:
\begin{eqnarray}
H_1 &=& \tilde{H}_1 \cos (\beta )+\tilde{H}_2 \sin (\beta ), \quad
H_2 = \tilde{H}_2 \cos (\beta )- \tilde{H}_1 \sin (\beta ),  \\
\eta_2 &=& \tilde{\eta} _2 \cos (2 \beta )-\tilde{\eta}_3 \sin (2 \beta ) \quad
\eta_3 = \tilde{\eta} _2 \sin (2 \beta )+ \tilde{\eta} _3 \cos (2 \beta ) 
\end{eqnarray}
The Effective operator for the top Yukawa couplings reads
\begin{eqnarray}
\mathcal{L}_{Yuk} =  f (q_{L}t_{R}^c)_{\delta}^{\dagger}\bigg[Y_{t1}Tr[P_{t1}^{\delta}.\Sigma(x)] + Y_{t2}Tr[P_{t2}^{\delta}.\Sigma(x)]\bigg] + \mbox{h.c.}
\end{eqnarray}   
After expanding to the first order in the pNGB fields, we obtain
\begin{eqnarray}
\mathcal{L}_{Yuk} &= & -\frac{f}{\sqrt{2}} \sin (2\alpha )  Y_{\text{t1}} \overline{t_L} t_R  - Y_{\text{t2}} \bigg( -i A_0
+ H \cos (\alpha ) + i \eta _2 \sin (\alpha ) \bigg)  \overline{t_L} t_R \nonumber \\ 
&-& Y_{\text{t1}}\bigg( h \cos (2 \alpha ) - \frac{i \eta _1 \sin (2 \alpha ) }{2 \sqrt{6}} + \frac{i \eta _3 \sin (2 \alpha )}{2 } - \frac{ i \varphi _0 \sin (2 \alpha )}{2 \sqrt{2}} \nonumber \\ &-&  \frac{i}{2 \sqrt{2}}  \big(\sin (2 \alpha )-2 \sin (\alpha )\big) \Lambda_{0}- \frac{i}{2 \sqrt{2}}  \big(\sin (2 \alpha )+2 \sin (\alpha )\big) \Lambda_{0}^{*}\bigg)   \overline{t_L} t_R \nonumber \\
&-& \left(  -\sqrt{2} Y_{\text{t2}}  H_{-} +  i ~ Y_{\text{t1}}   \sin (\alpha ) \left(\Lambda _- - \varphi _- \right)   \right)\overline{b_L} t_R\,.
\end{eqnarray}
Top mass is, therefore, proportional to $Y_{t1}$ only, which is the coupling to the doublet combination that carries the vacuum expectation value $\alpha$, leading to 
\begin{equation}
m_{t}=\frac{1}{\sqrt{2}} Y_{t1} f \sin(2 \alpha)\,.
\end{equation}
The second coupling, $Y_{t2}$, only characterises the couplings to DM candidate, e.g. the second doublet as well as singlet $\eta_2$.
The mass of the bottom quark is generated by a very similar operator
\begin{eqnarray}
\mathcal{L}_{Yuk} = f(q_{L}b_{R}^c)_{\delta}^{\dagger}\bigg[Y_{b1}Tr[P_{b1}^{\delta}.\Sigma(x)]+Y_{b2}Tr[P_{b2}^{\delta}.\Sigma(x)]\bigg] + \mbox{h.c.}
\end{eqnarray}
with
\begin{eqnarray}
\mathcal{L}_{Yuk} & = &-  \frac{f}{\sqrt{2}} \sin(2 \alpha)  Y_{b1} \overline{b_L} b_R - Y_{\text{b2}} \bigg( i A_0 
+ H \cos (\alpha ) + i \eta _2 \sin (\alpha )\bigg) \overline{b_L} b_R  \nonumber \\
&-& Y_{\text{b1}}\bigg(h \cos (2 \alpha ) -\frac{i \eta _1 \sin (2 \alpha ) }{2 \sqrt{6}}+\frac{i \eta _3 \sin (2 \alpha )}{2 }
- \frac{i  \varphi _0 \sin (2 \alpha ) }{2 \sqrt{2}} \nonumber \\  &-& \frac{i}{2 \sqrt{2}} \big(\sin (2 \alpha )+2 \sin (\alpha )\big) \Lambda_{0} 
- \frac{i}{2 \sqrt{2}} \big(\sin (2 \alpha )-2 \sin (\alpha )\big) \Lambda_{0}^{*} \bigg) \overline{b_L} b_R \nonumber \\ 
&-& \left( \sqrt{2} Y_{\text{b2}} H_{+} - i ~ Y_{\text{b1}} \sin (\alpha )  \left(\Lambda _+  -  \varphi _+  \right) \right) \overline{t_L} b_R\,. 
\end{eqnarray}
A loop of top and bottom quarks generates a potential terms for Goldstone fields, which reads
\begin{eqnarray}
\mathcal{V}_{t} &=& f^{4}C_{t}\bigg[\sum_{\delta}\bigg|Y_{t1}Tr[P_{t1}^{\delta}.\Sigma(x)]+Y_{t2}Tr[P_{t2}^{\delta}.\Sigma(x)]\bigg|^{2}  \nonumber \\ &+& \sum_{\delta}\bigg|Y_{b1}Tr[P_{b1}^{\delta}.\Sigma(x)]+Y_{b2}Tr[P_{b2}^{\delta}.\Sigma(x)]\bigg|^{2}\bigg]\,,
\end{eqnarray}
with $C_t$ being a from factor from the strong dynamics, including the loop factor.

To understand the effect of this operator on the vacuum alignment, we need to expand it in the pNGB fields, and ensure that no tadpoles for the pNGBs remains. This would ensure that the vacuum defined by $\alpha$ is a minimum, and fix the value of the angles $\alpha$ and $\beta$.
The tadpoles read:
\begin{eqnarray}
\mathcal{V}_{t} &\supset& 
f^{4} C_{t}\bigg[
2 \left( \mathcal{I}(Y_{t1} Y_{t2}^{*} ) - \mathcal{I}(Y_{b1} Y_{b2}^{*} ) \right)\frac{ i  A_0\sin (2 \alpha )}{\sqrt{2} f} + \frac{h (|Y_{t1}|^{2}+|Y_{b1}|^{2}) \sin (2 \alpha ) \cos (2 \alpha )}{\sqrt{2} f} \nonumber \\ 
&+& 2 \left(\mathcal{R}(Y_{t1} Y_{t2}^{*}) + \mathcal{R} (Y_{b1} Y_{b2}^{*} ) \right) \frac{H  \sin (2 \alpha ) \cos (\alpha )}{ \sqrt{2} f} + i ~  (|Y_{t1}|^{2} -|Y_{b1}|^{2})\sin (\alpha ) \sin (2 \alpha ) \Lambda _{0} \nonumber \\
&-&  i ~ (|Y_{t1}|^{2}-|Y_{b1}|^{2}) \sin (\alpha ) \sin (2 \alpha ) \Lambda _{0}^{*}
- 2 \left( \mathcal{I}(Y_{t1} Y_{t2}^{*} )+ \mathcal{I}( Y_{b1} Y_{b2}^{*} ) \right) \frac{ i \eta _2 \sin (\alpha ) \sin (2 \alpha )}{ \sqrt{2}} \nonumber \\
&+& \frac{1}{2}( |Y_{t1}|^{2} +|Y_{b1}|^{2})\sin ^2(2 \alpha ) \bigg] \label{yukawa}
\end{eqnarray}
where $\mathcal{R}$  and $\mathcal{I}$ indicate the real and Imaginary parts respectively. 
The potential generates tadpoles for many neutral pNGBs: some of them, however, are easily accounted for.
The tadpole for the Higgs-like state $h$ vanishes once the value of $\alpha$ is chosen, and for $C_t < 0$ (as expected for fermion loops) it leads to $\alpha = \pi/2$. The $H$ vanishes by fixing $\beta$, which is implicitly contained in the Yukawa couplings.  Note that if the Yukawa couplings in the up and dow sectors are aligned (as required to avoid flavour changing neutral currents), the minimum can be determined by the vanishing of the second Yukawa, $Y_{t2} = Y_{b2} = 0$, leading to:
\begin{equation}
\tan \beta = \frac{\tilde{Y}_{t2}}{\tilde{Y}_{t1}} = \frac{\tilde{Y}_{b2}}{\tilde{Y}_{b1}} \,.
\end{equation}
In this aligned scenario, the dark matter parity is preserved, as the effect of $\beta$ completely vanish, i.e. we get one inert copy of Higgs doublet.
The tadpoles for $A_{0}$ and $\eta_{2}$ vanish in the aligned scenario (consistently with the $\mathbb{Z}_2$ parity), but can also be removed once the Yukawas are assumed CP-conserving (i.e. real). The only problematic tadpole arises for $\mathcal{I} (\Lambda_0)$: it is proportional to difference of top and bottom physical Yukawa couplings, i.e. to the violation of custodial symmetry in the quark masses. As such, it cannot be removed, but it will generate a dangerous custodial-violating vacuum misalignment, which is strongly constrained by the $\rho$ parameter.

This is the main reason why, in the remaining of the paper, we will focus on fermion partial compositeness that allows to avoid this issue without fine tuning~\cite{Agugliaro:2018vsu}.


\subsubsection*{Hyper-fermion mass}

The hyper-fermion mass breaks explicitly the global symmetry $SU(6)$, thus it will generate masses (and a potential for the pNGBs), in the form 
\begin{eqnarray}
\mathcal{V}_{m} = - \frac{B f^3}{2 \sqrt{2}} Tr[M^\dagger \cdot \Sigma] + \mbox{h.c.} 
\end{eqnarray}
where $M$ is the mass term in eq.~\eqref{eq:HFmass} and $B$ is a form factor that can be determined on the lattice, if an underlying theory is specified.
It is convenient to define 
\begin{equation}
m = \frac{m_1+ m_2}{2}\,, \quad \delta= \frac{m_2- m_1}{m_2 + m_1}\,,
\end{equation}
so that in the limit $\delta \to 0$ the full $SO(6)$ is recovered.
Expanding the potential up to linear terms in the fields, we obtain
\begin{equation}
\mathcal{V}_{m}  \supset  -B \bigg[ \mbox{const} + \sqrt{2} f^3 m \cos (2 \alpha )  - 2 \sqrt{2} h f^2 m \sin (2 \alpha ) \bigg]\,.
\end{equation}
We see that the only tadpole involves the Higgs $h$ and that it will vanish once the angle $\alpha$ is fixed at the minimum.
The  contribution to the pNGB masses at the quadratic order can be found in the Appendix~\ref{a:HFmass}, where the relevant cubic interactions are listed as well, which  conserve both  the DM $\mathbb{Z}_2$  parity and the CP property.

\subsubsection*{Gauge loops}

The contribution of gauge loops to the pNGB potential can be written as~\cite{Peskin:1980gc, Preskill:1980mz}:
\begin{equation}
\mathcal{V}_{g} = \frac{C_{g}f^{4}}{4}\{g^{2}Tr[T_{L}^{i}\Sigma(T_{L}^{i}\Sigma)^{*}]+g^{'2}Tr[Y\Sigma(Y\Sigma)^{*}]\} \,,
\end{equation}
where $C_{g}$ is an low energy constant fixed by Lattice simulations. Expending up to linear terms in pNGB fields, we get
\begin{eqnarray}
\mathcal{V}_{g}&=& \frac{C_{g} g^2 f^{4}}{4}  \left(3+ \tan^2 \left(\theta_W \right) \right) \bigg(- \cos (\alpha)^{2} + h ~\frac{ \sin (2 \alpha ) }{ f} + ....  \bigg) 
\end{eqnarray}
It is observed that the only contribution from gauge potential would lead to minimum at $\alpha=0$ for $C_{g} > 0$ ($C_{g}$ remains positive for any vector like confining gauge theory \cite{Witten:1983ut}). Hence, the radiative contribution from gauge fields always tends to align the vacuum
in the direction that preserves the EW gauge symmetry.  At higher order, the gauge potential contributes to the pNGB mass and cubic interaction terms, which are described in Appendix.\ref{a:Gaugepot}.

\section{Top Partial Compositeness}\label{PartialComp}

We showed that the potential from the bilinear operator always generates a tadpole term for the CP-even component, $\mathcal{I}m (\Lambda_0)$, in the complex triplet, as illustrated in eq.~\eqref{yukawa}. For this reason we analyse the case of top quark mass generated by partial compositeness, where the bilinear couplings are replaced by linear couplings of $q_L$ and $t_R^c$ to fermionic operators in the strong sector. In an underlying theory, this implies an extension of the model with a second species of hyper-fermions~\cite{Ferretti:2013kya} or scalars carrying Hypercolor charge~\cite{Sannino:2016sfx}. 

We will analyse the potential by writing effective operators containing the spurions from partial compositeness: as shown in Ref.~\cite{Agugliaro:2018vsu} for $SU(5)/SO(5)$ coset that shares the same triplet structure as our model.  We thus  couple the top fields to composite operators transforming as the adjoint of  the global $SU(6)$, so that  the single trace operator with 2 sigma fields and 2  adjoint  spurions can generate quartic interaction for the Higgs potential.  For a correctly defined vacuum,  the tadpole terms for all pNGB fields should be removed after we choose appropriate combinations of spurions. Particularly for the fields developing VEVs,  the corresponding tadpole terms  will automatically vanish at the minimum of the potential.  After studying the pattern of couplings allowed by the absence of tadpoles, we will briefly study the possible underlying models leading to such scenario.

\subsection{Adjoint Representation}

In order to assist  the  embedding into the $SU(6)$ representation, we need to assume the existence of an additional $U(1)_X$ charge, so that the electric charge of the top partners is $Q = S_{L,3} + S_{R, 3} + X$, with $X = \frac{2}{3}$. 
The origin of this charge will be clear in the underlying models detailed in the next section.
For  $SU(6)$, the adjoint representation is a {\bf 35}-plet,  which decomposes under the unbroken $SO(6)$ as ${\tt 20} \oplus {\tt 15}$, respectively the symmetric and adjoint (anti-symmetric).  Their respective decomposition under the $SU(2)_L\times SU(2)_R$ custodial symmetry is:
\begin{eqnarray}
{\tt 20} &\to& (3,3) \oplus (2,2) \oplus (2,2) \oplus 3 \times (1,1) \,,  \nonumber  \\
{\tt 15} &\to&   (2,2)\oplus (2,2)\oplus (1,3)\oplus (3,1) \oplus (1,1)\,.
\end{eqnarray}
In the following we will write the operators in an explicitly $SU(6)$ invariant way, which was shown to be equivalent to writing $SO(6)$ invariant ones in Ref.~\cite{Agugliaro:2018vsu}. Nevertheless, as we will see later, it is important to distinguish the $SO(6)$ representation the spurions belong to.

Thus, $q_L$ can be embedded in any of the 4  bi-doublets $(2,2)$, while $t_R$ will be put in the 4 singlets $(1,1)$ and in the $T_R^3=0$ component of  the $SU(2)_R$ triplet $(1,3)$. Hence, for the left-handed quark doublet, the four possible embeddings read
\begin{equation} \label{eq:DL}
\footnotesize
\begin{array}{ccc}
D_{L,A}^{1} =  \left(
\begin{array}{cccc|cc}
 &  &  &  & 0 & 0 \\
 &  &  &  & 0 & 0 \\
 &  &  &  & \frac{t_{L}}{\sqrt{2}} & 0 \\
&  &  &  & \frac{b_{L}}{\sqrt{2}} & 0 \\ \hline
-\frac{b_{L}}{\sqrt{2}} & \frac{t_{L}}{\sqrt{2}} & 0 & 0 &  &  \\
0 & 0 & 0 & 0 &  & 
\end{array} \right)\,,
&\phantom{xx}&
D_{L,S}^{2} = \left(
\begin{array}{cccc|cc}
&  &  &  & 0 & 0 \\
&  & &  & 0 & 0 \\
&  &  &  & \frac{t_{L}}{\sqrt{2}} & 0 \\
&  &  &  & \frac{b_{L}}{\sqrt{2}} & 0 \\ \hline
\frac{b_{L}}{\sqrt{2}} & -\frac{t_{L}}{\sqrt{2}} & 0 & 0 &  &  \\
0 & 0 & 0 & 0 & & 
\end{array} \right)\,, \\
\phantom{\frac{1}{2}} & & \\
D_{L,A}^{3} = \left(
\begin{array}{cccc|cc}
&  &  &  & 0 & 0 \\
&  &  &  & 0 & 0 \\
&  & &  & 0 & \frac{t_{L}}{\sqrt{2}} \\
&  & &  & 0 & \frac{b_{L}}{\sqrt{2}} \\ \hline
0 & 0 & 0 & 0 &  &  \\
-\frac{b_{L}}{\sqrt{2}} & \frac{t_{L}}{\sqrt{2}} & 0 & 0 &  & 
\end{array} \right)\,,
&\phantom{xx}&
D_{L,S}^{4} = \left(
\begin{array}{cccc|cc}
&  &  &  & 0 & 0 \\
& &  &  & 0 & 0 \\
&  & &  & & \frac{t_{L}}{\sqrt{2}} \\
&  &  &  & 0 & \frac{b_{L}}{\sqrt{2}} \\ \hline
0 & 0 & 0 & 0 &  &  \\
\frac{b_{L}}{\sqrt{2}} & -\frac{t_{L}}{\sqrt{2}} & 0 & 0 &  & 
\end{array} \right)\,.
\end{array}
\end{equation}	
In the above basis, we can see that the two spurions in the right column, with $i=2,4$, belong to the symmetric representation of $SO(6)$ as $D_{L, S}^{2,4}\cdot\Sigma_{\rm EW} = \Sigma_{\rm EW}\cdot  (D_{L, S}^{2,4})^T$, while the ones on the left, with $i=1,3$, belong to the adjoint (anti-symmetric) as $D_{L, A}^{1,3}\cdot \Sigma_{\rm EW} = -\Sigma_{\rm EW}\cdot (D_{L, A}^{1,3})^T$.

The right-handed singlet top can be embedded in five possible ways: the  3 singlets $(1,1)$ in the $\tt 20$ symmetric and the  $(1,1)$  and  $(1,3) $ in the $\tt 15$ antisymmetric. For the latter,  they correspond  to the generator of the $U(\beta)$ rotation and  the $S_{R,3}$ generator respectively. In analogy to the left-handed scenario, we can find that $D_{R, S}^{1,2,3}\cdot \Sigma_{\rm EW} = \Sigma_{\rm EW}\cdot (D_{R, S}^{1,2,3})^T$ and $D_{R, A}^{4,5}\cdot \Sigma_{\rm EW} = -\Sigma_{\rm EW}\cdot (D_{R, A}^{4,5})^T$.
Explicitly, we have:
\begin{equation}\label{eq:DR}
\begin{array}{c}
D_{R,S}^{1} = t_{R}  \left(
\begin{array}{ccc}
\frac{1}{2 \sqrt{3}} \mathbbm{1}_2 &  &  \\
 & \frac{1}{2 \sqrt{3}} \mathbbm{1}_2 &  \\
  &  & -\frac{1}{\sqrt{3}} \mathbbm{1}_2 
\end{array} \right)\,, \\ \phantom{\frac{1}{2}} \\
D_{R,S}^{2} = t_{R}\left(
\begin{array}{ccc}
0_2 & & \\
 & 0_2 &  \\ &  & \frac{1}{\sqrt{2}} \sigma_3
\end{array} \right)\,, \quad
%
D_{R,S}^{3} = t_{R} \left(
\begin{array}{ccc}
0_2 &  &   \\
 & 0_2 &  \\
  &  & \frac{1}{\sqrt{2}} \sigma_1
\end{array} \right)\,, \\ \phantom{\frac{1}{2}} \\
D_{R,A}^{4} = t_{R} \left(
\begin{array}{ccc}
\frac{1}{2} \mathbbm{1}_2 &  &    \\
&  -\frac{1}{2} \mathbbm{1}_2 &    \\
  &  &  0_{2}
\end{array} \right)\,,
\quad
D_{R,A}^{5} = t_{R} \left(
\begin{array}{ccc}
0_2 & &  \\
& 0_2 &  \\
&  & \frac{i }{\sqrt{2}} \sigma_2 \\
\end{array} \right)\,.
\end{array}
\end{equation}
In general, the spurions $D_{L}$ and $D_{R}$ are linear combinations of the above matrices:
\begin{eqnarray}
D_{L} &=& Q_{A1} D_{L,A}^{1} + Q_{S2} D_{L,S}^{2} + Q_{A3} D_{L,A}^{3} + Q_{S4} D_{L,S}^{4}\,,\\
D_{R} &=& R_{S1} D_{R,S}^{1} + R_{S2} D_{R,S}^{2} + R_{S3} D_{R,S}^{3} + R_{A4} D_{A,S}^{4} + R_{A5} D_{R,A}^{5} \,,
\end{eqnarray}
where, without loss of generality, all the coefficients $Q_{A/Si}$ and $R_{A/Si}$ are assumed to be complex. The top mass is generated by the operator~\cite{Golterman:2017vdj,Alanne:2018wtp}:
\begin{multline}
C_t f ~ Tr [\bar{D}_{L}^{T}.\Sigma^{\dagger}.D_{R}.\Sigma] 
\supset i \frac{C_t f }{4 \sqrt{2}}  ~ \bigg(2 \sin (2 \alpha ) \left(\sqrt{2} R_{\text{A4}}
   Q_{\text{S2}} {}^*+2 R_{\text{A5}} Q_{\text{S4}}{}^*  -  2  R_{\text{S3}}  Q_{\text{A3}}{}^*\right)  +  \\
    \sin (4 \alpha )
    \left(\sqrt{6} R_{\text{S1}}-2 R_{\text{S2}}\right) Q_{\text{A1}}{}^* \bigg) \overline{t_L} t_R\,.
\end{multline}
The formula above shows that the top mass is only generated by the product of spurions belonging to different representations of $SU(6)$, thus we can envision two minimal options: I) left-handed $q_L$ in the anti-symmetric $\tt 15$ and right-handed $t_R$ in the symmetric $\tt 20$, II) vice-versa.

The pNGB potential generated by top interactions can be constructed by using the same spurions, stripped of the fields and with the SM indices properly contracted~\cite{Golterman:2017vdj,Agugliaro:2018vsu}. For the adjoint, the leading order potential contains two operators: 
\begin{eqnarray}
\mathcal{V}_{\rm top} = \frac{C_{LL} f^4 }{4}\,\,  Tr[\bar{D}_{L}^{T}\cdot \Sigma^{\dagger}\cdot D_{L}\cdot \Sigma] + \frac{C_{RR} f^4 }{4}\,\, Tr[\bar{D}_{R}^{T}\cdot \Sigma^{\dagger}\cdot D_{R}\cdot \Sigma] \,.
\end{eqnarray}
To simplify the analysis, we will focus on the two minimal scenarios giving rise to the top mass, but our conclusions will be general.
Focusing on the scenario I), where  
the left-handed top and bottom are in the anti-symmetric and the right-handed top in the symmetric, the potential up to tadpole terms generated by the left-handed spurions reads
\begin{eqnarray}
\mathcal{V}_{\rm top\; LL}^{(I)} &=& -  \frac{C_{LL} f^4}{8} \left( \cos (4 \alpha ) \left |
   Q_{A1}\right |^2 +   \cos (2 \alpha ) (3 \left |
   Q_{A1}\right |^2+\left | Q_{A3}\right | {}^2 \right)\nonumber \\
& + & \frac{C_{LL} f^{3}}{4}  \bigg(  h \sin (2 \alpha ) \left((4 \cos (2 \alpha )+3) \left |
   Q_{A1}\right |^2+\left | Q_{A3}\right |^2\right) \nonumber \\ 
  & +&  2 H \sin (\alpha ) \left(\cos(2\alpha )+2 \right) \left(Q_{A1}
   \left(Q_{A3}\right)^*+Q_{A3} \left(Q_{A1}\right)^*\right) \nonumber \\ 
   &+& i  A_0 \sin (2 \alpha ) \left(Q_{A1} \left(Q_{A3}\right)^*-Q_{A3} \left(Q_{A1}\right)^*\right)  \bigg)\,,
\end{eqnarray}
while for the right-handed spurions we have:
\begin{eqnarray}
\mathcal{V}_{\rm top\; RR}^{(I)} &=& \frac{C_{RR} f^4}{4} \left(\cos (2 \alpha ) \left | R_{S3}\right |^2 +  \frac{1}{8}\cos (4 \alpha ) \left | \sqrt{3} R_{S1} - \sqrt{2} R_{S2} \right |^2 \right)  \nonumber \\ 
  &-& \frac{C_{RR} f^3}{2} \bigg(  h  \sin (2 \alpha ) \left | R_{S3}\right |^2 +  \frac{1}{4} h  \sin (4 \alpha ) \left | \sqrt{3} R_{S1} - \sqrt{2} R_{S2} \right |^2  \nonumber \\  &-&  \sqrt{6} H \sin (\alpha ) \cos ^2(\alpha ) \left(R_{S1}
   \left(R_{S3}\right)^*+R_{S3} \left(R_{S1}\right)^*\right) \nonumber \\ &-& 2
  H \sin ^3(\alpha ) \left(R_{S3} \left(R_{S2}\right)^*+R_{S2} \left(R_{S3}\right)^*\right)  \bigg)\,.
\end{eqnarray}
Requiring  the second Higgs to be inert, the tadpole conditions are:
\begin{eqnarray}
  Q_{A3} \left(Q_{A1}\right)^*-Q_{A1}
   \left(Q_{A3}\right)^* &=& 0 ,  \quad   Q_{A3} \left(Q_{A1}\right)^*+ Q_{A1}
   \left(Q_{A3}\right)^* = 0\,,    \nonumber \\    R_{S1}
    \left(R_{S3}\right)^*+ R_{S3} \left(R_{S1}\right)^* &=& 0 , \quad R_{S3}
   \left(R_{S2}\right)^* + R_{S2} \left(R_{S3}\right)^* =0  \,.
\end{eqnarray}
There are two solutions that preserve a non-vanishing top mass: however, the choice $Q_{A1} = R_{S1} = R_{S2} = 0$ would lead to a potential $\mathcal{V}_{\rm top} \propto \cos (2\alpha)$ that allows for a minimum at $\alpha = 0$ or $\pi/2$. The only viable solution is therefore $R_{S3} = Q_{A3}=0$, for which the potential contains both  $\cos (2 \alpha)$ and $\cos (4\alpha)$ terms. Interestingly we find out  that only after imposing the tadpole vanishing condition,  the cubic pNGB interactions induced by the top spurion potential will conserve the DM parity defined in section~\ref{parities}.

For the scenario II), where the left-handed top and bottom are in the symmetric while the right-handed top in the anti-symmetric, the potential generated by the left-handed spurion reads
\begin{eqnarray}
\mathcal{V}_{\rm top\; LL}^{(II)} &=& \frac{C_{LL}  f^4}{4}  \left( \cos (2 \alpha ) \left( \frac{3}{2} \left | Q_{S2}\right |^2+ \frac{1}{2} \left | Q_{S4}\right |^2\right) \right)  \nonumber \\ & - & \frac{C_{LL}  f^3}{4}  \bigg(  h \sin (2 \alpha ) \left(3 \left | Q_{S2}\right |^2+\left | Q_{S4}\right |^2\right)  \nonumber \\ &+&   2 H  \sin (\alpha )  \left(Q_{S2} \left(Q_{S4}\right)^*+Q_{S4}\left(Q_{S2}\right)^*\right)  \nonumber \\
    &- &  i  A_{0} \sin (2 \alpha ) \left(Q_{S2} \left(Q_{S4}\right)^*-Q_{S4} \left(Q_{S2}\right)^*\right) \bigg) \,,
\end{eqnarray}
while for the right-handed spurion:
\begin{eqnarray}
\mathcal{V}_{\rm top\; RR}^{(II)} &=& - \frac{C_{RR}  f^4}{4}   \left(  \cos (2 \alpha ) \left(\frac{1}{2} \left | R_{A4}\right |^2+ \left | R_{A5}\right |^2\right) \right) \nonumber \\ &+&  \frac{C_{RR}  f^3}{4}   \bigg( h \sin (2 \alpha )  \left(\left | R_{A4}\right |^2+2 \left | R_{A5}\right |^2\right)   \nonumber \\ &+& i A_0 \sqrt{2}  \sin (2 \alpha ) \left(R_{A5} \left(R_{A4}\right)^*-R_{A4}\left(R_{A5}\right)^*\right)  \bigg)\,.
\end{eqnarray}
Requiring the second doublet to be inert, the general  tadpole conditions are:
\begin{eqnarray}
& R_{\text{A5}} \left(R_{\text{A4}}\right){}^*-R_{\text{A4}}
   \left(R_{\text{A5}}\right)^* =0\,, & \nonumber  \\
    Q_{\text{S2}}
  &  \left(Q_{\text{S4}}\right){}^*-Q_{\text{S4}} \left(Q_{\text{S2}}\right){}^* =0 , \quad Q_{\text{S2}}
   \left(Q_{\text{S4}}\right){}^* + Q_{\text{S4}} \left(Q_{\text{S2}}\right){}^* =0\,, &
\end{eqnarray}
which can be fulfilled by requiring $Q_{S2}=0$ or $Q_{S4}=0$,  plus the condition of $R_{A4}$ and $R_{A5}$ with the same phase. However, in either case the potential will be $\propto \cos (2 \alpha)$, thus not allowing for a physically acceptable vacuum.

In summary we find that the only viable choice  to preserve the Dark Matter parity requires that the  non-vanishing spurions are: $Q_{A1}$, $R_{S1}$ and $R_{S2}$.  Hence the top mass operator, expanded up to linear order in the pNGB fields, reads:
\beq \label{eq:topmass}
C_t f ~ Tr [\bar{D}_{L}^{T}.\Sigma^{\dagger}.D_{R}.\Sigma] 
\supset i\ m_{\rm top} \left( 1+  \frac{4}{f} \frac{\cos(4\alpha)} {\sin(4\alpha)}  h + \mathcal{O}(1/f^2) \right) \overline{t_L} t_R  
\eeq
with
\beq
m_{\rm top} = \frac{C_t f }{4}   \left(\sqrt{3} R_{S1}- \sqrt{2}  R_{S2}\right)  Q_{A1}^\ast\  \sin (4 \alpha )\,.
\eeq

\subsection{Underlying theories based on fermion-gauge interactions} \label{sec:underlying}
 
 Models of composite Higgs with fermion partial compositeness can be described in terms of underlying gauge-fermion theories~\cite{Ferretti:2013kya,Barnard:2013zea}. These theories consist of a confining HC gauge group, and two species of hyper-fermions belonging to different representations of the HC group. Requiring that the theories confine without falling inside the infra-red conformal window strongly limits the number of available HC groups~\cite{Ferretti:2016upr,Belyaev:2016ftv} and only allows for generating (light) partners for the top quark. Here we will identify models from the list in Ref.~\cite{Ferretti:2013kya} that can be extended to accommodate the $SU(6)/SO(6)$ coset in the Higgs sector.
 
\begin{table}[tb] 
\begin{center}
\begin{tabular}{l|c|c|c|c|}
  & $\mathcal{G}_{\rm HC}$ & $SU(3)_c$ & $SU(2)_L$ & $U(1)_Y$ \\
\hline
$\psi_{1/2}$ & \multirow{4}{*}{$R_1$} & $\bf 1$ & $\bf 2$ & $1/2$ \\
$\psi_{-1/2}$ &  & $\bf 1$ & $\bf 2$ & $-1/2$ \\
$\psi_{0}$ &  & $\bf 1$ & $\bf 1$ & $0$ \\
$\tilde \psi_{0}$ &  & $\bf 1$ & $\bf 1$ & $0$ \\
\hline
$\chi_t$ & $R_2$  & $\bf 3$ & $\bf 1$ & $2/3$ \\
$\tilde \chi_t$ & $\overline{R}_2$  & $\bf \bar{3}$ & $\bf 1$ & $-2/3$ \\
\hline
\end{tabular} \end{center}
\caption{Template for the hyper-fermions giving rise to $SU(6)/SO(6)$ and partial compositeness for the top quark. Under the HC interactions, $R_1$ is a real representation, while $R_2$ can be complex, real or pseudo-real. The representations allows for a bound state of two $\psi$ with one $\chi$.} \label{tab:Umodels}
\end{table} 
 
 The template for the models is illustrated in Table~\ref{tab:Umodels}, where we list the minimal set of hyper-fermions and their quantum numbers under the SM gauge symmetries. The table clearly shows the origin of the $U(1)_X$ we assumed in the previous section, as it is simply the hypercharge carried by the $\chi$ hyper-fermions. To determine the allowed $\mathcal{G}_{\rm HC}$ and representation, we will start from the list in Ref.~\cite{Ferretti:2013kya}, add one additional $\psi$ and verify that the model is still outside of the conformal window. For the latter, we made use of the conjectured all-orders beta function of Refs~\cite{Pica:2010mt}: this would be a very conservative estimate as this method is the most constraining one~\cite{Dietrich:2006cm,Sannino:2009aw,Ryttov:2009yw}.
We are thus left with the following choices:
\beq
\mathcal{G}_{\rm HC} = SO(7)\;\; \mbox{or}\;\; SO(9)\;\; & \mbox{with} & R_1 = \mbox{\bf Spin}\,, \quad R_2 = \mbox{\bf F}\;\; \mbox{(real)}\,.
\eeq 
Thus, for both models the complete global symmetry of the strong dynamics is $SU(6) \times SU(6)_\chi \times U(1)/SO(6) \times SO(6)_\chi$, where $U(1)_X$ is embedded in $SO(6)_\chi$. The global $U(1)$, which is spontaneously broken by the hyper-fermion condensates, leads to a potentially light pseudo-scalar, whose properties have been extensively studied in Refs~\cite{Belyaev:2016ftv,Cacciapaglia:2019bqz}.
 
 Defining this underlying theory allows us to write explicitly the 4-fermion interactions that give rise to the 3 spurions we need in the minimal model with Dark Matter. For the left-handed top doublet, we have:
 \beq
 D_{L,A}^1 \to \frac{y_L}{\Lambda_t^2}\ \bigg( (q_L \psi_{1/2}) (\overline{\chi}_t \overline{\psi}_0) - (q_L \psi_{0}) (\overline{\chi}_t \overline{\psi}_{-1/2}) \bigg)\,.
 \eeq
 For the two right-handed top spurions, we have
\beq
 D_{R,S}^1 &\to& \frac{y_{R1}}{\Lambda_t^2}\ \bigg(  (t_R^c \psi_{1/2})(\overline{\tilde{\chi}}_t \overline{\psi}_{1/2}) + (t_R^c \psi_{-1/2})(\overline{\tilde{\chi}}_t \overline{\psi}_{-1/2}) + \nonumber \\
  & & \phantom{xxxxx} - 2 \left(  (t_R^c \psi_0)(\overline{\tilde{\chi}}_t \overline{\psi}_0) + (t_R^c {\tilde \psi_0})(\overline{\tilde{\chi}}_t \overline{\tilde \psi}_0)  \right)\bigg)\,, \nonumber \\
 D_{R,S}^2 &\to& \frac{y_{R2}}{\Lambda_t^2}\ \bigg( (t_R^c \psi_0)(\overline{\tilde{\chi}}_t \overline{ \psi}_0) - (t_R^c {\tilde \psi_0})(\overline{\tilde{\chi}}_t \overline{\tilde{\psi}}_0) \bigg)\,. 
 \eeq
 The $\psi$ fermions appearing in the above expressions follow the matrix expressions of the spurions in Eqs~\eqref{eq:DL} and~\eqref{eq:DR}. From the above expressions, we can deduce that all spurions can be made real by choosing the phase of the SM spinors $q_L$ and $t_R^c$, and the overall phase of the $\chi$ hyper-fermions. Also, a Baryon number charge can be assigned to both $\chi$'s, so that it remains preserved.
The origins of such couplings need to be traced to the ultra-violet completion of the model, where an extended HC group may be able to account for them. Note also that repeating the same mechanism for all SM fermions by introducing more $\chi$'s at different scales, can be used to complete the model to an ultra-violet interacting fixed point (asymptotic safety).

 As a final comment, we would like to recall that such underlying theory can be studied on the lattice, thus allowing to calculate the properties of the low energy states, including masses and couplings of the resonances. Studies of theories with two species of fermions are already available in the literature, but only based on $Sp(4)$~\cite{Bennett:2017ttu,Bennett:2017kga,Lee:2018ztv} and $SU(4)$~\cite{DeGrand:2016mxr,Ayyar:2017qdf,Ayyar:2018zuk,Ayyar:2018glg} gauge groups, while no result is available for the HC groups leading to our model.

\section{pNGB  mass spectrum}\label{pNGBMass}

We will proceed to  investigate the Higgs potential and mass spectrum of pNGBs. Following the discussion in the previous section, we will focus on the scenario where only 3 spurions in the top sector are non-vanishing: this ensures that the vacuum misalignment is described by a single angle $\alpha$, such that $v = f \sin \alpha$, the Dark Matter parity is preserved and a non trivial vacuum is allowed.
In the following, we will parametrise the 3 spurions as
\beq
Q_A = Q_{A1}\,, \quad R_S = \sqrt{3} R_{S1} - \sqrt{2} R_{S2}\,, \quad r R_{S} = R'_{S} = \sqrt{3} R_{S1} + \sqrt{2} R_{S2}\,.
\eeq
Note that $R_S$ and $R'_S$ combinations are not orthogonal, but we choose them for future convenience and simplicity of formulas.
As the coefficients $C_{LL}$ and $C_{RR}$ always multiply the spurions $Q_A$ and $R_S$, we can fix the latter $Q_A = R_S = 1$ without loss of generality.
The model has 4 additional parameters:
\beq
C_g\,, \quad Bm\,, \quad \delta\,, \quad \alpha\,.
\eeq
As we will see in the following section, two of them can be fixed by requiring that the misalignment reproduces the electroweak scale and the value of the Higgs mass. Thus, we are left with 5 free parameters describing the model.

\subsection{Higgs Potential and free parameters}

The pNGB potential, which is generated by gauge loops, the hyper-fermion mass term and top spurions, contains terms equivalent to the quadratic and quartic terms in the Higgs potential.  For this model, we can  conveniently  parametrise the Higgs potential as:
\begin{eqnarray}
\mathcal{V}(\alpha) = \frac{f^4}{4} \left(C_{2 \alpha}~ \cos (2 \alpha) + C_{4\alpha}~ \cos (4 \alpha) \right)\,,
\end{eqnarray}
where the $C_{4 \alpha}$ term is equivalent to the Higgs quartic term.  The minimum is given by $\cos (2 \alpha)|_{min} = - C_{2 \alpha} /(4~ C_{4 \alpha})$, thus a small misalignment angle would require $C_{2  \alpha} \lesssim -4~ C_{4  \alpha}$.

For our choice of top spurions, the two coefficients are given by
\begin{eqnarray}
C_{2 \alpha} &=& - \frac{1}{2} C_g\ ( 3 g_2^2 + g_1^2) - 4 \sqrt{2} \frac{Bm}{f} - \frac{3}{2}~ C_{LL}\,,   \\
C_{4 \alpha} & = & \frac{1}{8} \left( C_{RR}  - 4~C_{LL}\right)\,.
\end{eqnarray}
where $C_{LL/RR}$ characterise the spurion contribution and determined by the loop form factors as well as pre-Yukawa coefficients $y_{L,R}$ of top quarks from the partial compositeness, so that $C_{LL } \propto y_L^2$, $C_{RR} \propto y_R^2 $, and $m_t^2  \propto C_{LL} C_{RR} $. This result is  similar to the one obtained in the model $SU(5)/SO(5)$~\cite{Agugliaro:2018vsu}, except that  we get one more degree of freedom in the pNGB potential. In particular the additional right-hand spurion encoded in the  parameter $r$  does not appear in $C_{2 \alpha}$ and $C_{4 \alpha}$. The mass of the Higgs boson, which is proportional to the second derivative of the potential with respect to $\alpha$, is given by
\begin{eqnarray}
m_h^2 = -  \bigg( C_{2\alpha} \cos (2 \alpha) +4 ~ C_{4 \alpha} \cos (4 \alpha) \bigg) f^2 
= 4~ C_{4 \alpha} \sin^2 (2 \alpha)  f^2  = 16 ~C_{4 \alpha} \cos^2 ( \alpha)  v^2 \,.
\end{eqnarray}
Thus, we can re-express two of the 7 free parameters in terms of $m_h$ and $\sin \alpha = v/f$:
\begin{eqnarray}
C_{LL} &=& \frac{m_{h}^2 }{6 v^2} \frac{\cos (2 \alpha )}{\cos^2(\alpha )}  -\frac{8 \sqrt{2} B  m  \sin (\alpha )  }{3 v}   - \frac{ C_g g_{2}^2  \left(\cos
   \left(2 \theta _W\right)+2\right) }{3 \cos^2\left(\theta _W\right)}\,, \\
  C_{RR} &=& 4 ~ C_{LL} + \frac{m_h^2 }{2 v^2}  \sec^2(\alpha ) \,.
\end{eqnarray}
The remaining free parameters are thus $r$, $Bm$, $\delta$ and $C_g$. The last one, which is the coefficient of the gauge loops, can be computed on the lattice. In Ref.~\cite{Ayyar:2019exp}, in a model based on a HC group $SU(4)$, it was found that the coefficient is numerically very close to the value found in QCD and determining the mass splitting $m_{\pi^\pm} - m_{\pi^0}$. By comparing the formula of the QCD mass splitting~\cite{Das:1967it} with the typical pNGB mass contribution in our model (see Appendix~\ref{a:Gaugepot}):
\beq
m_{\pi^\pm} - m_{\pi^0} \approx \frac{3 e^2}{(4\pi)^2} \frac{C_{LR}}{f^2} \leftrightarrow \delta m_{\rm pNGB} \approx g^2 C_g f^2\,,
\eeq
with $C_{LR} \approx 30\ f^4$, we estimate
\beq \label{eq:Cgestimate}
C_g \approx \frac{1}{(4 \pi)^2} \frac{C_{LR}}{f^4} \approx 0.2\,.
\eeq
This value should be considered just as a reference point, and in the following we will consider it as a free parameter.

\subsection{pNGB mass spectrum}

\begin{figure}
	\centering 	{ \includegraphics[height=4.5 cm,  
		width=7.cm]{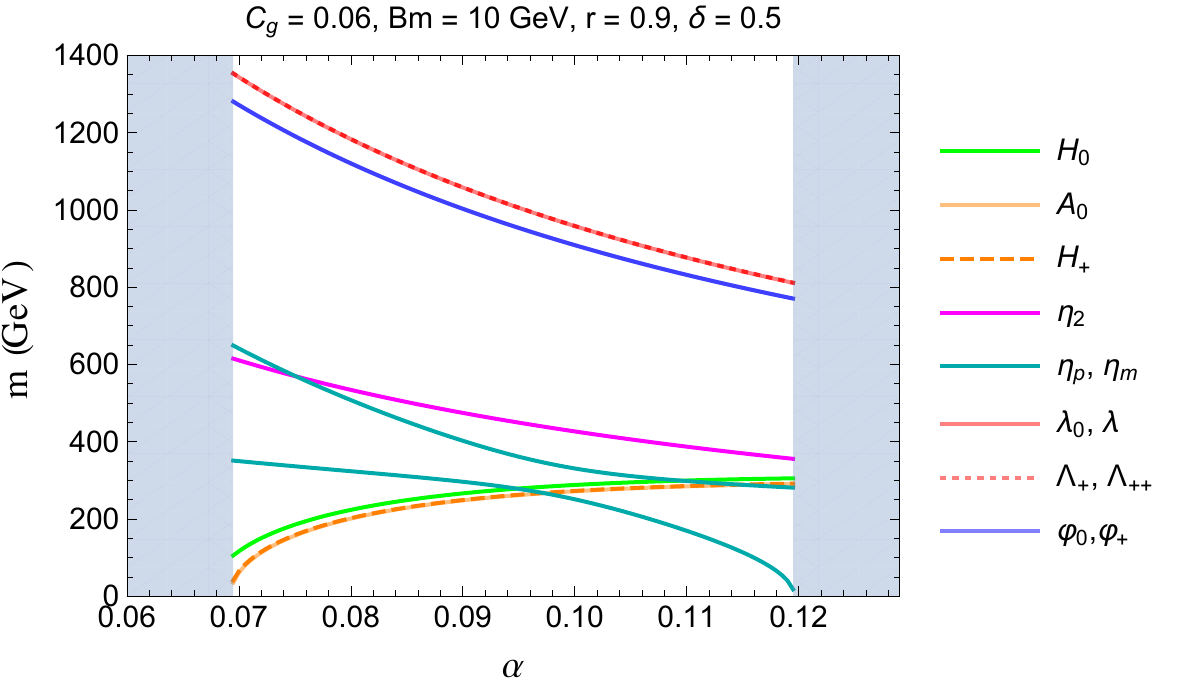}} \quad 
			 { \includegraphics[height=4.5 cm, 
		width=7.cm]{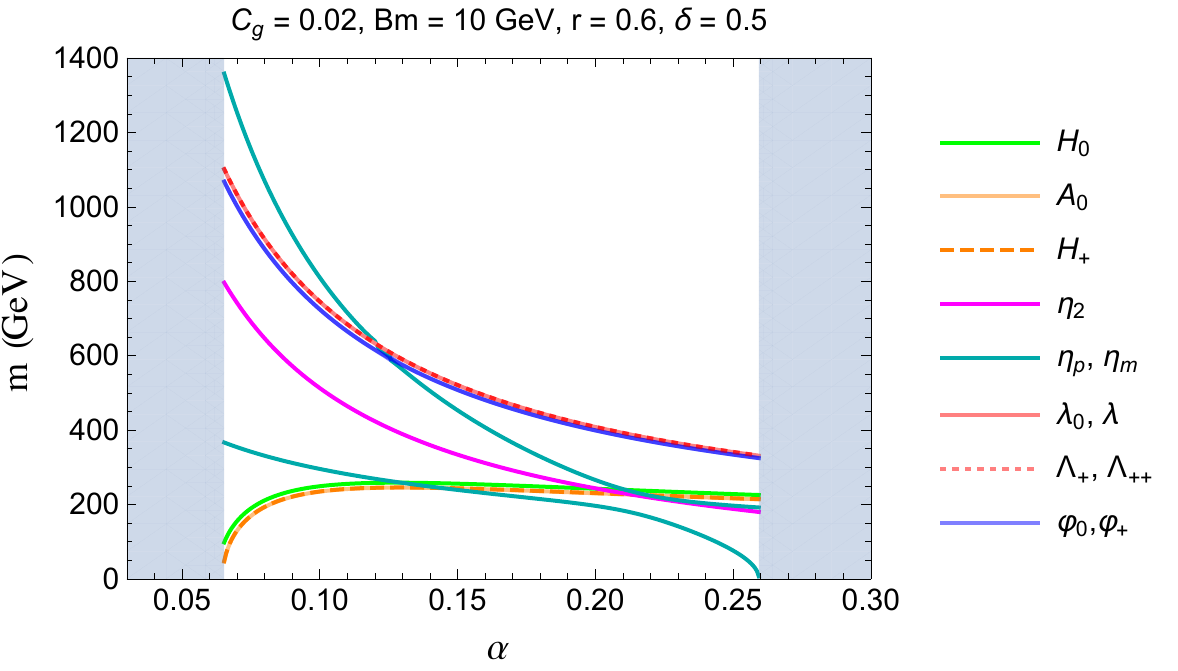}}
			{ \includegraphics[height=4.5 cm, 
		width=7.cm]{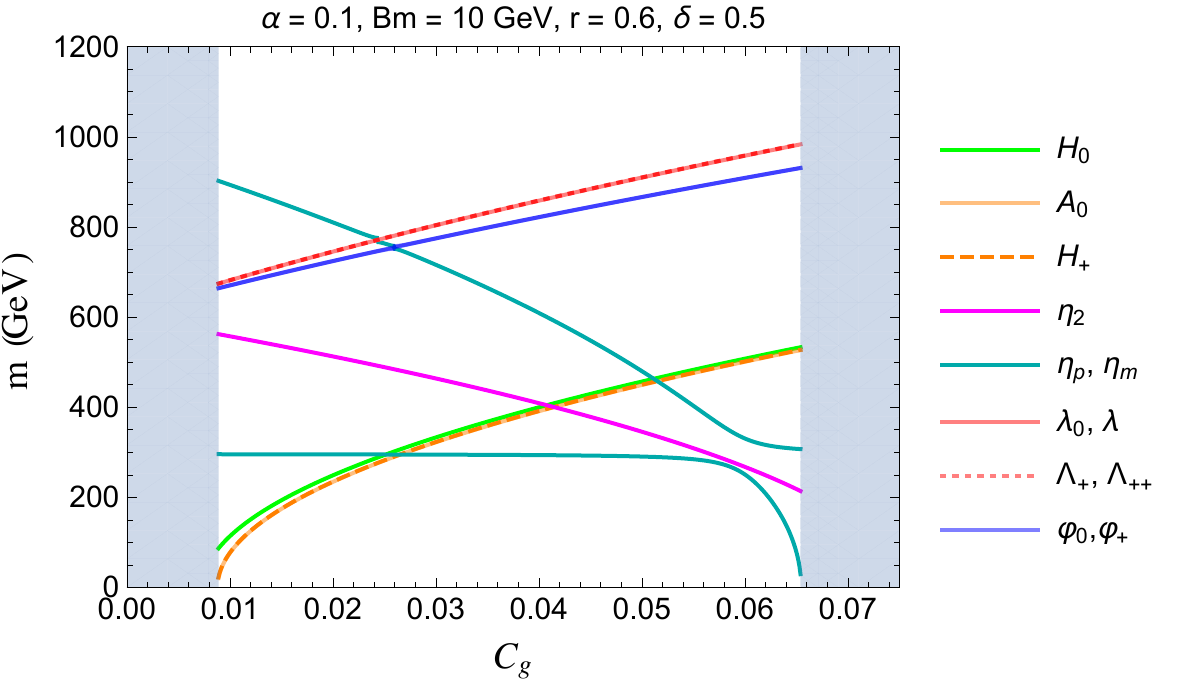}}  \quad 
			{ \includegraphics[height=4.5 cm, 
		width=7.cm]{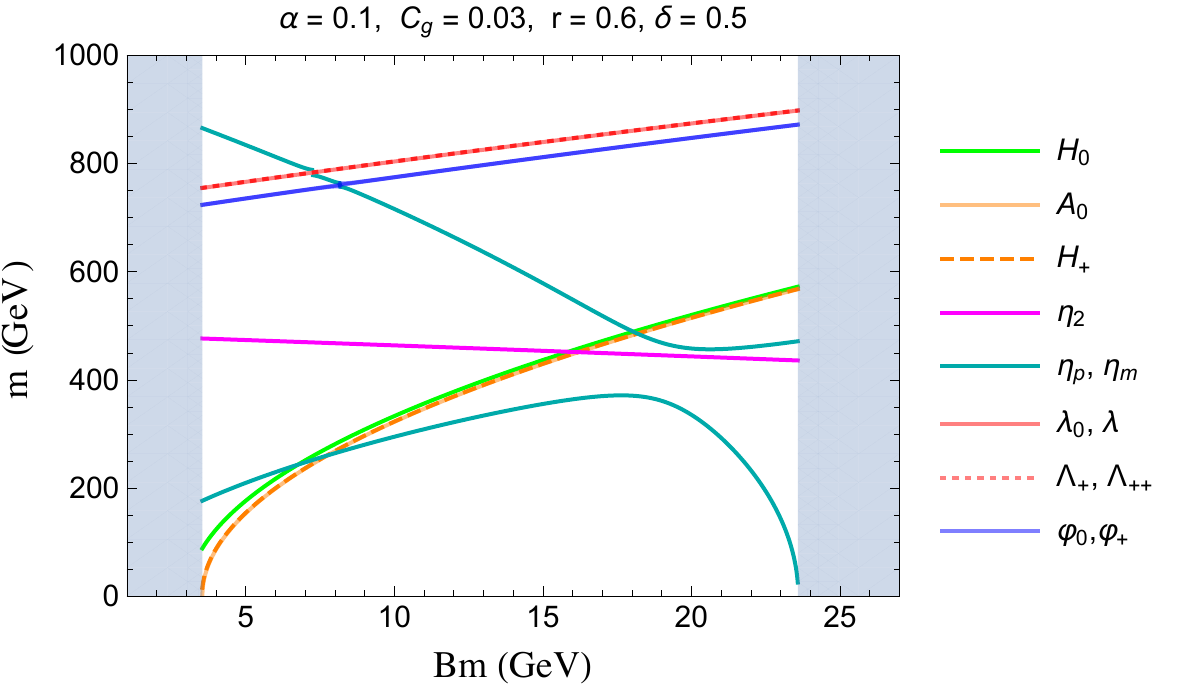}}
	\caption{The plots illustrate the mass spectra as  functions of $\alpha$, $C_g$ and $B m$ at selected  benchmark points. The $m_{A_0}$  becomes negative at the  lower limit of x-axis; however a singlet mostly aligned with $\eta_m$ turns to be tachyon for upper limit of x-axis.  The light blue area is theoretically unaccessible due to existence of  tachyons.  Either $m_{\eta_2}$ or  $m_{A_0}$  can be lighter depending on the parameter space.}
	\label{fig:mass1}
\end{figure}

We  will now analyse the masses of the non-Higgs pNGBs as a function of the remaining 5 free parameters: $\delta$, $\alpha$, $C_g$, $B m$,  and $r$.
First we observe that the parameter $\delta$, encoding the explicit breaking $SO(6) \to SO(4) \times SO(2)$ in the hyper-fermion mass term, will not generate any mass splitting within each multiplet of the custodial symmetry. On the other hand this $\delta$  does not enter into  masses of the second Higgs doublet and prevents the degeneracy between two CP-odd singlets, thus  crucial for the understanding of the mass patterns. The parameter $r$, encoding the additional spurion for the right-handed top, only enters the masses of  $\eta_2$, $H_{0, \pm}$, and $A_0$, not the spectra of  other pNGBs.  Following the analysis of the underlying models, we can safely assume that it is real. Results for the numerical spectra as a function of a single free parameter, fixing others  at selected  benchmark points, is shown in Figure~\ref{fig:mass1}, where the vertical light blue bands exclude the parameter space unaccessible due to the presence of a tachyonic state that is forbidden by vacuum consistency,   although  the lower limit band  might disappear for other  benchmark points.

We first focus on the $\mathbb{Z}_2$ odd states, which constitute candidates for Dark Matter in this model.
The analytic expressions for  the second Higgs doublet ($H_0$, $H_\pm$ and $A_0$) and the $\eta_2$ singlet can be expanded in the limit of small $\sin \alpha$, yielding
\begin{eqnarray}
m_{\eta_2}^2 &\simeq& \frac{1}{12} m_{h}^2 \left((-7 (r-2) r-3) \csc ^2(\alpha )+(r-16) r+7\right)  \nonumber \\ &+& \frac{8}{3} C_{g} M_{w}^2 \left((r-2) r \csc ^2(\alpha ) + 2r-1 \right) \left(\cos \left(2 \theta_W\right)+2\right) \sec ^2\left(\theta _W\right) \nonumber \\ &-&\frac{4}{3} \sqrt{2} B m (4 (r-2) r+\frac{3}{2}(2+\delta)) v \csc (\alpha )  \,, 
\end{eqnarray}

\begin{eqnarray}
m_{H_0}^2 &\simeq&\frac{1}{12} m_{h}^2 \left(\left(1-7 r^2\right) \csc ^2(\alpha )+r (r+14)-5\right) \nonumber \\ &+& \frac{4}{3} C_{g} M_{w}^2 \left( \left(2 r^2+1\right) \csc ^2(\alpha ) - 2 (2r -1) \right)\left(\cos
   \left(2 \theta _W\right)+2\right) \sec ^2\left(\theta _W\right) \nonumber \\&+& \frac{8}{3} \sqrt{2} B m \left(2 r^2+1\right) v \csc (\alpha )\,,
\end{eqnarray}

\begin{eqnarray}
m_{A_0}^2 &\simeq& \frac{1}{12} m_{h}^2 \left(\left(1-7 r^2\right) \csc ^2(\alpha )+r^2-3\right) 
\nonumber \\  &+& \frac{4}{3} C_{g} M_{w}^2 \left(\left(2 r^2+1\right) \csc ^2(\alpha )
   \left(\cos \left(2 \theta _W\right)+2\right) \sec ^2\left(\theta _W\right)-2 \tan
   ^2\left(\theta _W\right)\right)  \nonumber \\
   &+&\frac{8}{3} \sqrt{2} B m \left(2 r^2+1\right) v \csc (\alpha ) \,,
\end{eqnarray}

\begin{eqnarray}
m_{H_\pm}^2 &\simeq&\frac{1}{12} m_{h}^2 \left(\left(1-7 r^2\right) \csc ^2(\alpha )+r^2-3\right) \nonumber \\  &+&
\frac{4}{3} \text{Cg} M_{W}^2 \left( \left(2 r^2+1\right) \csc ^2(\alpha ) \left(\cos
   \left(2 \theta _W\right)+2\right)+1 \right) \sec ^2\left(\theta _W\right)   \nonumber \\ &+&
\frac{8}{3} \sqrt{2} B m \left(2 r^2+1\right) v \csc (\alpha ) \nonumber\\ 
&\simeq& m_{A_0}^2 + \frac{4}{3} C_{g} M_{W}^2 \left(3 \sec^2 \left(\theta _W\right) -2 \right)\,.
\end{eqnarray}
From the above equations, we can see that there is always a hierarchy $m_{H_\pm} \gtrsim m_{A_0}$, with mass splitting proportional to $C_g > 0$.  Furthermore, this mass splitting will remain small, being proportional to the $W$ mass. The mass difference $(m_{H_0}^2 - m_{H_\pm}^2)$, instead, depends on two parameters, $r$ and $C_g$, and for  $r>1/7$ and $0<C_g <0.1$  the mass  of $m_{H_0}$ is larger than the mass of charged Higgs in case of no tachyons. The bound on $r$ is  also constrained by  the condition of absence of tachyons, with the lower limit cut off by  $m_{\eta_2}^2>0$, and  the upper limit coming from $m_{A_0}^2>0$. This shows that $r$ plays a crucial role in ensuring the stability of the Dark Matter by preventing dangerous tadpoles in this sector. As an example, for a benchmark point $\alpha=0.1$, $C_g =0.05$,  $B m = 10$ GeV and $\delta =0.5$,   $r$ will roughly range between $[0.45, 0.9]$.  The numerical results in Fig.~\ref{fig:mass1} show already interesting features of the spectrum: the lightest state  is always an CP-odd neutral state, either  $A_0$ or one $\mathbb{Z}_2$-even singlet.   While for the DM candidate, there is a switch between $A_0$ (being almost degenerate with $H_\pm$) and the singlet $\eta_2$.  We can  recognise that  the real and complex triplets  are the heaviest states  provided that $\alpha$, $C_g$ and $B m$ are  not too  small.   In addition, the states belonging to the same $SU(2)_L$ multiplet tend to be degenerate (we will further explore this feature at the end of this section).

We now move to the  $\mathbb{Z}_2$ even states. There exist two unmixed states, $\lambda_0$ and $\Lambda_{++}$, that remain close in mass due to the custodial symmetry. Their mass expressions can be read off from the potential:
\begin{eqnarray}
m_{\lambda_0}^2 &=& \frac{2}{3} m_{h}^2 \cot^2(2 \alpha ) +  \frac{4}{3} \sqrt{2} B m v (3 (1+\delta)-2 \cos (2 \alpha )) \csc (\alpha )  
 \nonumber \\ &+&     \frac{4}{3} C_{g} M_{W}^2 \csc ^2(\alpha ) \left(2 \cos (2 \alpha ) \tan ^2\left(\theta _W\right)+3 \sec ^2\left(\theta _W\right)+6\right) \,, 
\end{eqnarray}

\begin{eqnarray}
m_{\Lambda_{++}}^2 &=& \frac{2}{3} m_{h}^2 \cot ^2(2 \alpha ) +   \frac{4}{3} \sqrt{2} B m v (3 (1+\delta)-2 \cos (2 \alpha )) \csc (\alpha )  \nonumber \\ &+& \frac{4}{3}
   C_{g} M_W^2 \csc ^2(\alpha ) \left(\cos (2 \alpha )-(\cos (2 \alpha )-6)
   \sec ^2\left(\theta _W\right)+3\right) \,.
   \end{eqnarray}
The other components of the triplets and remaining singlets feature non-trivial mixing.
For the two singly-charged states, $\Lambda_{\pm}$ and $\varphi_{\pm}$, the $2\times2$ mass matrix can be parametrised as
\begin{eqnarray}
\left(\begin{array}{cc} \Lambda_+ , \varphi_+ \end{array} \right) \left( \begin{array}{cc} N_{\Lambda_+ \Lambda_-} & N_{\Lambda_+ \varphi_-} \\ N_{\varphi_+ \Lambda_-} & N_{\varphi_+ \varphi_-} \end{array} \right)  \left(\begin{array}{c} \Lambda_- \\  \varphi_- \end{array} \right) \,,
\end{eqnarray}
with elements
\begin{eqnarray}
N_{\Lambda_+ \Lambda_-} &=&  \frac{2}{3}  m_h^2 \cot ^2(2 \alpha )  +   \frac{C_g M_W^2}{6} \csc ^2(\alpha ) \sec ^2 \left(\theta _W\right) \left(2 (12 \cos (\alpha )-\cos (2 \alpha )) \sin ^2\left(\theta _W\right) \right.  \nonumber \\    &+ &27  \left. \cos \left(2 \theta _W\right)+45 \right)  +  \frac{4}{3} \sqrt{2} B m v (3 (1+\delta) - 2 \cos (2 \alpha )) \csc (\alpha ) \,, \nonumber  \\  
N_{\varphi_+\varphi_-}& =& N_{\Lambda_+ \Lambda_-} -8 C_{g} M_{W}^2 \cot (\alpha ) \csc (\alpha ) \tan ^2\left(\theta _W\right)\,,   \nonumber \\
 N_{\Lambda_+ \varphi_-} &=& N_{\varphi_+ \Lambda_-}  = - 2 C_{g} M_{W}^2 \tan ^2\left(\theta _W\right)\,.
\end{eqnarray}
Neglecting correction at the order $\sin^4 \alpha$, the mass eigenvalues are approximated as
\begin{eqnarray}
m_{\Lambda_+} &=& N_{\Lambda_+ \Lambda_-} +  \frac{1}{2} C_{g} M_{W}^2 \tan^2 (\theta_W) \sin (\alpha) \tan (\alpha)\,,   \\
m_{\varphi_+} &=& N_{\varphi_+\varphi_-} - \frac{1}{2} C_{g} M_{W}^2 \tan^2 (\theta_W) \sin (\alpha) \tan (\alpha) \,.
\end{eqnarray}
This implies that, in the charged sector, the mass spectra are dominated by the diagonal elements.  The mass splitting between $\Lambda_+$ and $\varphi_+$  can be  remarkably small for small $\alpha$ since  the dominant part  $(N_{\Lambda_+ \Lambda_+} -N_{\varphi_+ \varphi_+})$ is proportional to   $C_g /\sin^2 \alpha$. This characteristics is visualised in  the bottom-left plot  in  Figure~\ref{fig:mass1},  where   the separation between the red dotted line and  the blue solid line becomes larger as  $C_g$ increases.

In the neutral sector, the mixing is more complicated by the presence of 4 CP-odd states, $\lambda = \tfrac{1}{\sqrt{2}}(\Lambda_0+ \Lambda_0^*)$,  $\varphi_0$ $,\eta_1$ and $\eta_3$, which mix via a $4 \times 4$ mass matrix. In order to simplify the analysis,  we can first rotate the two singlets as follows:
\begin{eqnarray}
\eta_m &=& \frac{1}{\sqrt{5}} \left(\sqrt{3 }\eta_1 -\sqrt{2} \eta_3  \right)\,, \nonumber \\ 
\eta_p &=& \frac{1}{\sqrt{5}} \left(\sqrt{2} \eta_ 1 + \sqrt{3} \eta_3 \right) \,.
\end{eqnarray}
This choice is motivated by the fact that only $\eta_m$ appears in the gauge and top spurion potential, while terms involving $\eta_p$ only appear in the potential generated by the hyper-fermion masses. Thus, for $Bm = 0$, $\eta_p$ would remain exactly massless, and a non-zero hyper-fermion mass is therefore required to avoid this.
The neutral mixing can be parametrised as:
\begin{eqnarray}
 \left( \begin{array}{cccc}  \lambda & \varphi_0 & \eta_m &  \eta_p  \end{array} \right) VN_{4\times 4} \left( \begin{array}{c} \lambda \\  \varphi_0\\ \eta_m \\ \eta_p \end{array} \right)\,,
\end{eqnarray}
where we split the matrix $VN$ in 3 terms:
\begin{eqnarray}
 VN  = m_h^2 ~ M_1 + C_g M_W^2  ~ M_2 +  B m v ~ M_3 
  \end{eqnarray}
proportional to different mass scales.
The explicit expressions for $M_1$, $M_2$ and $M_3$ are listed below: 
 {\footnotesize
\begin{eqnarray}
 M_1 &=&    
\frac{1}{24}  \left(
\begin{array}{cccc}
2 \left(\frac{2} {\sin^2(\alpha )}+\frac{5- 11 \cos^2(\alpha )}{\cos^2(\alpha )}\right) & \frac{6 \tan^2(\alpha )}{\sqrt{2}} & - \frac{\sqrt{5} (5 \cos (2 \alpha)+3)}{\sqrt{2} \cos^2(\alpha ) }& 0 \\
 \frac{6 \tan ^2(\alpha )}{\sqrt{2}} &\left( \frac{4}{\sin^2(\alpha )}+\frac{7- 19 \cos^2(\alpha )}{\cos^2(\alpha )} \right) & - \frac{\sqrt{5}  (5 \cos (2 \alpha )+3)}{2 \cos^2(\alpha )} & 0 \\ - \frac{\sqrt{5} (5 \cos (2 \alpha )+3) }{\sqrt{2} \cos^2(\alpha )} &
   - \frac{\sqrt{5} (5 \cos (2 \alpha )+3) }{2 \cos^2(\alpha )} & \frac{5
   (20 \cos (2 \alpha )+3 \cos (4 \alpha )+9) }{2 \sin^2(2 \alpha ) }& 0 \\
 0 & 0 & 0 & 0 \\
\end{array}
\right)\,,  \\
 M_2 & =&  \left(
\begin{array}{cccc}
 \frac{4 \left(3 \sec ^2\left(\theta _W\right)+2 \cos (2
   \alpha ) \tan ^2\left(\theta _W\right)+6\right) }{3 \sin ^2(\alpha )} & 2 \sqrt{2} \tan
   ^2\left(\theta _W\right) & -\frac{2 \sqrt{10}}{3}  \tan ^2\left(\theta _W\right)
   & 0 \\ 2 \sqrt{2} \tan ^2\left(\theta _W\right) & -\frac{4 \left(\cos (2 \alpha ) \tan ^2\left(\theta _W\right)-9\right)}{3 \sin^2(\alpha )}
 & \frac{4  \sqrt{5} }{3} \tan ^2\left(\theta _W\right) & 0 \\
 -\frac{2\sqrt{10}}{3}  \tan ^2\left(\theta _W\right) & \frac{4 \sqrt{5}}{3}  \tan^2\left(\theta _W\right) & -\frac{20 \left(\cos \left(2 \theta_W\right)+2\right)}{3  \sin^2(\alpha ) \cos^2\left(\theta _W\right)}  & 0 \\
 0 & 0 & 0 & 0 \\
\end{array}
\right)\,, \\
  M_3 &=&   \left(
\begin{array}{cccc}
 \frac{4 \sqrt{2} (3(1+\delta)-2 \cos (2 \alpha )) }{3  \sin(\alpha )} & 0 & \frac{32 \sin (\alpha )}{3 \sqrt{5}} &  \frac{4 \sqrt{2} \sin (\alpha ) }{\sqrt{15}}  \\
 0 &  \frac{4 \sqrt{2} (3 (1+\delta)-2 \cos (2 \alpha )) }{3  \sin(\alpha )} &\frac{16 \sqrt{2}  \sin (\alpha )}{3 \sqrt{5}} &\frac{4 \sin (\alpha )}{\sqrt{15}} \\
 \frac{32 \sin (\alpha )}{3 \sqrt{5}} &\frac{16 \sqrt{2}  \sin (\alpha )}{3 \sqrt{5}}  & -\frac{4 \sqrt{2}(6 \cos (2 \alpha )+29 +9 \delta)}{15 \sin (\alpha )}   
  &  \frac{16 \delta + 12 \sin^2 (\alpha)}{5
   \sqrt{3}  \sin (\alpha )}  \\
 \frac{4 \sqrt{2} \sin (\alpha ) }{\sqrt{15}} &\frac{4 \sin (\alpha )}{\sqrt{15}} &
 \frac{16 \delta + 12 \sin^2 (\alpha)}{5
   \sqrt{3}  \sin (\alpha )} & \frac{4 \sqrt{2}(\cos (2 \alpha )+14 -11 \delta)}{15 \sin (\alpha )}  \\
\end{array}
\right)  \,.  
\end{eqnarray}}
We observe that  the  two singlets $\eta_m$ and $\eta_p$ can  mix  significantly   due to the $M_3$ term, proportional to $Bm  \delta $.  While the mixing between the triplets tends to be small.  Hence at the leading order, one can approximate
\begin{eqnarray}
m_{\lambda} \simeq  m_{\lambda_0}  \simeq m_{\Lambda_+} \simeq m_{\Lambda_{++}} \,, \quad \mbox{and} \quad  m_{\phi_0} \simeq m_{\phi_+} \,.
\end{eqnarray}
This pattern is clearly illustrated in Fig.~\ref{fig:mass1}.  Note that we  label the mass eigenstate with the name of the gauge eigenstate that constitutes its major component.  Concerning the two singlets, $\eta_m$ and $\eta_p$,  their masses are rather different as only one of them receives a mass from the top and gauge interactions. In particular, the mixing effects from the $B m \delta$ term  are  only important in the region where the two masses  are close, i.e. $m_{\eta_m} \simeq m_{\eta_p} $. In Fig.~\ref{fig:mass1} this can be  observed from the two separate cyan lines (for $\delta=0$ they will be two intersecting lines), where we see a level flip where the two lines approach each other, as a sign of discontinuous spectra for $\eta_m$ and $\eta_p$ due to mixing effects.  And the upper limit for the parameters in the figure is  always set by  $m_{\eta_m}^2 >0$.  


\begin{figure}
	\centering 	\subfigure[]{\includegraphics[height=5. cm, 
		width=6cm]{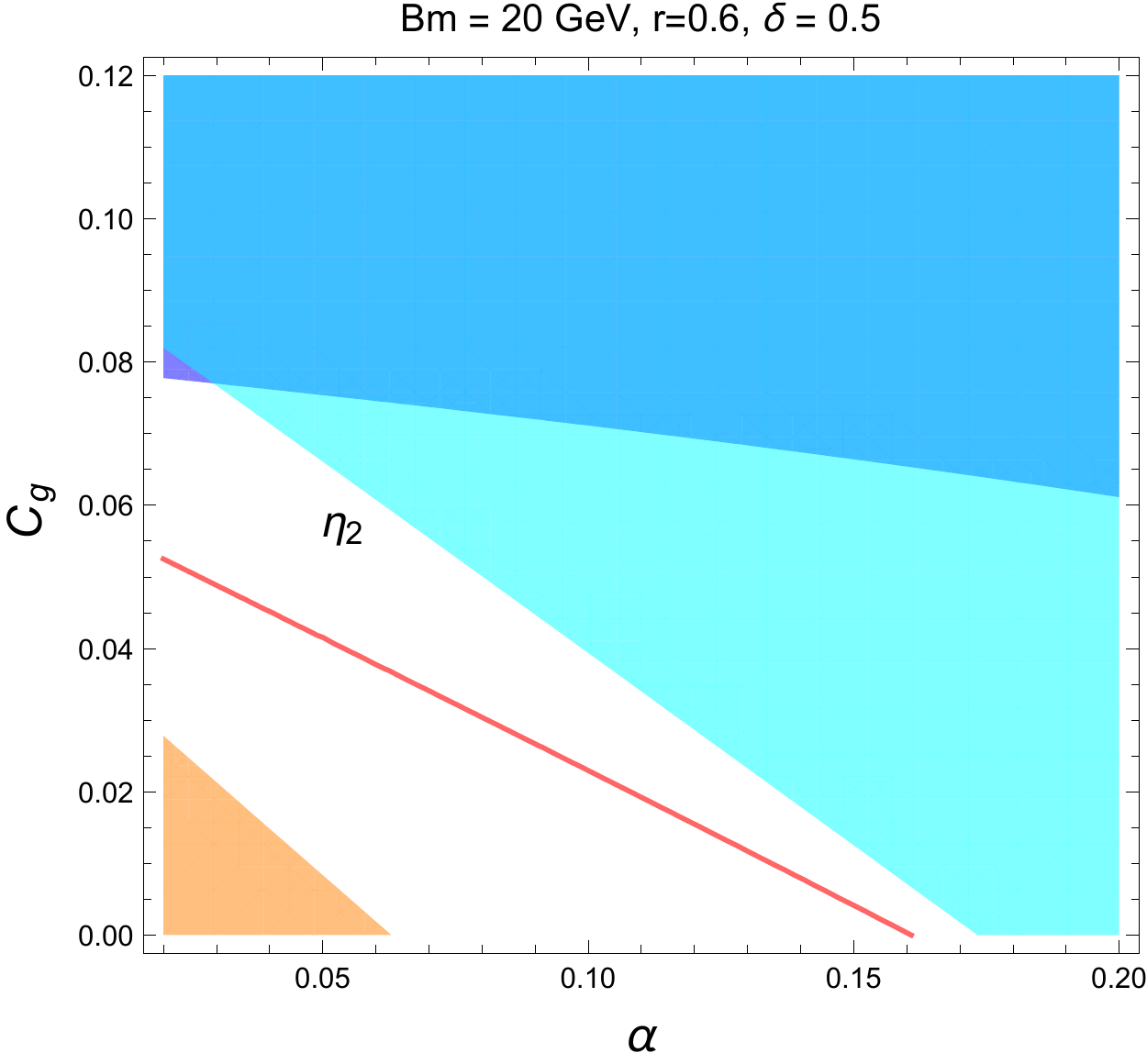}} \qquad
		\subfigure[]{\includegraphics[height=5. cm, 
		width=6cm]{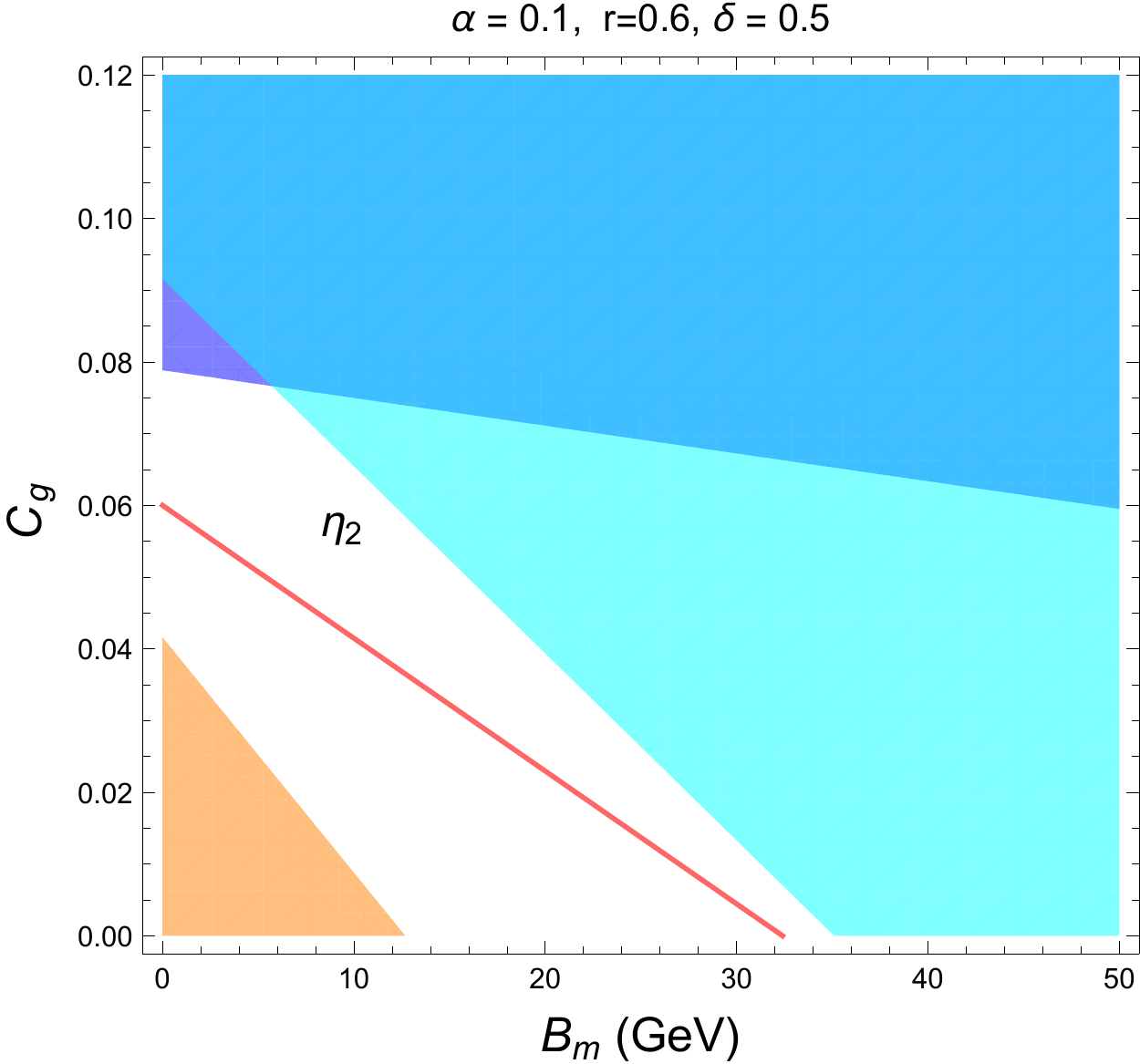}}
		 \subfigure[]{\includegraphics[height=5. cm, 
		width=6cm]{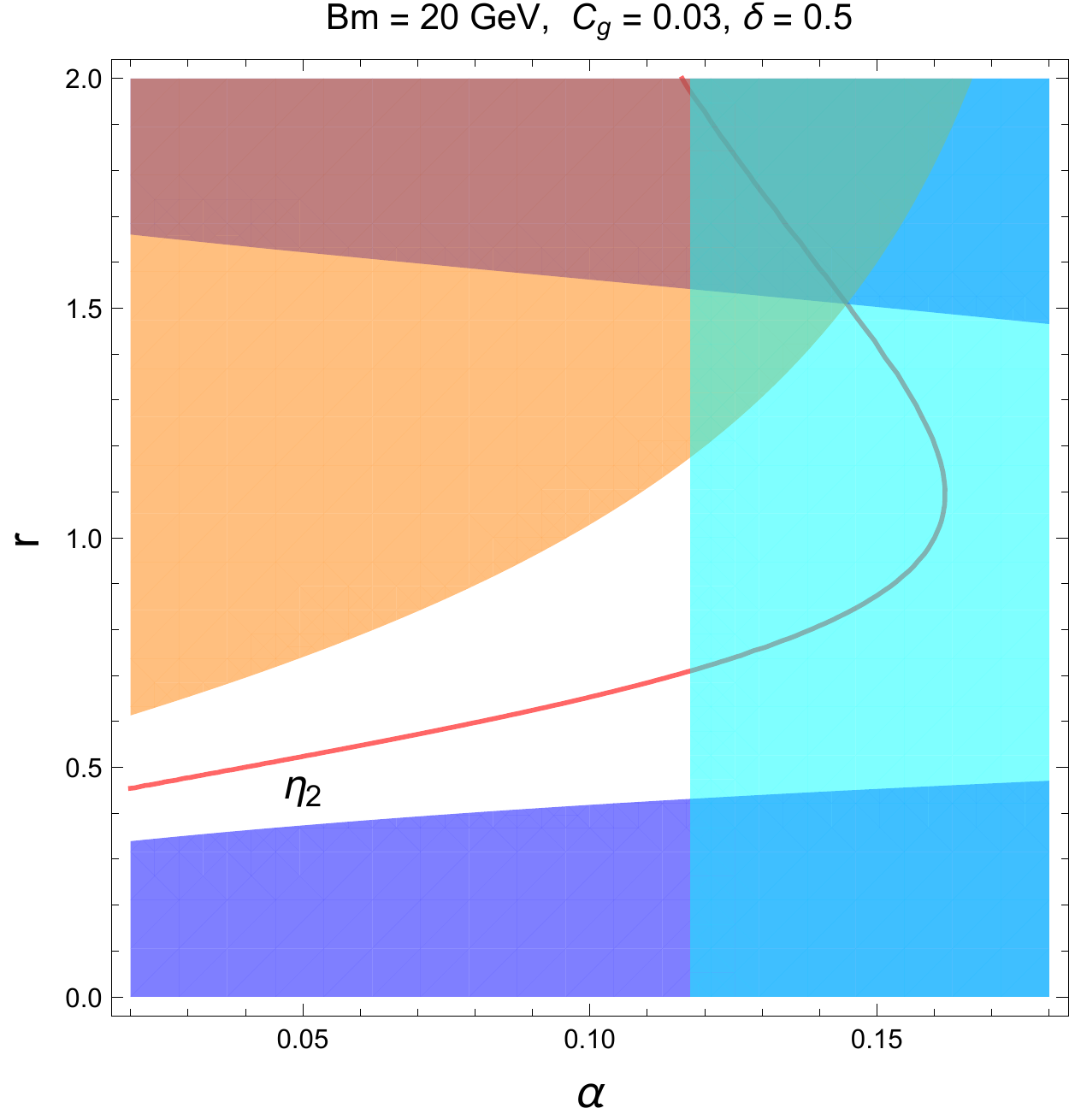}} \qquad
		\subfigure[]{\includegraphics[height=5. cm, 
		width=6cm]{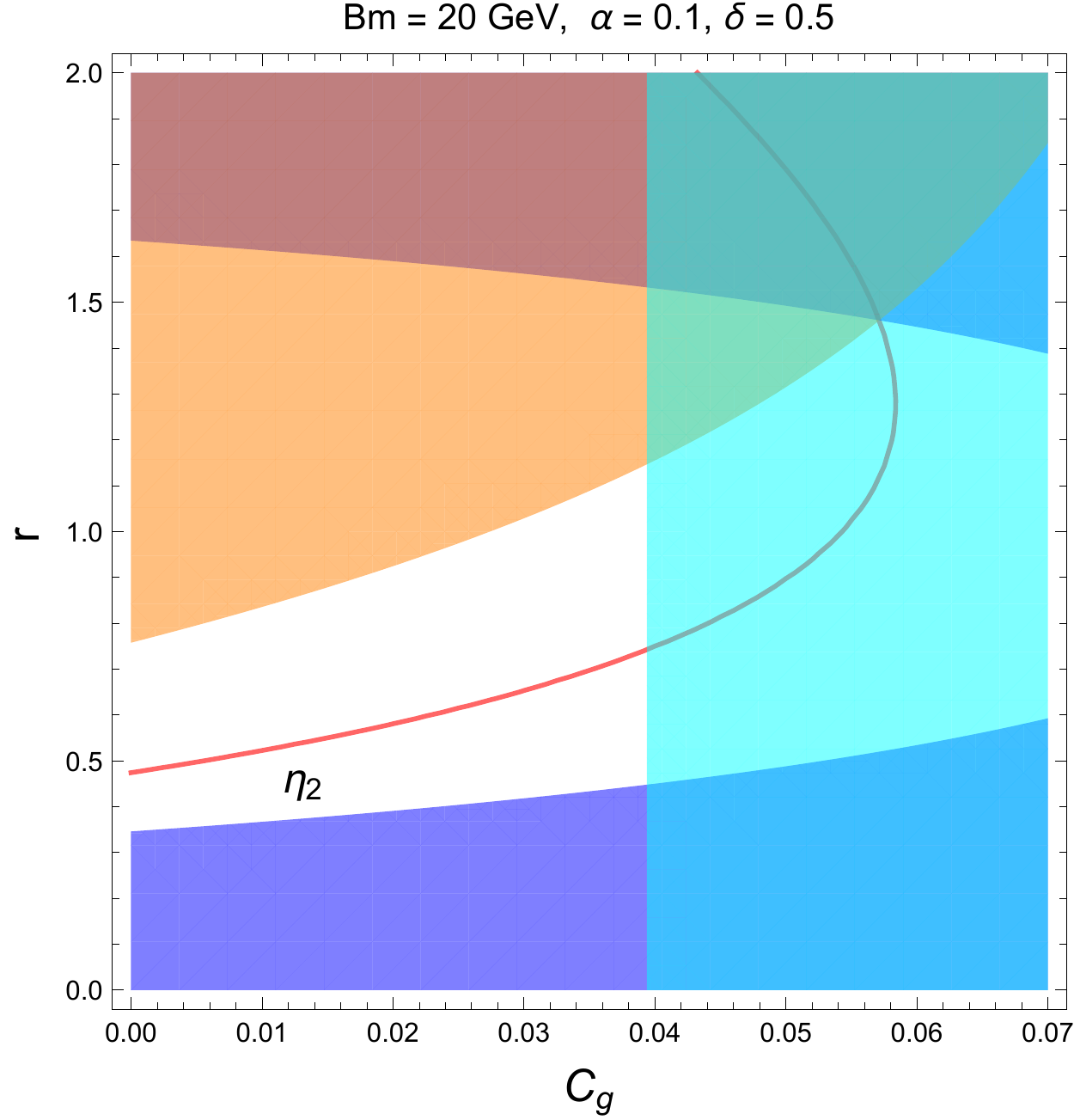}}
		\caption{The tachyonic region is in specific color in the planes of (a) $C_g -\alpha$ (b) $C_g - Bm$  (c) $r- \alpha$ and (d) $r -C_g$, where the  blue shading  indicates constraint  from the $\eta_2$ scalar,   the cyan region from mixing neutral states,  and  the orange  region from the second doublet.  The red line stands for  the contour  with $m_{\eta_2} = m_{A_0}$. }
	\label{fig:tachyon}
\end{figure}

\begin{figure}
	\centering
	\subfigure[]{
		\includegraphics[height=5. cm, 
		width=7cm]{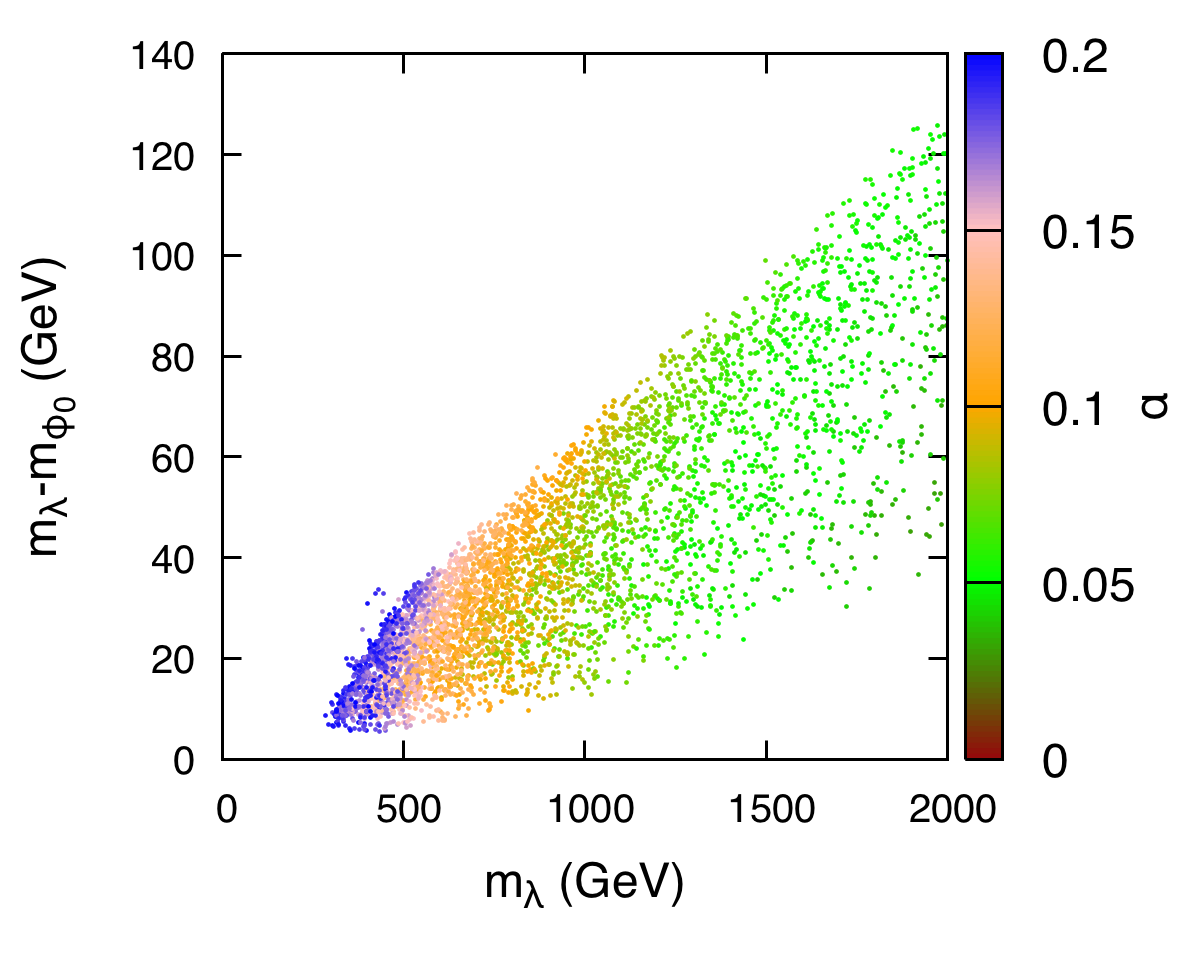}}
	\subfigure[]{
		\includegraphics[height=5. cm, 
		width=7cm]{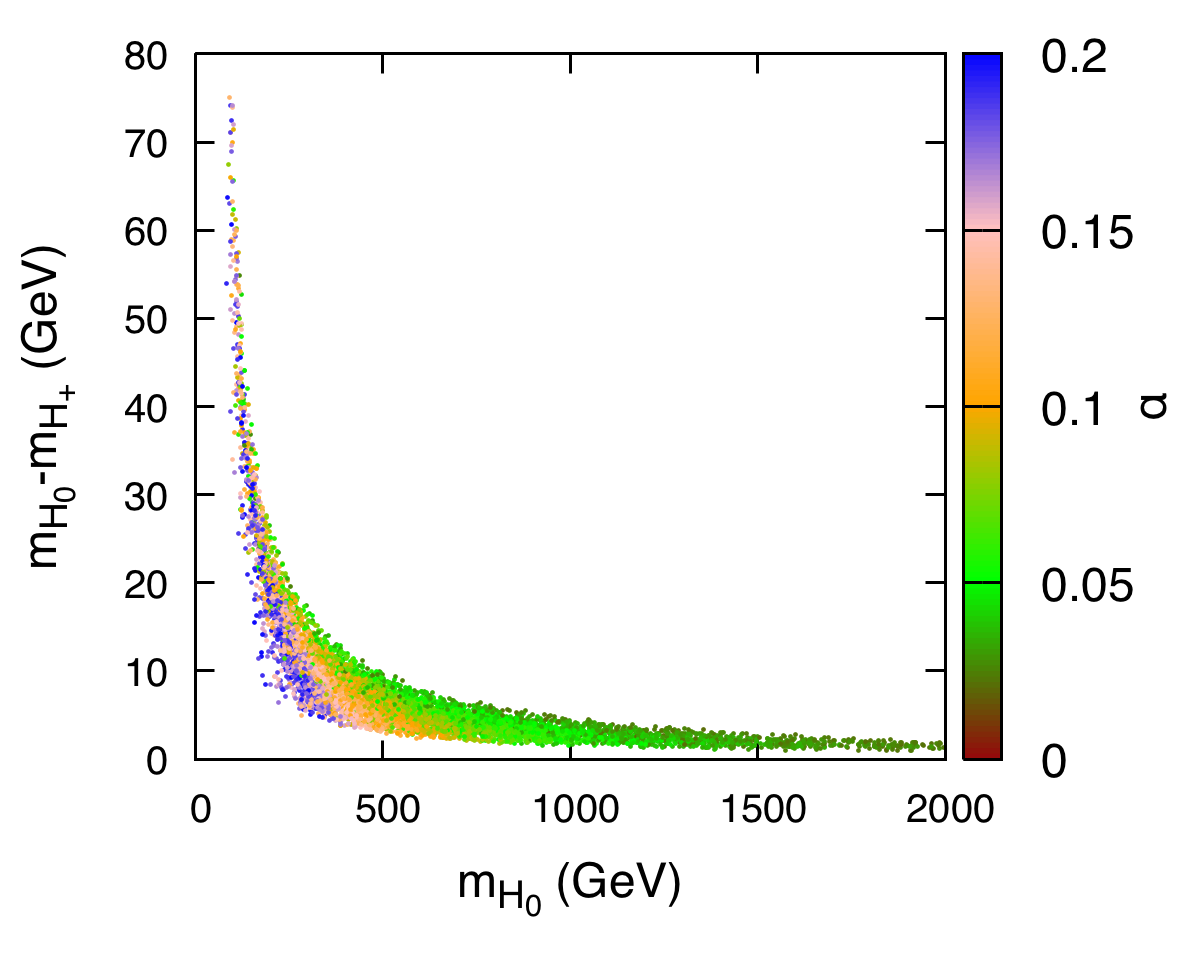}}
		\subfigure[]{
		\includegraphics[height=5. cm, 
		width=7cm]{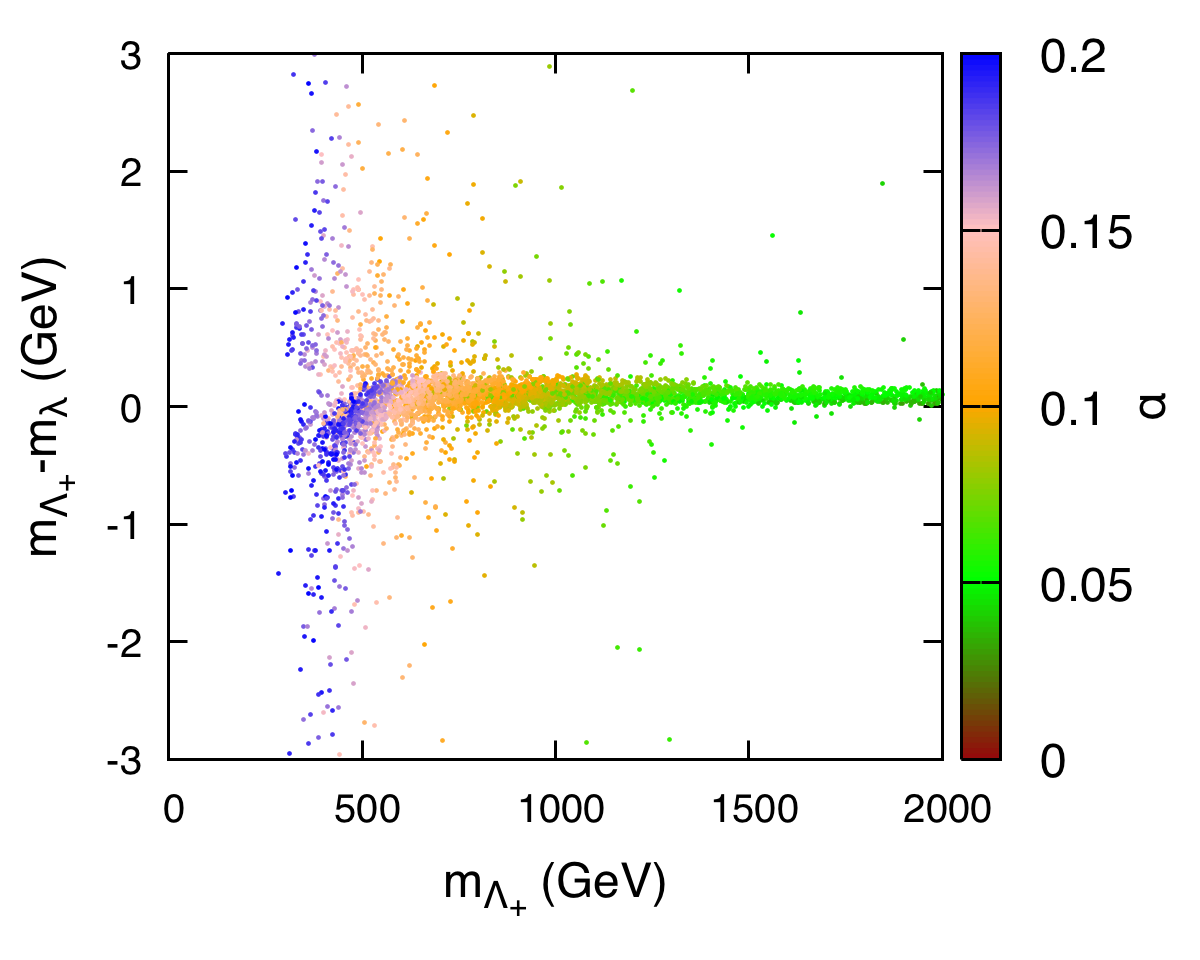}}
			\subfigure[]{
		\includegraphics[height=5. cm, 
		width=7cm]{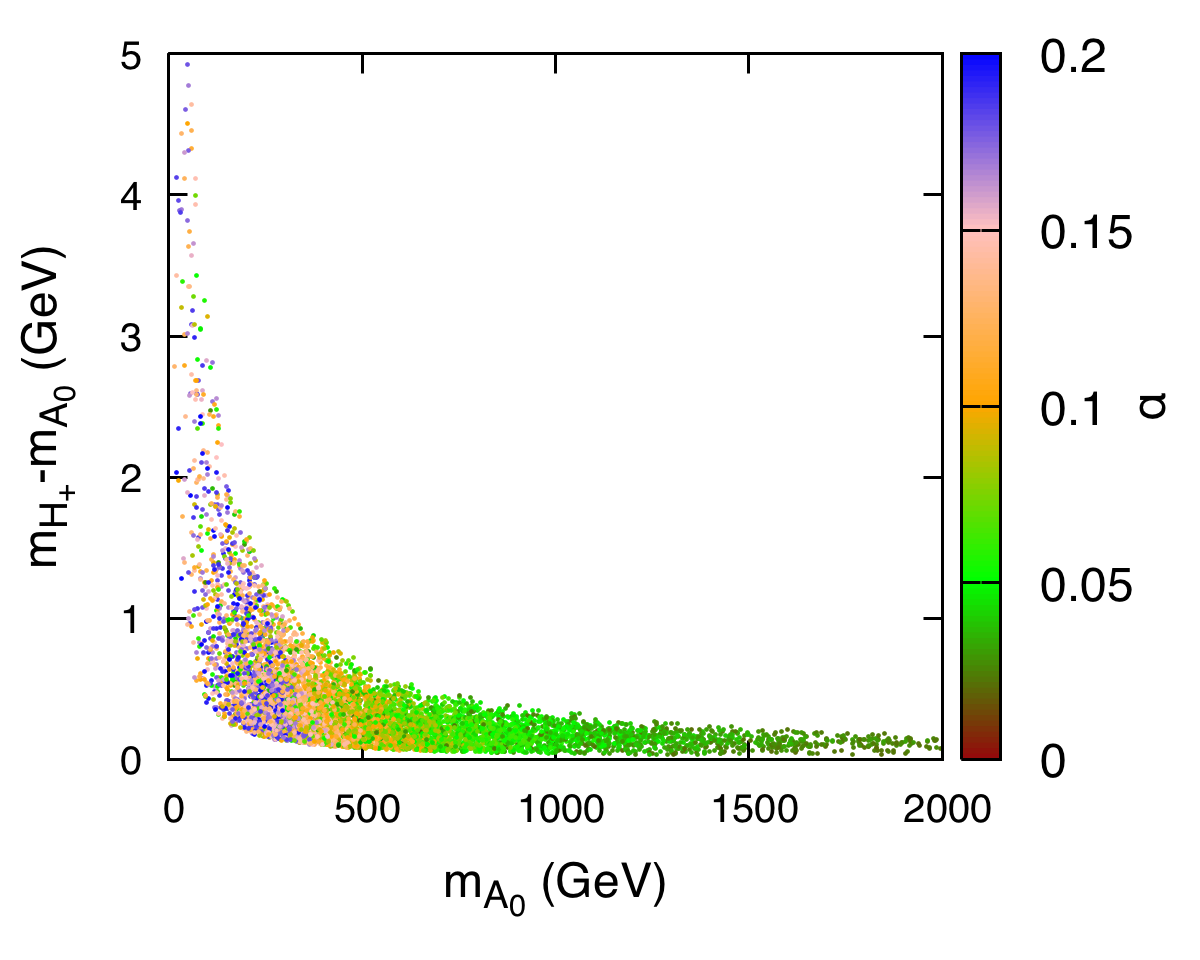}}
		 \subfigure[]{
		\includegraphics[height=5. cm, 
		width=7cm]{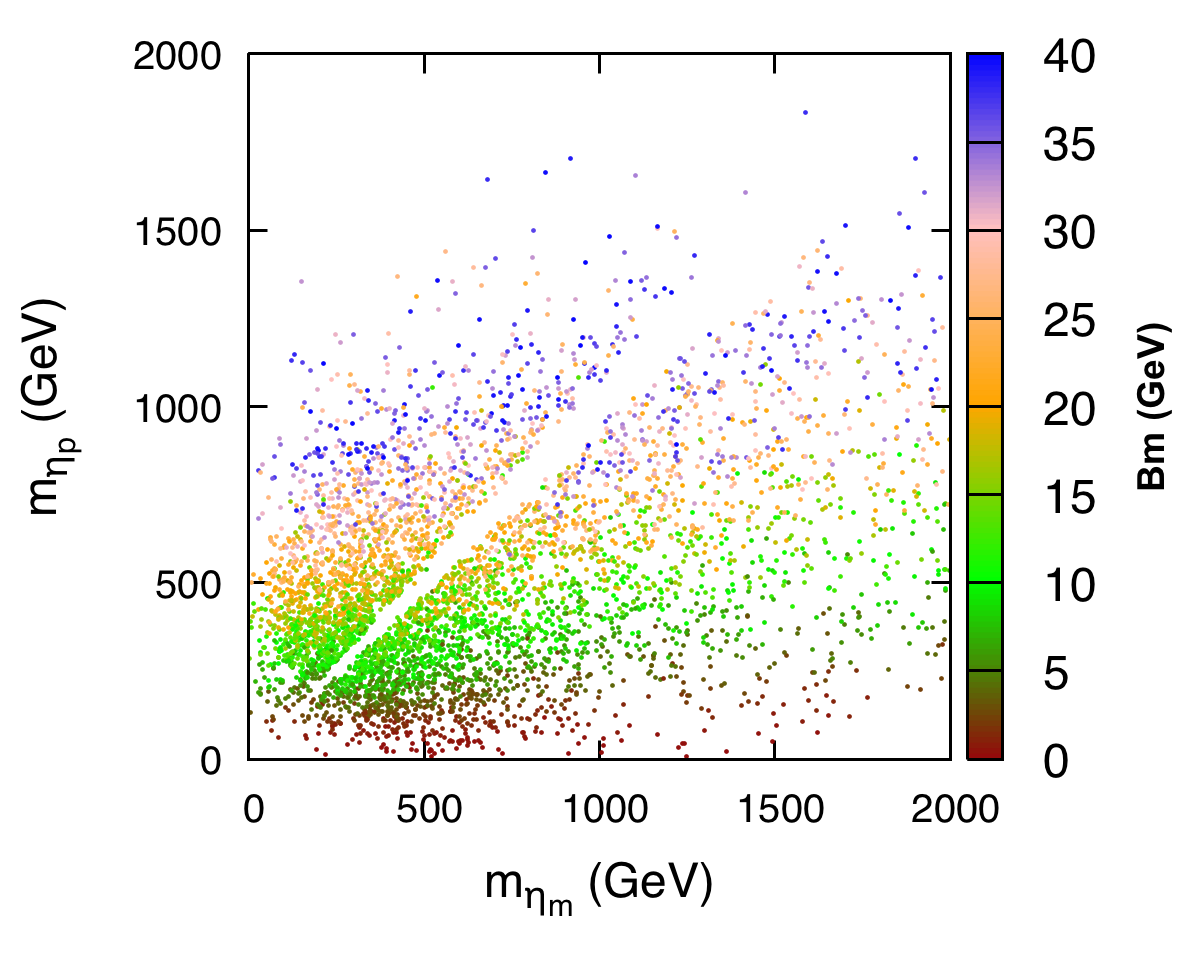}}
	\subfigure[]{
		\includegraphics[height=5. cm, 
		width=7cm]{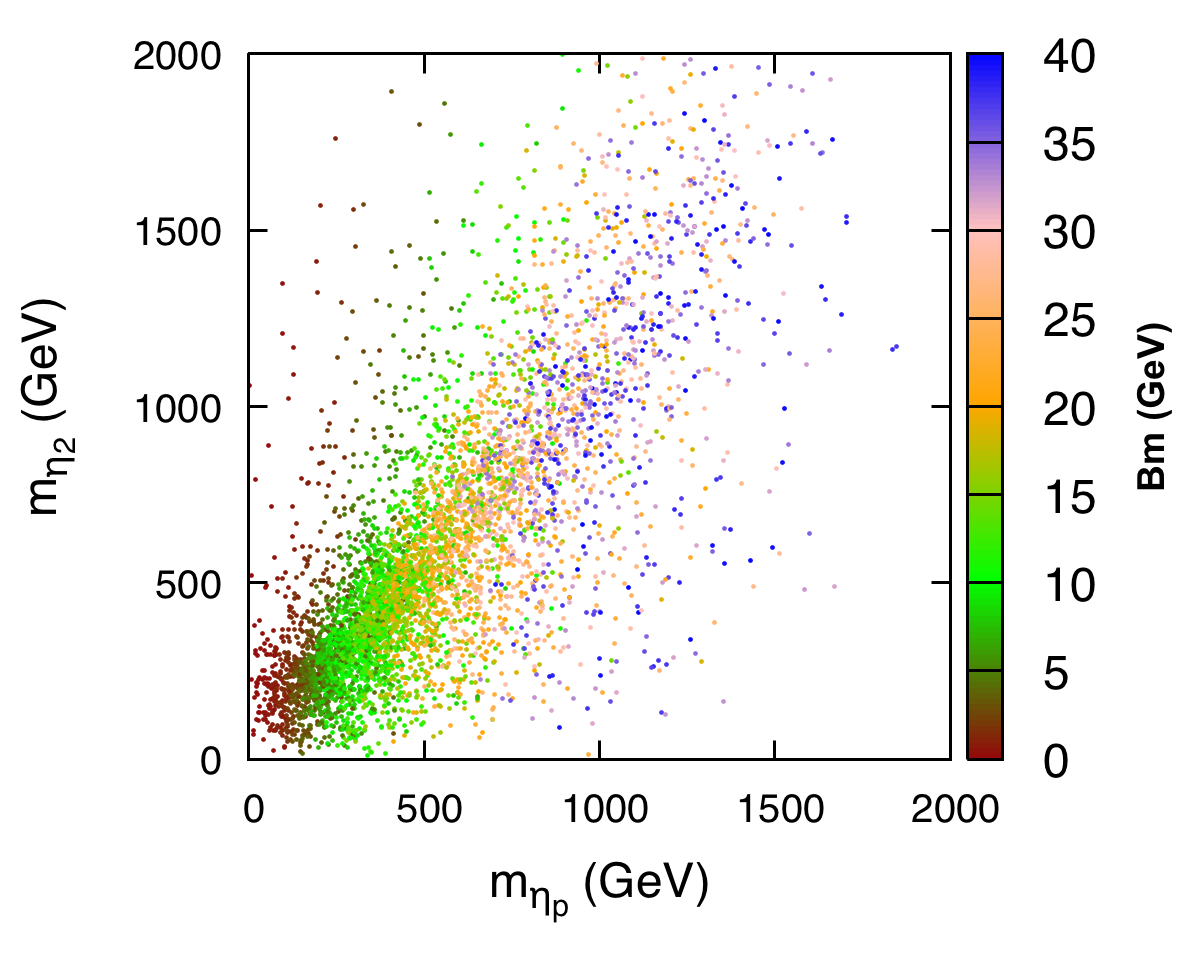}}
\caption{Mass splitting (a)-(d) in second Higgs doublet,  real and complex triplets and mass correlation  (e)-(f) among 3 singlets  for $0.02<\alpha<0.2$, $0 < B m  < 40$ GeV, $0.01<C_g<0.09$,   $0.1<r<1.5$, and $\delta$ in $(-0.5, -0.2)$ or $(0.2, 0.5)$  after imposing no tachyon condition. } \label{fig:mass2}
\end{figure}

Now that we know the general features of the spectrum, it is needed to better understand the region of parameter space where our choice of vacuum alignment, where the Dark Matter parity is preserved, remains consistent. To do so, we need to impose that no tachyon appears in the pNGB spectrum.
We thus sliced the parameter space by fixing two parameters and scanning over the remaining two.
In Figure~\ref{fig:tachyon}, we show the allowed  parameter space for four sample choices. In these  plots,  the blue coloured  region indicates where $m_{\eta_2}^2 <0 $,  the cyan  region marks  ${\rm det}(VN) <0$, while the orange region stands for  $m_{A_0}^2 <0 $, leaving the white band to be the non-tachyonionic consistent region. We remark that there is no constraint from the charged states, so that no vacua breaking the electromagnetic charge arise. 
In  the top-right plane, $C_g$--$Bm$, values $Bm <0$ are excluded by the presence of a tachyon for the pseudo-scalars (corresponding to $\eta_m$). The two plots, c) and d) in the lower row, confirm that $r>0$ and a minimal value is necessary in order for  $\eta_2$ to be non-tachyonic. 
We also see that the white band always interpolates between a tachyon in $\eta_2$ and one in $A_0$, meaning that the lightest odd state flips between the two in between: we thus mark the demarcation line in red, where $m_{\eta_2} = m_{A_0}$.  The section tagged with ``$\eta_2$'', close to the blue colour edge, corresponds to  $m_{\eta_2} < m_{A_0}$  so that $\eta_2$ is the DM candidate.  Note that in a larger portion of allowed  space, the CP-odd Higgs $A_0$ is more likely  the  lightest $\mathbb{Z}_2$-odd particle. The plots in the top row also show that there is always an upper bound for $C_g$, above which a pseudo-scalar becomes tachyonic. Numerically, we find $C_g \lesssim 0.08$ for the benchmarks we considered, value that is at odds with the lattice estimate in Eq.~\eqref{eq:Cgestimate}. Thus, a realistic underlying description for this model should somehow feature a smaller value. We also see an upper bound on $\alpha$: as we will see in the next section, this upper bound is actually competitive with the typically strong bounds coming from electroweak precision observables.

To complete our  understanding of the mass spectra, we performed a numerical scan for the 5 free parameters in the ranges $0.02<\alpha<0.2$, $0 < B m  < 40$ GeV, $0.01<C_g<0.09$,   $0.1<r<1.5$, and $0.2<|\delta| <0.5$.  Furthermore, we  require  the mass squared of all pNGBs to be positive so that all points shown are consistently tachyon-free.  In Figure~\ref{fig:mass2}, we  show the mass splitting between the neutral and charged states as well as the mass correlation among the 3 singlets.  In  plot a),  the correlation between $(m_{\lambda}-m_{\varphi_0})$ and $m_{\lambda}$ is displayed, which indicates that the mass gap between the two neutral states  can be relatively large, $\sim 100$ GeV  for  a  $m_{\lambda}$ in TeV scale, consistent with the analytic result.   The same  applies to  $(m_{\Lambda_+} - m_{\phi_+})$   due to the mass degeneracy inside each multiplet.  However  the situation is opposite for the doublet. As illustrated by plot b), the mass difference among $H_0$ and $H_{+}$ is  notably large  for small  $m_{H_0}$  because of  the $r$ dependence,  while the plots c) and d) show that  $(m_{\Lambda_+} -m_\lambda)$ and $(m_{H_+}-m_{A_0})$ are  mainly in sub-GeV scale.  The mass correlation among  3 singlets, $\eta_2$, $\eta_m$ and $\eta_p$,  is presented in plots e) and f):  we can clearly see  that $m_{\eta_p}$ goes to zero for very tiny $B m$, and  $\eta_m$, $\eta_2$ can also be arbitrarily light. Thus, the presence of light CP-odd states can be probed by typical axion-like-particle (ALP) searches coupling to two photons, as we will discuss in the next section. In particular, in plot e)  we  observe two branches while a small wedge along the diagonal line $m_{\eta_p}= m_{\eta_m}$ is unpopulated. This  signals  the fact that the mixing terms spoil the degeneracy for $\delta \neq 0$, since the mass eigenstates are labelled with the gauge eigenstate that matches its largest component.

\section{Phenomenology}\label{pheno}

In this section we will briefly comment on some phenomenological aspects of this model. A complete study is well beyond the scope of this paper, and we leave that to further studies.

\subsection{Electroweak precision tests}

Electroweak precision tests  receive corrections due to the composite nature of the Higgs and to the presence of the composite sector. They can be expressed in terms of the oblique parameters~\cite{Peskin:1990zt,Altarelli:1990zd} $S$ and $T$. Following Ref.~\cite{Arbey:2015exa}, we  split the contribution in 3 pieces encoding, respectively, the effect of the modified Higgs couplings, the presence of light states (additional pNGBs) and the strong dynamics:
\beq
\Delta S &=& \frac{\sin^2 \alpha}{6\pi} \ln \frac{\Lambda}{m_h} + \Delta S_{\rm pNGBs} + \Delta S_{\rm HC}\,, \\
\Delta T &=& -\frac{3 \sin^2 \alpha}{8\pi \cos^2 \theta_W} \ln \frac{\Lambda}{m_h} + \Delta T_{\rm pNGBs} + \Delta T_{\rm HC}\,.
\eeq
The first term for both is universal to all composite Higgs models as it only depends on the modification of the Higgs couplings to $W$ and $Z$, while the cut-off dependence should be related to the mass of the lowest non-pNGB resonances (we can safely assume $\Lambda \approx 4 \pi f$). The last term encodes the effect of the strong sector itself. As the custodial symmetry is embedded in it, we expect that $\Delta T_{\rm HC} \approx 0$, while the $S$ parameter can be sizeable. Taking the approximation in Ref.~\cite{Peskin:1991sw}, appropriately corrected by a factor of $\sin^2 \alpha = v^2/f^2$, it can be expressed as
\beq
\Delta S_{\rm HC} \approx \frac{\sin^2 \alpha}{6 \pi} 2 d_\psi\,,
\eeq
where $d_\psi$ is the dimension of the fermions $\psi$ under the confining interactions, and the factor of $2$ counts the fact that there are 2 SU(2)$_L$ doublets in our model. For the underlying models discussed in Section~\ref{sec:underlying}, we have $d_\psi = 8$ for $SO(7)_{\rm HC}$ and $d_\psi = 16$ for $SO(9)_{\rm HC}$. This contribution can also be computed on the lattice, provided that the double-counting between the light pNGBs (including the Higgs) and the strong interactions is properly taken into account~\cite{Foadi:2012ga}. Note that calculations in terms of loops of the lowest-lying resonances (vectors and top partners) are also available~\cite{Contino:2015mha}, however they are only trustable if those states are light and weakly coupled to the pNGBs.

In the remaining, we will focus on the contribution of the pNGBs, which is the one better under control in the effective description. In our model, besides the Higgs, only the triplets and the second doublet will contribute. The effect of the triplets is similar to the Georgi-Macachek model~\cite{Georgi:1985nv}, and it has been discussed extensively in Ref.~\cite{Hoang:2013jfa}.  The results show that the correction to the $S$ parameter is always small and well within the experimental error, while the contribution to the $T$ parameter can be sizeable if a largish mass splitting between the triplet components is generated. 
The case of the inert second Higgs doublet has also been extensively studied~\cite{Barbieri:2006dq}, leading to similar conclusions: while the contribution to $S$ is always small, the one to $T$ can be sizeable.
We can thus express the contribution to $T$ as
\beq
\Delta T_{\rm pNGBs} \approx \frac{1}{24 \pi \sin^2 \theta_W} \frac{\delta m^2}{m_W^2}\,, \qquad \delta m^2 = \left. \delta m^2 \right|_{\rm triplets} + \left. \delta m^2 \right|_{\rm doublet}\,.
\eeq
The contribution of the doublet can be conveniently be written as~\cite{Barbieri:2006dq}
\beq
\left. \delta m^2 \right|_{\rm doublet} = -  (m_{H_0} - m_{H^\pm}) (m_{H^\pm} - m_{A_0}) < 0\,,
\eeq
which is always negative because of the hierarchy $m_{H_0} > m_{H^\pm} \gtrsim m_{A_0}$, and always small because of the near degeneracy of $H^\pm$ and $A^0$.
For the triplets the expression is more complicated (see Ref.~\cite{Hoang:2013jfa}), however it will always vanish when the components in each multiplet are near-degenerate: this is exactly the case in our model where the mass splits within each triplet are always within a few GeV. The only remarkable property is that $\left. \delta m^2 \right|_{\rm triplets}$ can be either positive or negative. 
 As we expect $|\delta m^2| \approx$~few~GeV$^2$, the effect of the pNGBs is always negligible and the bound on $\sin \alpha$ is dominated by the contribution of the Higgs couplings and the HC sector. 
For the two models we find 
\beq
\sin \alpha \lesssim 0.16\;\; (SO(7)_{\rm HC})\,, \quad \sin \alpha \lesssim 0.14\;\; (SO(9)_{\rm HC})\,,
\eeq

To conclude this section, we would like to mention another mechanism that may reduce the constraint on $\alpha$ from electroweak physics, namely the presence of a light $0^{++}$ resonance, possibly associated to an IR conformal phase right above the condensation scale. This case has been studied in detail in Refs~\cite{Arbey:2015exa, BuarqueFranzosi:2018eaj}.

\subsection{Light singlet $\eta_p$ phenomenology}

The pNGB spectrum contains many new scalars besides the Higgs (and the Dark Matter candidate). Among them, we identified one pseudo-scalar singlet which can receive a mass only from the hyper-fermion mass term $B m$, thus it can be arbitrarily light. This is a peculiar feature of the $SU(6)/SO(6)$ model that did not appear in the more minimal $SU(5)/SO(5)$ case. We will thus briefly establish the constraints on such a light state.

Notably, it only features single-couplings to the electroweak gauge bosons via the WZW anomaly, while couplings to the top are absent. If we assume the same type of operator as responsible for the other quark and lepton masses, it is safe to assume that couplings to all SM fermions are absent.
It is also simpler to study the couplings in the limit of small $\alpha$, as required by electroweak precision tests. From Section~\ref{sec:WZW} we see that excluding the heavy triplet, only $\eta_1$ has couplings at the leading order, i.e. not suppressed by powers of $\alpha$. In the following, we will adopt the same notation used for axion-like particles (ALPs)~\cite{Bauer:2017ris}, and use the results from Refs~\cite{Bauer:2017ris,Bauer:2018uxu} at current experiments and future colliders.

In the limit $\alpha \to 0$, the couplings from the WZW term can be written in terms of gauge eigenstates as
\beq
\mathcal{L}_{\rm ALP} = g_2^2 \frac{C_{WW}}{\Lambda}\ \eta_1 W_{\mu \nu} \tilde{W}^{\mu \nu} + g_1^2 \frac{C_{BB}}{\Lambda}\ \eta_1 B_{\mu \nu} \tilde{B}^{\mu \nu}\,,
\eeq
where in our model (see Table~\ref{tab:wzw})
\beq
\frac{C_{WW}}{\Lambda} = \frac{C_{BB}}{\Lambda} = \sqrt{\frac{3}{2}} \frac{d_\psi}{48 \pi^2 f}\,. 
\eeq
In the neutral sector, the couplings can be rewritten in terms of mass eigenstates (the photon and the $Z$) as follows
\beq
\mathcal{L}_{\rm ALP} \supset  e^2 \frac{C_{\gamma \gamma}}{\Lambda}\ \eta_1 F_{\mu \nu} \tilde{F}^{\mu \nu} + \frac{4 e^2}{\sin (2 \theta_W)} \frac{C_{\gamma Z}}{\Lambda}\ \eta_1 Z_{\mu \nu} \tilde{F}^{\mu \nu} + \frac{4 e^2}{\sin^2 (2 \theta_W)} \frac{C_{ZZ}}{\Lambda}\ \eta_1 Z_{\mu \nu} \tilde{Z}^{\mu \nu} \,,
\eeq
with
\beq
\frac{C_{\gamma \gamma}}{\Lambda} = 2 \frac{C_{WW}}{\Lambda}\,, \quad \frac{C_{\gamma Z}}{\Lambda} = \frac{\cos (2\theta_W)}{2} \frac{C_{\gamma \gamma}}{\Lambda}\,, \quad \frac{C_{ZZ}}{\Lambda} = \frac{1+\cos^2 (2 \theta_W)}{4} \frac{C_{\gamma \gamma}}{\Lambda}\,.
\eeq
For the light mass eigenstate, $\eta_p$, an additional factor of $\sqrt{2/5}$ appears due to the change of basis. We will thus express all the constraints in terms of the coupling to photons:
\beq
\frac{C_{\gamma \gamma}^{\eta_p}}{\Lambda} = \sqrt{\frac{3}{5}} \frac{d_\psi}{24 \pi^2 f} = \frac{0.017}{\mbox{TeV}} \left( \frac{d_\psi}{8} \right) \left( \frac{\sin \alpha}{0.16} \right)\,,
\eeq
where we have used $v = f \sin \alpha$, and used the reference values for the model based on $SO(7)_{\rm HC}$.
For light masses, below $m_Z$, the only open decay channel is in $\gamma \gamma$. A comprehensive collection of current bounds can be found in Ref.~\cite{Bauer:2017ris}: for this value of the coupling to photons, the leading constraint is coming from beam dump experiments~\cite{Riordan:1987aw,Bjorken:1988as}, leading to $m_{\eta_p} \gtrsim 1$~GeV. At higher masses, colliders provide the leading constraints, but reaching down to couplings of the order of $0.1/\mbox{TeV}$.
Projections for future colliders can be found in Ref.~\cite{Bauer:2018uxu}. The most sensitive machines are $e^+ e^-$ colliders: the FCC-ee option considered by the authors offers the best reach up to masses of $m_{\eta_p} \approx 100$~GeV, and reaching down to couplings $|C_{\gamma \gamma}^{\eta_p}|/\Lambda \lesssim 10^{-4}/\mbox{TeV}$. The latter would correspond to $\sin \alpha \lesssim 10^{-3}$, implying a severe fine-tuning in the misalignment. On the other hand, CLIC seems to fall short of the typical value of the coupling, thus it would not be able to improve significantly the limits.
Hadron colliders can also probe this state via decays of the $Z$ boson, $Z \to \gamma a \to 3 \gamma$, via the coupling $C_{\gamma Z}$. Rescaling the results of Ref.~\cite{Bauer:2018uxu}, we estimate that the high-luminosity LHC run (with integrated luminosity of 3/ab) could probe $|C_{\gamma \gamma}^{\eta_p}|/\Lambda \approx 5\times 10^{-3}/\mbox{TeV}$ for $m_{\eta_p} < m_Z$. On the other hand, a 100 TeV collider FCC-hh with 15/fb integrated luminosity 
would be enough to reach below $10^{-3}/\mbox{TeV}$.

Other potentially sensitive channels include $h \to \eta_p \eta_p$ and $h \to Z \eta_p$, via couplings generated by loops. However, due to the absence of a coupling to tops in our models, these effective couplings are very suppressed and not leading to significant bounds. Similarly, couplings to fermions can be generated by loops of the WZW interactions, however leading to very small couplings.

%

\section{Conclusions and outlook}\label{Conclusion}

The breaking of the electroweak symmetry by means of a confining strong dynamics is still a valid alternative to the Higgs sector of the SM.
Generically, models based on strong dynamics are natural (thanks to the absence of light elementary scalars) and predict heavier new physics states compared to weakly coupled options like, for instance, supersymmetry.

Motivated by gauge-fermion underlying  descriptions, which allow to study the strong dynamics on the lattice, we focused in this paper on a non-minimal coset $SU(6)/SO(6)$. While a large number of light pNGBs are predicted, the model can be described by a minimal underlying model consisting of 6 Weyl fermions in a real representation of the confining gauge group. This model is an extension of the original Georgi {\it et al.} model based on $SU(5)/SO(5)$, which contains electroweak triplets. In our extension, a second Higgs doublet appears, accompanied by two additional singlets. The main motivation behind this extension is the fact that the second doublet and one singlet can be made stable thanks to a conserved $\mathbb{Z}_2$ symmetry.

We also introduced partial compositeness for the generation of the SM fermion masses: besides the usual considerations based on flavour, this set up is required in this model by the absence of a misalignment of the vacuum along a custodial-breaking triplet direction. Thus, we showed that partial compositeness with the fermions embedded in the adjoint of $SU(6)$ can avoid this issue while successfully generate fermion masses and the correct Higgs potential.  This set up can be achieved by two underlying models based on confining gauge groups $SO(7)_{\rm HC}$ and $SO(9)_{\rm HC}$, respectively.

In this work we studied in detail the vacuum structure of the model and the typical spectra of the pNGBs. We find that the typical mass split between components of the same electroweak multiplet are very small, thus breaking of the custodial symmetry at loop level is suppressed. Typical electroweak precision bounds thus force the misalignment angle to be $\sin \alpha \lesssim 0.16$, with the main contribution to the $S$ parameter coming from the strong sector (this may be reduced thanks to the resonance contribution). The pNGBs have some remarkable properties that make this model very interesting from the phenomenological point of view.

First, the $\mathbb{Z}_2$--odd states always contain a lightest neutral state that may be a good Dark Matter candidate. We identify two regions of the parameter space: one where the lightest is a CP-odd singlet, $\eta_2$, and another where the lightest is part of the second doublet, $A_0$. In the latter case, the charged Higgs remains near-degenerate, with mass splits of the order of a few GeV. On the other hand, the other neutral state $H_0$ is always heavier, thus suppressing potential direct detection bounds via the $Z$ boson exchange. This patterns is very distinctive and different from other non-composite 2HDMs.

A second interesting feature is the presence of a $\mathbb{Z}_2$--even pseudo-scalar, $\eta_p$, that does not receive a mass from the top and gauge couplings: the only mass is generated by a hyper-fermion mass, thus it can be arbitrarily small. It only couples to the electroweak gauge bosons via the WZW anomaly, while couplings to fermions are absent at linear level. We studied the bounds for light masses, which dominantly come via the couplings to photons. A lower limit is given by beam dump experiments, requiring $m_{\eta_p} \gtrsim 1$~GeV. The couplings are however too small to be reached by current collider experiments at the LHC. However, the high-luminosity run of the LHC with an integrated luminosity of 3/ab may be able to improve the bound on $\sin \alpha$ for masses $m_{\eta_p} < m_Z$ via the search for $Z \to \gamma \eta_p \to 3 \gamma$. A 100 TeV collider, however, has a much better reach, besides being able to directly produce heavier spin-1 and spin-1/2 resonances. 
Thanks to the coupling to photons, a future $e^+ e^-$ collider like FCC-ee has a very good reach for $m_{\eta_p} \lesssim 100$~GeV via production in association with a photon. Thus, a combination of $e^+ e^-$ and hadronic colliders is needed to fully probe the model.

\section*{Acknowledgements}
G.C. and A.D. acknowledge partial support from the Labex Lyon Institute of the Origins - LIO. The research of H.C. is supported by the Ministry of Science, ICT \& Future Planning of Korea, the Pohang City Government, and the Gyeongsangbuk-do Provincial Government.
G.C., A.D.  and A.K. would like to acknowledge the support of the CNRS LIA (Laboratoire International Associ{\'e}) THEP (Theoretical High Energy Physics) and the INFRE-HEPNET (IndoFrench Network on High Energy Physics) of CEFIPRA/IFCPAR (Indo-French Centre for the Promotion of Advanced Research).

\section{Appendix}\label{Appendix}
\appendix
\section{Details of the Model}\label{a:GBosons}
Generators of SU(6) :
The unbroken generators $S^i$, $i = 1, \dots 15$ and broken generators $X^a$, $a = 1, \dots 20$ satisfy the following condition:
\begin{eqnarray}
S^i. \Sigma_{EW} + \Sigma_{EW}. S^{i T} = 0  \qquad  X^a. \Sigma_{EW} -  \Sigma_{EW}. X^{a T} =0  \label{commute}
\end{eqnarray}
\begin{eqnarray}
S_{L}^i=    \frac{1}{2} 
\left(
\begin{array}{c|c}
\mathbbm{1}_2  \otimes  \sigma_i   & \\ \hline
& 0_{2} \end{array} \right)
\qquad 
S_{R}^i=   \frac{1}{2} 
\left(
\begin{array}{c|c}
\sigma_i \otimes \mathbbm{1}_2  & \\ \hline
& 0_{2} \end{array} \right)\,,   \qquad  i = 1, 2, 3
\end{eqnarray} 
where $S_{L,R}$ are generators for the $SU(2)_L\times SU(2)_R$ respectively. The generators   for the  two Higgs doublets are:
{\footnotesize{
\begin{eqnarray}
X_{10} \equiv X_{h} = \frac{1}{2 }\left(
\begin{array}{cccc|cc}
&  &  &  & 0 & 0 \\
&  &  &  & 1 & 0 \\
&  &  &  & -1 & 0 \\
&  &  &  &  0 & 0 \\ \hline
0 & 1 & -1 & 0 &  &  \\
0 & 0 & 0 & 0 & &  \\
\end{array} \right)
\qquad
X_{11} \equiv X_{G_{0}}= \frac{1}{2 } \left(
\begin{array}{cccc|cc} 
&  &  &  & 0 & 0 \\
&  &  &  & i & 0 \\
&  &  &  & i & 0 \\
&  &  &  & 0 & 0 \\ \hline
0 & -i & -i & 0 &  &  \\
0 & 0 & 0 & 0 & &  \\
\end{array} \right) 
\end{eqnarray}
\begin{eqnarray}
X_{12} \equiv X_{G_1} = \frac{1}{2 } \left(
\begin{array}{cccc|cc} 
&  &  &  & 1 & 0 \\
&  &  &  & 0 & 0 \\
&  &  &  & 0 & 0 \\
&  &  &  & 1 & 0 \\ \hline
1 & 0 & 0 & 1 & &  \\
0 & 0 & 0 & 0 & &  \\
\end{array} \right)
\qquad
X_{13} \equiv X_{G_2} = \frac{1}{2 } \left(
\begin{array}{cccc|cc}
&  &  &  & i & 0 \\
&  &  &  & 0 & 0 \\
&  &  &  & 0 & 0 \\
&  &  &  & -i & 0 \\ \hline
-i & 0 & 0 & i & &  \\
0 & 0 & 0 & 0 &  &  \\
\end{array} \right)
\end{eqnarray} 
\begin{eqnarray}
X_{14} \equiv X_{H_0} = \frac{1}{2 } \left(
\begin{array}{cccc|cc}
&  &  &  & 0 & 0 \\
&  &  &  & 0 & 1 \\
&  &  &  & 0 & -1 \\
&  &  &  & 0 & 0 \\ \hline
0 & 0 & 0 & 0 &  &  \\
0 & 1 & -1 & 0 & &  \\
\end{array} \right)
\qquad
X_{15} \equiv X_{{A}_{0}} = \frac{1}{2 } \left(
\begin{array}{cccc|cc}
&  &  &  & 0 & 0 \\
&  &  &  & 0 & i \\
&  &  &  & 0 & i \\
&  &  &  & 0 & 0 \\ \hline
0 & 0 & 0 & 0 &  &  \\
0 & -i & -i & 0 &  &  \\
\end{array} \right)
\end{eqnarray}
\begin{eqnarray}
X_{16} \equiv X_{{H}_{1}} = \frac{1}{2 } \left(
\begin{array}{cccc|cc}
&  &  &  & 0 & 1 \\
& &  & & 0 & 0 \\
& &  &  & 0 & 0 \\
&  &  &  & 0 & 1 \\ \hline
0 & 0 & 0 & 0 &  &  \\
1 & 0 & 0 & 1 & &  \\
\end{array} \right)
\qquad
X_{17} \equiv X_{{H}_{2}} = \frac{1}{2} \left(
\begin{array}{cccc|cc} 
&  &  &  & 0 & i \\
&  &  &  & 0 & 0 \\
&  &  &  & 0 & 0 \\
& &  &  & 0 &- i \\ \hline
0 & 0 & 0 & 0 &  &  \\
-i & 0 & 0 & i &  &  \\
\end{array} \right)
\end{eqnarray}}}
the  other generators associated with  other pNGBs:
\begin{eqnarray}
X_{i} = \frac{1}{2 }  \left(
\begin{array}{c|c}
\sigma_1 \otimes \sigma_i  &   \\ \hline
& 0_{2} 
\end{array} \right)
\qquad
X_{i+3} = \frac{1}{2 } 
\left(
\begin{array}{c|c}
\sigma_2 \otimes \sigma_i  & \\ \hline
& 0_{2} 
\end{array} \right)
\qquad
X_{i+6} = \frac{1}{2 }
\left(
\begin{array}{c|c}
\sigma_3 \otimes \sigma_i  & \\ \hline
& 0_{2} 
\end{array} \right) \, \label{triplet}
\end{eqnarray} 
with $i=1,2,3$, where $X_i$, $X_{i+3}$ are generators of the complex triplet $\Lambda$, and  $X_{i+6}$ are for the real triplet $\varphi$.
Three remaining broken generators $X_{18}$, $X_{19}$ and $X_{20}$ correspond  to  the singlets $\eta_{1,2,3}$:
\begin{eqnarray}
\footnotesize
X_{18} = \frac{1}{2 \sqrt{3}} \left( \begin{array}{cc|c}
\mathbbm{1}_{2} &   &  \\
& \mathbbm{1}_{2} &   \\ \hline
&   & - 2 \mathbbm{1}_{2}  \\
\end{array} \right) 
\qquad 
X_{19}  = \frac{1}{\sqrt{2}} \left( \begin{array}{cc|c}
0_2 &   &  \\
& 0_2 &   \\ \hline
&   &  \sigma_{1}  
\end{array} \right)
\qquad
X_{20}  = \frac{1}{ \sqrt{2}} \left( \begin{array}{cc|c}
0_2 &   &  \\
& 0_2 &   \\ \hline
&   &  \sigma_{3}  
\end{array} \right)\end{eqnarray}
using the definition of $\sigma_{\pm} = \frac{1}{2}( \sigma_1 \pm i \sigma_2) $ and $\widetilde{\sigma}_{\pm} = \frac{1}{2} (\mathbbm{1}_2 \pm \sigma_3) $,  where $\sigma_i$, $i=1,2,3$ are the standard Pauli matrices.The additional $9$ unbroken generators can be expressed as:
\begin{eqnarray}
\footnotesize
S_7 =   \frac{1}{2} \left(
\begin{array}{cc|c}
&  &  i \sigma_-  \\
&  & -i  \widetilde{\sigma}_+  \\ \hline
-i \sigma_+  & i \widetilde{\sigma}_+ &   
\end{array} \right) 
\quad 
S_8 =   \frac{1}{2} \left(
\begin{array}{cc|c}
&  &   \sigma_-  \\
&  &  \widetilde{\sigma}_+  \\ \hline
\sigma_+  &  \widetilde{\sigma}_+ &   
\end{array} \right)
\quad 
S_9 =   \frac{1}{2} \left(
\begin{array}{cc|c}
&  &   i\widetilde{\sigma}_+  \\
&  &  i\sigma_-  \\ \hline
- i \widetilde{\sigma}_+  &  -i\sigma_+ &   
\end{array} \right)
\end{eqnarray}
\begin{eqnarray}
\footnotesize
S_{10} =   \frac{1}{2} \left(
\begin{array}{cc|c}
&  &   \widetilde{\sigma}_+  \\
&  & - \sigma_-  \\ \hline
\widetilde{\sigma}_+  & - \sigma_+ &   
\end{array} \right) \quad 
S_{11} =   \frac{1}{2} \left(
\begin{array}{cc|c}
&  &  i  \widetilde{\sigma}_-  \\
&  & - i \sigma_+  \\ \hline
-i  \widetilde{\sigma}_-  & i \sigma_- &   
\end{array} \right)
\quad 
S_{12} =   \frac{1}{2} \left(
\begin{array}{cc|c}
&  &   \widetilde{\sigma}_-  \\
&  &  \sigma_+  \\ \hline
\widetilde{\sigma}_-  &  \sigma_- &   
\end{array} \right)
\end{eqnarray}
\begin{eqnarray}
\footnotesize
S_{13} = \frac{1}{2} \left(
\begin{array}{cc|c}
&  &  i \sigma_+  \\
&  & i  \widetilde{\sigma}_-  \\ \hline
-i \sigma_-  & -i \widetilde{\sigma}_- &   
\end{array} \right)  \qquad
S_{14} =  \frac{1}{2} \left(
\begin{array}{cc|c}
&  &   \sigma_+  \\
&  & -  \widetilde{\sigma}_-  \\ \hline
\sigma_-  & - \widetilde{\sigma}_- &   
\end{array} \right)  
\quad S_{15} = \frac{1}{\sqrt{2}} \left(
\begin{array}{cc|c}
0_{2} &    &   \\
&  0 _{2} &     \\ \hline
&  & \sigma_2   
\end{array}
\right)
\end{eqnarray}
\subsection{Projectors for 4-fermi operators}\label{a:4fermioperator}
$P_{1,2}^{\delta}$ for generating top quark mass are defined as: 
{\footnotesize{
\begin{eqnarray}
P_{t1}^{1} = - \frac{i}{2} \left(
\begin{array}{cccc|cc}
&  &  &  & 0 & 0 \\
&  &  &  & 0 & 0 \\
&  &  &  & 1 & 0 \\
&  &  &  & 0 & 0 \\ \hline
0 & 0 & 1 & 0 &  &  \\
0 & 0 & 0 & 0 &  & 
\end{array} \right) \qquad
P_{t2}^{1} = - \frac{i}{2}\left(
\begin{array}{cccc|cc}
&  &  &  & 0 & 0 \\
&  &  &  & 0 & 0 \\
&  &  &  & 0 & 1 \\
&  &  &  & 0 & 0 \\ \hline
0 & 0 & 0 & 0 & 0 & 0 \\
0 & 0 & 1 & 0 & 0 & 0
\end{array} \right)
\end{eqnarray}
\begin{eqnarray}
P_{t1}^{2} = - \frac{i}{2} \left(
\begin{array}{cccc|cc}
&  &  &  & 0 & 0 \\
&  &  &  & 0 & 0 \\
&  &  &  & 0 & 0 \\
&  &  &  & 1 & 0 \\ \hline
0 & 0 & 0 & 1 &  &  \\
0 & 0 & 0 & 0 & & 
\end{array} \right)
\qquad
P_{t2}^{2} = - \frac{i}{2} \left(
\begin{array}{cccc|cc}
&  &  &  & 0 & 0 \\
&  &  &  & 0 & 0 \\
&  &  &  & 0 & 0 \\
&  &  &  & 0 & 1 \\ \hline
0 & 0 & 0 & 0 &  &  \\
0 & 0 & 0 & 1 & & 
\end{array} \right)
\end{eqnarray}}}
$P_{1,2}^{\delta}$ for generating bottom quark mass are defined as: 
{\footnotesize{
\begin{eqnarray}
P_{b1}^{1} = \frac{i}{2} \left(
\begin{array}{cccc|cc}
&  &  &  & 1 & 0 \\
&  &  &  & 0 & 0 \\
&  &  &  & 0 & 0 \\
&  &  &  & 0 & 0 \\ \hline
1 & 0 & 0 & 0 &  &  \\
0 & 0 & 0 & 0 & & 
\end{array} \right)
\qquad
P_{b2}^{1} = \frac{i}{2} \left(
\begin{array}{cccc|cc}
&  &  &  & 0 & 1 \\
&  &  &  & 0 & 0 \\
&  &  &  & 0 & 0 \\
&  &  &  & 0 & 0 \\ \hline
0 & 0 & 0 & 0 & &  \\
1 & 0 & 0 & 0 & & 
\end{array} \right)
\end{eqnarray}
\begin{eqnarray}
P_{b1}^{2} =  \frac{i}{2} \left(
\begin{array}{cccc|cc}
&  &  &  & 0 & 0 \\
&  &  &  &1 & 0 \\
&  &  &  & 0 & 0 \\
&  &  &  & 0 & 0 \\ \hline
0 & 1 & 0 & 0 & 0 & 0 \\
0 & 0 & 0 & 0 & 0 & 0 
\end{array} \right)
\qquad
P_{b2}^{2} =  \frac{i}{2} \left(
\begin{array}{cccc|cc}
&  &  &  & 0 & 0 \\
&  &  &  & 0 & 1 \\
&  &  &  & 0 & 0 \\
&  &  &  & 0 & 0 \\ \hline
0 & 0 & 0 & 0 &  &  \\
0 & 1 & 0 & 0 &  & \end{array} \right)
\end{eqnarray}}}

\subsection{Gauge Potential}\label{a:Gaugepot}
The contribution of gauge loops to pNGB potetial at leading order. This would contribute to the masses to the corresponding pNGB.
\begin{eqnarray}
\mathcal{V}_{g}&=& \frac{C_{g} g^2 f^{2} }{4}   \bigg[  \left(3+ \tan^2 \left(\theta_W \right) \right) \bigg(  h^2~  \cos (2 \alpha )  +  H^2 ~ \cos ^2(\alpha ) -    \left( \frac{1}{2} \left( \sqrt{3} \eta _1- \sqrt{2} \eta _3\right)^2  +  \eta_2^2 \right)  \sin ^2(\alpha )    \bigg) 
\nonumber \\ &+&   A_0^2 ~   \sec ^2\left(\theta _W\right) \left(\cos (2 \alpha )+\cos \left(2 \theta_W\right)+1\right)   + 2 H_- H_+   \left(\cos (2 \alpha )+\tan ^2\left(\theta _W\right)+2\right)  \nonumber \\ &+&   \Lambda _0 \Lambda _0^* ~  \left(\cos (2 \alpha ) \left(\cos \left(2 \theta _W\right)+6\right) \sec ^2\left(\theta _W\right)+3 \left(\tan^2\left(\theta_W\right)+3\right)\right) \nonumber \\  &+& \frac{1}{4} \varphi_0^2 ~  \left(11 \cos (2 \alpha )-2 \sin ^2(\alpha ) \tan ^2\left(\theta_W\right)+21\right)  -  \varphi _0 \left(\sqrt{3} \eta _1-\sqrt{2} \eta _3\right)  \sin ^2(\alpha ) \left(1- \tan ^2\left(\theta _W\right) \right) \nonumber \\ &-&   \left( \sqrt{3} \eta _1 - \sqrt{2} \eta _3\right) \left(  \Lambda _0  + \Lambda_0^* \right) \sin ^2(\alpha ) \sec^2\left(\theta _W\right)  - \varphi _0  \left(\Lambda _0+\Lambda _0^* \right)  \sin ^2(\alpha ) \left(3- \tan^2 \left( \theta _W\right) \right) \nonumber \\ 
&+& 4 \Lambda_{++}\Lambda_{--}   \left(\cos (2 \alpha )+2 \tan ^2\left(\theta _W\right)+3\right)   +  \left(\Lambda _+ \varphi_-+\Lambda _- \varphi _+\right) \sin ^2(\alpha ) \left(3- \tan^2 \left(\theta_W \right) \right)   \nonumber \\ &+& \frac{1}{2} \varphi _- \varphi _+   \left(11 \cos (2 \alpha )+(8 (1-  \cos (\alpha ))-6 \sin^2 ( \alpha )) \tan ^2\left(\theta _W\right)+21\right) \nonumber \\ &+& \frac{1}{2}  \Lambda _- \Lambda _+  \left(11 \cos (2 \alpha )+(8 (1+  \cos (\alpha ))-6 \sin^2 ( \alpha )) \tan ^2\left(\theta _W\right)+21\right) \bigg] 
\end{eqnarray}
Expanding the gauge potential into the cubic order,  we find the trilinear interaction related to DM candidates of  $\eta_2$ and the second Higgs doublet:
\begin{eqnarray}
\mathcal{V}_{g}(\eta_2) & =& \frac{C_{g} g^2 f}{\sqrt{2} } \eta_2 \bigg[ \frac{1}{3}  \sin (2 \alpha ) \left(2 \tan^2 \left(\theta_W \right)+3 \right) ~ i  A_0  \left(\Lambda
	_{0}^*-\Lambda _0\right)   \nonumber \\ 
&+&  \frac{1}{3}    \sin (2 \alpha )~  H_0   \left(\sqrt{3} \left(\tan^2 \left(\theta_W \right)+3 \right)   \eta _1+ \left( 2\tan^2 \left(\theta_W \right)+3 \right) \left(\Lambda_0+\Lambda _{0}^*\right)\right) \nonumber \\ 
	&-&  \frac{1}{3}    \sin (2 \alpha )\left(\tan^2 \left(\theta_W \right)-3 \right) ~  H_0   \varphi_0  + \bigg( \sin (\alpha)  \tan^2 \left(\theta_W \right) ~  H_+ \left(\Lambda _-+\varphi _-\right)  \nonumber \\ 
&+& \frac{1}{6} \sin (2 \alpha ) \left(\tan^2 \left(\theta_W \right)+6 \right) ~   H_+ \left(\Lambda _--\varphi
	_-\right)  + h.c.  \bigg) \bigg] 
\end{eqnarray}
as well as the first  Higgs interacting with  other scalars:
\begin{eqnarray}
\mathcal{V}_{g}(h) & =&- \frac{C_{g}g^2 f}{2} h \bigg[   \sin (2\alpha ) 
\left(\tan^2 \left(\theta_W \right)+ 1 \right)  A_0^2  + \frac{1}{3 }\sin (2 \alpha )  \left(\tan^2 \left(\theta_W \right)+3 \right)  ( h^2+ H^2_0)  \nonumber \\ &+&  \frac{1}{6} \sin (2 \alpha ) \left(\tan^2 \left(\theta_W \right)+15 \right)   \varphi _0^2  + \frac{1}{3}  \sin (2\alpha )  \left(\tan^2 \left(\theta_W \right)+3 \right) \eta _2^2  \nonumber \\ &+& \frac{1}{6 }    \sin (2 \alpha ) \left(\tan^2 \left(\theta_W \right)+3\right) (\sqrt{3} \eta_1 - \sqrt{2} \eta_3)^2 + \frac{2}{3}   \sin (2 \alpha ) \left(7 \tan^2 \left(\theta_W \right)+9 \right) \Lambda _0  \Lambda_{0}^* \nonumber \\
&+& \frac{1}{3}\sin (2 \alpha ) \left(\sqrt{3} \eta _1- \sqrt{2} \eta _3\right)  \left( \left(2 \tan^2 \left(\theta_W \right)+3\right)   (\Lambda _{0}+ \Lambda_0^*) -  \left(\tan^2 \left(\theta_W \right)-3 \right)  \varphi _0   \right) \nonumber \\ 
&+&\frac{1}{3 }  \sin (2\alpha ) \left(2 \tan^2 \left(\theta_W \right)-3 \right)
\left( \left(\Lambda _+ \varphi _-+\Lambda _- \varphi _+\right) -  \varphi _0 \left(\Lambda _0+\Lambda _{0}^* \right) \right) 	 \nonumber \\ &+&  4 \sin (2\alpha ) \Lambda _{--} \Lambda _{++} + \sqrt{2}   \sin (2 \alpha )  H_- H_+   + 2 \sin (\alpha ) \left( \tan^2 \left(\theta_W \right)\right) \left(\Lambda _+ \Lambda _-  - \varphi _- \varphi _+ \right)  \nonumber \\ &+&  \frac{1}{3 }  \sin (2\alpha )  \left(4 \tan^2 \left(\theta_W \right)+15 \right)  \left( \Lambda _+ \Lambda _- +   \varphi _+ \varphi _- \right)  \bigg]
\end{eqnarray}

\subsection{HF Mass potenial} \label{a:HFmass}
The HF mass term will generate mass terms for the pNGB: 
\begin{eqnarray}
\mathcal{V}_{m} &=& \frac{ B f }{\sqrt{2}} \bigg[
2  h^2 \left(m_1+m_2\right) \cos (2 \alpha ) + 2   A_0^2 \left(m_1+m_2\right)+ 2  H_0^2 \left(m_1+m_2\right) \cos ^2(\alpha ) \nonumber \\
&-& \frac{1}{2}    \varphi _0^2 \left( 2 m_1 \sin ^2(\alpha )- m_2 (\cos (2 \alpha )+7)\right) - \varphi _- \varphi _+ \left(  2 m_1 \sin ^2(\alpha )- m_2 (\cos (2 \alpha )+7)\right) \nonumber \\ &-& 2  \Lambda _0 \Lambda _{0}^* \left(2 m_1 \sin ^2(\alpha )-m_2 (\cos (2 \alpha )+3)\right) -   \Lambda _- \Lambda _+ \left(2 m_1 \sin ^2(\alpha ) - m_2 (\cos (2 \alpha )+7)\right) \nonumber \\ &+& 8  m_2 \Lambda _{--} \Lambda _{++} + 2   \left(\Lambda _+ \varphi _-+\Lambda _- \varphi _+\right)  \sin ^2(\alpha ) \left(m_1+m_2 \right)  + 4  H_- H_+ \left(m_1+m_2\right)   \nonumber \\ &+& \frac{2  }{\sqrt{3}} \eta _1 \sin ^2(\alpha ) \left(m_1+m_2 \right) \left(\Lambda _0+\Lambda_{0}^*\right) - 2 \varphi _0 \sin ^2(\alpha ) \left(m_1 + m_2 \right) \left(\Lambda _0+\Lambda
_{0}^*\right) \nonumber \\& + & \frac{2  }{\sqrt{3} } \eta _1 \varphi _0 \sin ^2(\alpha ) \left(m_1+ m_2 \right) 
+ \frac{1}{6}  \eta _1^2 \left(\left(m_1+m_2\right) (5 \cos (2 \alpha )+3)+8 m_1\right) \nonumber  \\
&+& 2  \left(\eta _2^2 +\eta_3^2\right) \left(m_1 \left(\cos ^2(\alpha )+1\right)-m_2 \sin ^2(\alpha )\right)   + 4 \sqrt{\frac{2}{3}} \eta _1 \eta _3 \sin ^2(\alpha ) \left(m_1+m_2\right)  \bigg]  
\end{eqnarray}
Expanding to the cubic order, we can find the following interactions:
\begin{eqnarray}
\mathcal{V}_m ({\eta_2}) &=&   \frac{ 4 B m \sin (2 \alpha )  }{3}  \eta _2 \bigg(-i A_0 \left(\Lambda
   _0-\Lambda _{0}^*\right)+ \left( H_+ \left(\Lambda _--\varphi _-\right) + h.c. \right) \nonumber \\ &+& H_0 \left(\sqrt{3} \eta _1+\Lambda _0+\Lambda _{0}^*+\varphi
   _0\right)\bigg)
\end{eqnarray}

\begin{eqnarray}
\mathcal{V}_m({h} ) &=&- \frac{ 2 B m \sin (2 \alpha )  }{3 } h \bigg( 2 \sqrt{2}  \left(h^2 + H_0^2 \right) 
+   \sqrt{2} \eta _1^2-2 \sqrt{3} \eta _3 \eta _1 + 2 \sqrt{2} \eta _3^2  \nonumber \\ &+& 2 \sqrt{2} \left(\eta_2^2+\left(\Lambda _--\varphi _-\right) \left(\Lambda _+-\varphi_+\right)\right) - 2 \eta _3 \left(\Lambda _0+\Lambda _{0}^*+\varphi _0\right) \nonumber \\ &+& \sqrt{2} \left(2
   \Lambda _0+\varphi _0\right) \left(2 \Lambda _{0}^*+\varphi _0\right)\bigg)
\end{eqnarray}

\section{An alternative complex basis} \label{complex basis}
In this section, we modify the real basis by changing  the broken generators related to $H_1$ and $H_2$ and 2 singlets, so that the two Higgs mix in $\Pi$.  We can call it  complex basis since this translates the $SO(2)$ symmetry into a $U(1)$ symmetry. We choose a slightly different  $\Sigma_{\rm EW}$  for this new basis:
\begin{eqnarray}
\Sigma_{\rm EW}  =\left(
\begin{array}{cc|cc}
& -i \sigma_2 &  \\
i \sigma_2 &  & \\ \hline
& & \sigma_1
\end{array} \right)
\end{eqnarray}
so that the broken and unbroken generators are defined with respect to this reference vacuum according to Eq.{\ref{commute}}. For the 2 Higgs doublets: 
{\footnotesize{
\begin{eqnarray}
X_{10} \equiv X_h = \frac{1}{2 \sqrt{2}} \left(
\begin{array}{cccc|cc}
&  &  &  & 0 & 0 \\
&  &  &  & 1 & -1 \\
&  &  &  & -1 & 1 \\
&  &  &  & 0 & 0 \\ \hline
0 & 1 & -1 & 0 &  &  \\
0 & -1 & 1 & 0 & & 
\end{array} \right) \,, \, 
X_{11} \equiv X_{G_0 }= - \frac{1}{2 \sqrt{2}} \left(
\begin{array}{cccc|cc}
& &  &  & 0 & 0 \\
&  &  &  & i & -i \\
&  &  &  & i & -i \\
&  &  &  & 0 & 0 \\ \hline
0 & -i & -i & 0 &  & \\
0 & i & i & 0 &  & 
\end{array} \right)
\end{eqnarray}
\begin{eqnarray}
X_{12}  \equiv X_{G_1} = - \frac{1}{2 \sqrt{2}} \left(
\begin{array}{cccc|cc}
&  &  &  & 1 & -1 \\
&  &  &  & 0 & 0 \\
&  &  &  & 0 & 0 \\
&  &  &  & 1 & -1 \\ \hline
1 & 0 & 0 & 1 &  &  \\
-1 & 0 & 0 & -1 &  & 
\end{array} \right) \,, \, 
X_{13}  \equiv X_{G_2} = - \frac{1}{2 \sqrt{2}} \left(
\begin{array}{cccc|cc}
&  &  &  & -i & i \\
&  &  &  & 0 & 0 \\
&  &  &  & 0 & 0 \\
&  &  &  & i & -i \\ \hline
i & 0 & 0 & -i & &  \\
-i & 0 & 0 & i &  & 
\end{array} \right)
\end{eqnarray} 
\begin{eqnarray}
X_{14} \equiv X_{H_0} = \frac{1}{2 \sqrt{2}} \left(
\begin{array}{cccc|cc}
 &  &  &  & 0 & 0 \\
 &  &  &  & - i & -i \\
 &  &  &  & i & i \\
&  &  &  & 0 & 0 \\ \hline
0 & i & -i & 0 &  &  \\
0 & i & -i & 0 & & 
\end{array} \right) \,, \, 
X_{15} \equiv X_{A_0} = - \frac{1}{2 \sqrt{2}} \left(
\begin{array}{cccc|cc}
& &  &  & 0 & 0 \\
&  &  &  & 1 & 1 \\
&  &  &  & 1 & 1 \\
&  &  &  & 0 & 0 \\ \hline
0 & 1 & 1 & 0 &  & \\
0 & 1 & 1 & 0 &  & 
\end{array} \right)
\end{eqnarray}
\begin{eqnarray}
X_{16} \equiv X_{H_1} = \frac{1}{2 \sqrt{2}} \left(
\begin{array}{cccc|cc}
 &  &  &  & -i & -i \\
 &  &  &  & 0 & 0 \\
 &  &  &  & 0 & 0 \\
&  &  &  & -i & -i \\ \hline
i &  &  & i &  &  \\
i &  &  & i & & 
\end{array} \right) \,, \, 
X_{17} \equiv X_{H_2} = - \frac{1}{2 \sqrt{2}} \left(
\begin{array}{cccc|cc}
& &  &  & -1 & -1 \\
&  &  &  & 0 & 0 \\
&  &  &  & 0 & 0 \\
&  &  &  & 1 & 1 \\ \hline
-1 &  &  & 1 &  & \\
-1 &  &  & 1 &  & 
\end{array} \right)
\end{eqnarray}
}}
The broken generators $X_{i}$, $i=1,\cdots,  9$ associated with the complex  and real triplets  are the same as Eq.\ref{triplet} in the real basis,  so is the generator for  the $\eta_{1}$ singlet, while  the generators for $\eta_2$ and $\eta_3$ switch with each other :
{\footnotesize{
\begin{eqnarray}
\footnotesize
X_{18} = \frac{1}{2 \sqrt{3}} \left( \begin{array}{cc|c}
\mathbbm{1}_{2} &   &  \\
 & \mathbbm{1}_{2} &   \\ \hline
&   & - 2 \mathbbm{1}_{2}  \\
\end{array} \right) 
\quad 
X_{19}  = -\frac{1}{\sqrt{2}} \left( \begin{array}{cc|c}
 0_2 &   &  \\
 & 0_2 &   \\ \hline
&   &  \sigma_{2}  
\end{array} \right)
\quad
X_{20}  = -\frac{1}{ \sqrt{2}} \left( \begin{array}{cc|c}
 0_2 &   &  \\
 & 0_2 &   \\ \hline
&   &  \sigma_{1}  
\end{array} \right)\end{eqnarray}
}}
For the unbroken generators,  the EW  $SU(2)_L \times SU(2)_R$  generators and the one associated with $\beta$ rotation are:
\begin{eqnarray}
S_{L}^i=    \frac{1}{2} 
\left(
\begin{array}{c|c}
\mathbbm{1}_2  \otimes  \sigma_i   & \\ \hline
& 0_{2} \end{array} \right) \,, 
\quad 
S_{R}^i=   \frac{1}{2} 
\left(
\begin{array}{c|c}
\sigma_i \otimes \mathbbm{1}_2  & \\ \hline
& 0_{2} \end{array} \right)\,,   \quad
S_{15}=    \frac{1}{\sqrt{2}}
\left(
\begin{array}{c|c}
0_4    & \\ \hline
& - \sigma_{3} \end{array} \right)
\end{eqnarray} 
with  $i = 1, 2, 3 $.  The remaining  unbroken generators  will be omitted due to irrelevance for Lagrangian.  Note that in this complex basis,  EW symmetry is embedded  in  an appropriate approach so that  the electric charge  equals  $Q= S_L^3 + S_R^3$. The vacuum misalignment can  be achieved by a $SU(6)$ rotation $U(\alpha, \beta) = U(\beta)U(\alpha) U(\beta)^\dagger$, with $U(\alpha)$ and $U(\beta)$ expressed as:
{\footnotesize{
\begin{eqnarray}
U(\alpha)= \left(
\begin{array}{cccccc}
 1 & 0 & 0 & 0 & 0 & 0 \\
 0 & \cos ^2\left(\frac{\alpha }{2}\right) & \sin ^2\left(\frac{\alpha }{2}\right) & 0 &
   \frac{1}{2} i \sin (\alpha ) & -\frac{1}{2} i \sin (\alpha ) \\
 0 & \sin ^2\left(\frac{\alpha }{2}\right) & \cos ^2\left(\frac{\alpha }{2}\right) & 0 &
   -\frac{1}{2} i \sin (\alpha ) & \frac{1}{2} i \sin (\alpha ) \\
 0 & 0 & 0 & 1 & 0 & 0 \\
 0 & \frac{1}{2} i \sin (\alpha ) & -\frac{1}{2} i \sin (\alpha ) & 0 & \cos
   ^2\left(\frac{\alpha }{2}\right) & \sin ^2\left(\frac{\alpha }{2}\right) \\
 0 & -\frac{1}{2} i \sin (\alpha ) & \frac{1}{2} i \sin (\alpha ) & 0 & \sin
   ^2\left(\frac{\alpha }{2}\right) & \cos ^2\left(\frac{\alpha }{2}\right) \\
\end{array}
\right)\,, 
\end{eqnarray}
\begin{eqnarray}
U(\beta) = e^{- i \sqrt{2} \,\beta \, \bf{S_{15}} } = \left(
\begin{array}{cccccc}
 1 & 0 & 0 & 0 & 0 & 0 \\
 0 & 1 & 0 & 0 & 0 & 0 \\
 0 & 0 & 1 & 0 & 0 & 0 \\
 0 & 0 & 0 & 1 & 0 & 0 \\
 0 & 0 & 0 & 0 & e^{+i \beta } & 0 \\
 0 & 0 & 0 & 0 & 0 & e^{-i \beta } \\
\end{array}
\right)
\end{eqnarray}}}
The $\Pi$ is parameterised in terms of Goldstone fields  as:
\begin{eqnarray}
\tiny
\Pi=\left(
\begin{array}{cccccc}
\frac{1}{6} \left(\sqrt{3} \eta _1+3 \varphi _0\right) & \frac{\varphi _+}{\sqrt{2}} &
\frac{\Lambda _+}{\sqrt{2}} & \Lambda _{++} & \frac{1}{2} \left(G_+-i H_+\right) &
\frac{1}{2} \left(-G_+-i H_+\right) \\
\frac{\varphi _-}{\sqrt{2}} & \frac{1}{6} \left(\sqrt{3} \eta _1-3 \varphi _0\right) &
\Lambda _0 & -\frac{\Lambda _+}{\sqrt{2}} & \frac{h+A_0+i G_0-i H_0}{2 \sqrt{2}} &
\frac{A_0-i \left(-i h+G_0+H_0\right)}{2 \sqrt{2}} \\
\frac{\Lambda _-}{\sqrt{2}} & \Lambda _{0}^* & \frac{1}{6} \left(\sqrt{3} \eta
_1-3 \varphi _0\right) & -\frac{\varphi _+}{\sqrt{2}} & \frac{A_0+i \left(i
	h+G_0+H_0\right)}{2 \sqrt{2}} & \frac{h+A_0-i G_0+i H_0}{2 \sqrt{2}} \\
\Lambda _{--} & -\frac{\Lambda _-}{\sqrt{2}} & -\frac{\varphi _-}{\sqrt{2}} &
\frac{1}{6} \left(\sqrt{3} \eta _1+3 \varphi _0\right) & \frac{1}{2} \left(G_--i
H_-\right) & \frac{1}{2} \left(-G_--i H_-\right) \\
\frac{1}{2} \left(G_-+i H_-\right) & \frac{h+A_0-i G_0+i H_0}{2 \sqrt{2}} & \frac{A_0-i
	\left(-i h+G_0+H_0\right)}{2 \sqrt{2}} & \frac{1}{2} \left(G_++i H_+\right) &
-\frac{\eta _1}{\sqrt{3}} & \frac{i \left(\eta _2+i \eta _3\right)}{\sqrt{2}} \\
\frac{1}{2} \left(i H_--G_-\right) & \frac{A_0+i \left(i h+G_0+H_0\right)}{2 \sqrt{2}}
& \frac{h+A_0+i G_0-i H_0}{2 \sqrt{2}} & \frac{1}{2} \left(i H_+-G_+\right) &
\frac{-i \eta _2-\eta _3}{\sqrt{2}} & -\frac{\eta _1}{\sqrt{3}} \\
\end{array}
\right) \nonumber \\ \end{eqnarray}
\begin{eqnarray}
\footnotesize
2 ~\Pi = \left(\begin{array}{cccc}
\varphi + \frac{\eta _1}{\sqrt{3}} \mathbbm{1}_{2} &  \Lambda    &  {H}_{1} - i H_2  & - {H}_{1} - i H_2 \\
\Lambda^{\dagger}     & - \varphi + \frac{\eta _1}{\sqrt{3}} \mathbbm{1}_{2} & - \widetilde{{H}_{1}} - i \widetilde{{H}_{2}}  &  \widetilde{{H}_{1}} - i \widetilde{{H}_{2}} \\         
\left({H}_{1} - i H_2 \right)^{\dagger} & \left(-\widetilde{{H}_{1}} - i \widetilde{{H}_{2}} \right)^{\dagger} & -\frac{2\eta
	_1}{\sqrt{3}}   & - \sqrt{2} \left(\eta_3 - i \eta _2  \right)\\
\left(- {H}_{1} - i H_2 \right)^{\dagger} & \left( \widetilde{H_{1}} - i \widetilde{H_2} \right)^{\dagger}  &  -\sqrt{2}\left( \eta_3 + i \eta _2  \right) & - \frac{2 \eta
	_1}{\sqrt{3}}  \\
\end{array} \right)
\end{eqnarray}
where
\begin{eqnarray}
\varphi = \sigma^{a} \varphi^{a}  \equiv 
\begin{pmatrix}
\varphi^{0}         & \sqrt{2}\varphi^{+} \\
\sqrt{2}\varphi^{-} & - \varphi^{0} \\
\end{pmatrix}
\qquad 
\Lambda = \sigma^{a} \Lambda^{a}  \equiv 
\begin{pmatrix}
\sqrt{2}\Lambda^{+}         & 2 \Lambda^{++} \\
2\Lambda_{0} & - \sqrt{2}\Lambda^{+} \\
\end{pmatrix}
\end{eqnarray}
We have verify that all  interactions in the complex basis ($U(1)$ basis) are the same as the real basis ($SO(2)$ basis). The $\eta_2$ becomes the imaginary part  compared with the real basis.The  DM  and CP parities are  less trivial in this complex basis .
\begin{eqnarray}
\footnotesize
\Omega_{DM} =\left( \begin{array}{cc|c}
\mathbbm{1}_{2} &   &  \\
& \mathbbm{1}_{2} &   \\ \hline
&   & -  \sigma_{1}  \\
\end{array} \right) , \quad  \Omega_{\rm DM} \Sigma_\alpha (H_2, \eta_2)\Omega_{\rm DM} =  \Sigma_\alpha (- H_2, -\eta_2)
\end{eqnarray}
\begin{eqnarray}
\footnotesize
\Omega_{CP} =\left( \begin{array}{cc|c}
\mathbbm{1}_{2} &   &  \\
& \mathbbm{1}_{2} &   \\ \hline
&   &   \sigma_{1}  \\
\end{array} \right) , \quad  \Omega_{\rm CP} \Sigma_{\alpha, \beta} (\phi_{odd})\Omega_{\rm CP} =  \Sigma_{\alpha, \beta}^\dagger (- \phi_{odd})
\end{eqnarray}
For the partial compositeness,  the left handed top spurions in the symmetric and antisymmetric representation are simply linear combinations: $D_{S2} \pm D_{S4}$ and $D_{A1} \pm D_{A3}$. 

\subsection{Coupling vertices}\label{a:GBvertex}
The couplings of 2 gauge bosons with SM Higgs field can be written as:
\begin{eqnarray}
\mathcal{L}_{h} & = &\frac{g^2 }{8} f h \sin (2 \alpha ) \left(2 W_\mu^- W_\mu^+ + Z_\mu^2 \sec ^2(\theta_w)\right)
\end{eqnarray}
The couplings of 2 gauge bosons with 2 scalars can be written as:
\begin{eqnarray}
\mathcal{L}_{W^- W^-} & = & - \frac{1}{4} g^2 W^{-\mu } W_{\mu }{}^- \bigg(3 \Lambda _+ \varphi _+ \sin^2(\alpha )+2 \Lambda _+^2 \sin ^4\left(\frac{\alpha }{2}\right)+2 \varphi _+^2 \cos^4\left(\frac{\alpha }{2}\right)-H_+^2 \sin ^2(\alpha ) \nonumber \\  &+ & \Lambda _{++} \left(4 \left(\Lambda _0 \sin^4\left(\frac{\alpha }{2}\right)+ \Lambda_{0}^* \cos ^4\left(\frac{\alpha }{2}\right) \right) - \sin ^2(\alpha) \left(\sqrt{3} \eta _1-\sqrt{2} \eta _3+3 \varphi _0\right)\right)\bigg) \\
   \mathcal{L}_{W^+ Z} & = & \frac{1}{4} g^2  \sec (\theta_{w}) W^+_\mu Z^\mu   \bigg(H_- \left(-2 H_0 \cos (\alpha ) \sin^2 (\theta_{w}) + i A_0 (\cos (2 \theta_{w})-\cos (2 \alpha ))\right)  \nonumber \\ &+ & 4 \Lambda _{--} \varphi _+ \sin ^2\left(\frac{\alpha }{2}\right) (2 \sin^2 (\frac{\alpha}{2})-3 \cos^2 (\theta_{w})) + 4  \Lambda _{--}  \Lambda _+ \cos^2\left(\frac{\alpha }{2}\right)(2 \cos^2(\frac{\alpha}{2})-3 \cos^2(\theta_{w}))  \nonumber \\ &+&  \Lambda _- \bigg( 4 \Lambda _0 \cos ^2\left(\frac{\alpha }{2}\right) \left(\cos^2(\theta_{w})-2 \cos (\alpha )\right)+2 \varphi _0 \sin^2\left(\frac{\alpha }{2}\right) (\cos (\alpha )-\cos (2 \theta_{w}))\bigg) \nonumber \\ &+&  \varphi _-  \bigg(4 \Lambda _0^* \sin ^2\left(\frac{\alpha}{2}\right) (\cos^2( \theta_{w}) +2 \cos (\alpha ) )-2 \varphi _0 \cos^2\left(\frac{\alpha }{2}\right) (\cos (\alpha )+\cos (2 \theta_{w}))\bigg) \nonumber \\ & +  & ( \Lambda _- +  \varphi _- ) \left(\sqrt{3} \eta _1-\sqrt{2} \eta_3\right) \sin ^2(\alpha )  \bigg)
   \end{eqnarray}

\begin{eqnarray}
\mathcal{L}_{W^+ W^-} & = &   \frac{1}{16} g^2  W_\mu^+ W_\mu ^- \bigg(  4 \left( h^2 \cos (2 \alpha ) +  A_0^2 + H_0^2 \cos ^2(\alpha )  \right)+ \varphi_0^2 (5 \cos (2 \alpha )+11)   \nonumber \\ & - & 2 \sin ^2(\alpha )\left( \left(\sqrt{3} \eta _1 - \sqrt{2} \eta _3\right)^2+ 2 \eta_2^2 \right) - 4  \varphi _0 \left(\sqrt{3} \eta _1 - \sqrt{2} \eta _3 + 2 \left(\Lambda _0+ \Lambda _0^*\right) \right)  \nonumber \\ &+&  4 \Lambda _0 \Lambda _0^* (\cos (2 \alpha )+3)  + 8 H_+ H_- \cos ^2(\alpha )  + 16 \varphi _- \varphi _+ ( \sin^2 (\frac{\alpha}{2} )+\cos^2 (\alpha))   \nonumber \\  &+& 16 \left(\Lambda _- \Lambda _+  ( \cos^2 (\frac{\alpha}{2} )+\cos^2 (\alpha))  +  \frac{\sin ^2(\alpha )}{2} (\varphi _+  \Lambda _-+ \varphi_+\Lambda_-) +  \cos ^2(\alpha ) \Lambda _{++} \Lambda _{--} \right)     \bigg) \nonumber \\
   \end{eqnarray}

 \begin{eqnarray}
  \mathcal{L}_{Z Z}   &= & \frac{1}{16} g^2 \sec ^2(\theta_{w}) Z^{\mu } Z_{\mu } \bigg( 2  \cos (2 \alpha ) \left( h^2 + H_0^2 + A_0^2 \right) + 2 \Lambda_0 \Lambda _0^* (5 \cos (2 \alpha )+3) \nonumber \\ &-& \sin ^2(\alpha ) \left(\left(\sqrt{3} \eta _1-\sqrt{2} \eta _3\right){}^2 +2 \eta _2^2 +\varphi _0^2\right) + 2 \sin ^2(\alpha )  \varphi _0 \left(\sqrt{3} \eta _1-\sqrt{2} \eta _3+\Lambda_0+\Lambda _0^*\right)  \nonumber \\ &+& 2 \Lambda _- \Lambda _+ \left(3 \cos ^2(\alpha ) -4 \cos (\alpha )\cos (2 \theta_{w})   + \cos (4 \theta_{w})\right) + 4 \cos ^2(2 \theta_{w}) \left(H_- H_+ + 4 \Lambda _{++} \Lambda _{--}\right) \nonumber \\ &+& 2\varphi _- \varphi _+ \left(3 \cos^2(\alpha )+4 \cos (\alpha ) \cos (2 \theta_{w})+\cos (4 \theta_{w})\right) -2 \left( \Lambda _- \varphi _+ + \Lambda_+ \varphi_- \right) \sin ^2(\alpha ) \nonumber \\&-&  2 \sin ^2(\alpha )  \left(\sqrt{3} \eta _1-\sqrt{2} \eta _3\right) \left(\Lambda_0+\Lambda _0^*\right)  \bigg) 
\end{eqnarray}

\begin{eqnarray} 
 \mathcal{L}_{W^+ \gamma}  &= &  \frac{1}{2} g^2  \sin (\theta_{w}) W_+^{\mu } \gamma_{\mu } \bigg( H_- \left(H_0 \cos (\alpha ) + i A_0\right) -6  \sin^2(\frac{\alpha}{2}) \varphi _+ \Lambda _{--}- 6  \cos ^2(\frac{\alpha}{2}) \Lambda _+ \Lambda _{--}   \nonumber \\ &+&  2 \Lambda _- \left(\Lambda _0  \cos ^2(\frac{\alpha}{2}) -\varphi _0  \sin^2(\frac{\alpha}{2}) \right) + 2 \varphi _- \left( \Lambda _0^*  \sin^2(\frac{\alpha}{2})- \varphi _0 \cos ^2(\frac{\alpha}{2}) \right) \bigg) \\
\mathcal{L}_{Z \gamma} & = &  g^2 \tan (\theta_{w}) Z^{\mu } \gamma_{\mu }  \bigg(    \left(\cos (2\theta_w) + \cos (\alpha ) \right) \varphi _-
\varphi _+  +  \left(\cos (2\theta_w) - \cos (\alpha ) \right) \Lambda _- \Lambda _+  \nonumber \\ &+&\cos (2 \theta_{w}) \left(H_- H_+  + 4 \Lambda_{--}  \Lambda_{++}\right)\bigg)  
\\
\mathcal{L}_{\gamma \gamma} & = &   g^2 \sin ^2(\theta_{w})  \gamma ^{\mu }  \gamma_{\mu } \left(H_- H_++\Lambda _- \Lambda _++\varphi _- \varphi _++4 \Lambda_{--} \Lambda_{++}\right) 
\end{eqnarray}
using  the notation of  $ \pi_1 \overset{\leftrightarrow}{\partial}_ \mu \pi_2 = \pi_1 \partial_\mu \pi_2 - \partial_\mu \pi_1 \pi_2$,   the differential  gauge interactions are:
\begin{eqnarray}
\mathcal{L}_{W^+} & = &\frac{i}{2} g W_+^{\mu } \bigg(  H_{-}  \overset{\leftrightarrow}{\partial}_\mu \left(i A_0 +   \cos (\alpha ) H_0 \right)  + 2   \cos^2(\frac{\alpha}{2})   \left( \Lambda_{+} \overset{\leftrightarrow}{\partial}_\mu \Lambda_{--}+ \Lambda _- \overset{\leftrightarrow}{\partial}_\mu \Lambda_{0}   + \varphi _0 \overset{\leftrightarrow}{\partial}_\mu \varphi_{-}  \right)  \nonumber \\  &+&   2   \sin^2(\frac{\alpha}{2})  \left(  \varphi_{+} \overset{\leftrightarrow}{\partial}_\mu \Lambda_{--} +  \varphi _- \overset{\leftrightarrow}{\partial}_\mu \Lambda_{0}^* + \varphi _0 \overset{\leftrightarrow}{\partial}_\mu \Lambda_{-}  \right) \bigg)
\\ 
\mathcal{L}_{Z} & = & -\frac{1}{2} i g \sec (\theta_{w}) Z^{\mu } \bigg( \left(\cos (2 \theta_{w}) - \cos (\alpha )  \right)  \Lambda _+ \overset{\leftrightarrow}{\partial}_\mu \Lambda_{-}  + \left( \cos (2 \theta_{w}) +  \cos (\alpha )  \right) \varphi _+\overset{\leftrightarrow}{\partial}_\mu \varphi_{-}  \nonumber \\
&+& \cos (2 \theta_{w}) \left(H_+ \overset{\leftrightarrow}{\partial}_\mu  H_{-} + 2 \Lambda_{++} \overset{\leftrightarrow}{\partial}_\mu \Lambda_{--} \right) +  \cos (\alpha ) \left(i H_0 \overset{\leftrightarrow}{\partial}_\mu A_{0}+2 \Lambda _0^* \overset{\leftrightarrow}{\partial}_\mu \Lambda_{0} \right) \bigg)
\\
\mathcal{L}_{\gamma} & = & -i g  \sin (\theta_{w}) \gamma ^{\mu } \bigg(H_+ \overset{\leftrightarrow}{\partial}_\mu H_{-} + \Lambda_+  \overset{\leftrightarrow}{\partial}_\mu \Lambda_{-} +\varphi _+ \overset{\leftrightarrow}{\partial}_\mu \varphi_{-}  +  2 \Lambda_{++}  \overset{\leftrightarrow}{\partial}_\mu \Lambda_{--}  \bigg)
\end{eqnarray}

\subsection{Most general vacuum}\label{a:GeneralVac}
The most general vacuum structure is defined in terms of three parameter: two generators aligned along the two Higgs doublet and third one along the generator related to $ \lambda_0 =  - \frac{i}{\sqrt{2}} (\Lambda_0 - \Lambda_0^*)$. The new vacuum is defined in terms of three angles:  
\begin{eqnarray}
\sigma(x)_{\alpha,\beta,\gamma} = U(\alpha,\beta,\gamma) \cdot \sigma(x)\cdot  U(\alpha,\beta,\gamma)^{\dagger}
\end{eqnarray}
The most general CP conserving vacuum is defined as:
\begin{eqnarray}
U(a,b,c) = e^{i a X_{10}+ i b X_{14}+ i c T_{\lambda_0} }
\end{eqnarray}
\begin{eqnarray}
\footnotesize
T_{\lambda_0}= \left(
\begin{array}{cccc|c}
0 & 0 & 0 & 0 &     \\
0 & 0 & - \frac{i}{\sqrt{2}} & 0 &    \\
0 &  \frac{i}{\sqrt{2}}  & 0 & 0 &   \\
0 & 0 & 0 & 0 &    \\  \hline
&  &  &  &  0_2    \\
\end{array}
\right)
\end{eqnarray}
The generators closing under the SU(2) algebras are:
$(X_{10},X_{14},S_{15})$, $(X_{10},S_{8},T_{\lambda_0})$, $(X_{14}, S_{12},T_{\lambda_0})$  and $(S_{8},S_{12}, S_{15})$.
Hence the above matrix  in the complex basis can be parametrised as:
\begin{eqnarray}
&&U(\alpha,\beta,\gamma) = e^{- i \beta\sqrt{2} S_{15}}\cdot e^{ i \gamma \sqrt{2} S_{8}}\cdot e^{i \alpha \sqrt{2} X_{10}}\cdot e^{- i \gamma \sqrt{2} S_{8}} \cdot e^{i \beta\sqrt{2} S_{15}} 
\end{eqnarray}
\begin{eqnarray}
\footnotesize
\left(
\begin{array}{cccccc}
1 & 0 & 0 & 0 & 0 & 0 \\
0 & c_{\frac{\alpha}{2}}^2- s^2_\frac{\alpha}{2} s_\gamma^2  &  s^2_{\frac{\alpha}{2}} c_\gamma^2- s_\alpha s_\gamma  & 0 &  i e^{-i \beta } c_\gamma  (\frac{s_\alpha}{2}+  s^2_{\frac{\alpha}{2}} s_\gamma) & -i  e^{i \beta } c_\gamma (\frac{s_\alpha}{2}+ s^2_{\frac{\alpha}{2}} s_\gamma) \\
0 & s^2_{\frac{\alpha}{2}} c_\gamma^2+ s_\alpha  s_\gamma  & c_{\frac{\alpha}{2}}^2- s^2_\frac{\alpha}{2} s_\gamma^2 & 0 & - i  e^{-i \beta } c_\gamma (\frac{s_\alpha}{2} - s^2_{\frac{\alpha}{2}} s_\gamma) & i  e^{i \beta } c_\gamma (\frac{s_\alpha}{2} -  s^2_{\frac{\alpha}{2}} s_\gamma ) \\
0 & 0 & 0 & 1 & 0 & 0 \\
0 & i  e^{i \beta } c_\gamma  (\frac{s_\alpha}{2} -  s^2_{\frac{\alpha}{2}} s_\gamma) & -i  e^{i \beta } c_\gamma (\frac{s_\alpha}{2} + s^2_{\frac{\alpha}{2}} s_\gamma) & 0 &  c^2_{\frac{\alpha}{2}}+ s^2_{\frac{\alpha}{2}} s_\gamma^2  & e^{2 i \beta } c_\gamma^2  s^2_{\frac{\alpha }{2}} \\
0 & -i  e^{-i \beta } c_\gamma (\frac{s_\alpha}{2} - s^2_{\frac{\alpha}{2}} s_\gamma ) & i e^{-i \beta } c_\gamma  (\frac{s_\alpha}{2} + s^2_{\frac{\alpha}{2}} s_\gamma ) & 0 & e^{-2 i \beta } c^2_\gamma  s^2_{\frac{\alpha }{2}} & c^2_{\frac{\alpha}{2}}+ s^2_{\frac{\alpha}{2}} s_\gamma^2 \\
\end{array}
\right)
\end{eqnarray}
for  $\beta=0$, i.e. only $h$ and $\lambda_0$ develop VEV  with $H_0$ to be inert,  we can obtain the following expressions for the masses of W and Z gauge bosons.
\begin{eqnarray}
m_W^2 &=& \frac{1}{4} f^2 g^2 \sin ^2(\alpha ) \left(\sin ^2(\gamma )+1\right)  \\
m_Z^2  &=&  \frac{1}{4} f^2 g^2 \sin ^2(\alpha ) \sec ^2(\text{$\theta $w}) \left((2 \cos (2
\alpha )+1) \sin ^2(\gamma )+1\right) 
\end{eqnarray}
The coupling of $h$ and  $\lambda_0$ to $W^+W^-$ and $ZZ$ are:
\begin{eqnarray}
g_{h W^+ W^-} &=& \frac{f g^2 \sin (\alpha ) \cos (\gamma ) }{2}  ~ \left(\cos (\alpha ) \left(\sin ^2(\gamma
)+1\right)-\sin ^2(\gamma )\right) \end{eqnarray}
\begin{eqnarray}
g_{\lambda_0 W^+ W^-} &=&   - \frac{f g^2 \sin (\alpha ) \sin (\gamma ) }{2}~ \left(\cos (\alpha ) \left(\sin ^2(\gamma
)+1\right)+\cos^2(\gamma )\right) 
\end{eqnarray}
\begin{eqnarray}
g_{hZZ} &=&  \frac{f g^2 \sec ^2(\text{$\theta $w}) }{32}  ~ \bigg( (3 \sin(2 \alpha )-2 \sin (3 \alpha )+2 \sin (4 \alpha )) \cos (\gamma )  \nonumber \\ &+& 8 \sin ^3\left(\frac{\alpha
}{2}\right) \left(5 \cos \left(\frac{\alpha }{2}\right)+4 \cos \left(\frac{3 \alpha
}{2}\right)+2 \cos \left(\frac{5 \alpha }{2}\right)\right) \cos (3 \gamma )\bigg)
\end{eqnarray}
\begin{eqnarray}
g_{\lambda_0 ZZ} &=& -\frac{  f g^2 \sec ^2(\text{$\theta $w})}{32}~
   \bigg((\sin (2 \alpha )+2 \sin (3 \alpha )+6 \sin (4 \alpha )) \sin (\gamma ) \nonumber \\ &+&8 \sin
   ^3\left(\frac{\alpha }{2}\right) \left(5 \cos \left(\frac{\alpha }{2}\right)+4 \cos
   \left(\frac{3 \alpha }{2}\right)+2 \cos \left(\frac{5 \alpha }{2}\right)\right) \sin (3
   \gamma )\bigg)
\end{eqnarray}
As we can see, the Eq.~\ref{mass} and  Eq.~\ref{HWZ} will be recovered for $\gamma=0$, so is the custodial symmetry. The mixing  can be generated by operators with custodial symmetry breaking, such as  the gauge  and  top spurion potential, without violating the CP property. 
 \subsection{Neutrino mass}
The neutrino interaction with the complex triplet can be generated by bilinear operators.  We can write down the projectors which  select the isospin triplet $\Lambda$ with hypercharge of $1$,  so that it couples to SM left-handed lepton $L = (\nu_L, e_L)^T$  through a Yukawa term of $\bar{L}^c \Lambda L$.
\begin{eqnarray}
\footnotesize
P_{\Lambda_0+ \Lambda_0^*} =  \frac{1}{2} 
\left(\begin{array}{cc|c}
\widetilde{\sigma}_- &  &  \\
  & \widetilde{\sigma}_+  &     \\ \hline
 &  & 0_2
 \end{array} \right) \qquad 
 P_{\Lambda_0- \Lambda_0^*} =  \frac{1}{2} 
\left(\begin{array}{cc|c}
\widetilde{\sigma}_- &  &  \\
  & -\widetilde{\sigma}_+ &     \\ \hline
 &  & 0_2
 \end{array} \right) \end{eqnarray}
 
 \begin{eqnarray}
 \footnotesize
P_{+} =  \frac{1}{2} 
\left(\begin{array}{cc|c}
  \sigma_1 & &   \\ 
  & 0_2 & \\ \hline
 &  & 0_2
   \end{array} \right) \quad
P_{++} =   
\left(\begin{array}{cc|c}
 - \widetilde{\sigma}_+ &  &   \\
  & 0_2  &   \\  \hline
 &  &  0_2 
  \end{array} \right)
  \end{eqnarray}
The effective operator for this Yukawa coupling reads:
\begin{eqnarray}
\mathcal{L}_{yuk} &\supset  & f y_{\nu} \left(\nu_L \nu_L {\rm{Tr}} [\left(P_{\Lambda_0+ \Lambda_0^*}  + P_{\Lambda_0- \Lambda_0^*}  \right). \Sigma(x)] +  \nu_L e_L Tr[P_+. \Sigma (x) + e_L e_L {\rm{Tr}} [P_{++}.\Sigma(x) ]  + h.c.  \right) \nonumber \\
&= & \left(\sin ^2(\alpha ) \left(f - i  \frac{1}{3 \sqrt{2} }\left(\sqrt{3} \eta _1-3 \sqrt{2} \eta _3+3 \varphi
   _0\right)\right) + h \sin (2 \alpha )  \right)  y_{\nu} \nu_L \nu_L \nonumber \\ 
 &+ & \left( i  \frac{1}{2\sqrt{2} }(\cos (2 \alpha )+3) (\Lambda_0+\Lambda_0^* )  + i \sqrt{2} \cos (\alpha ) (\Lambda_0-\Lambda_0^* )  \right)y_{\nu} \nu_L \nu_L  \nonumber\\ &+& i \left(\cos (\alpha ) \left(\Lambda _+-\varphi _+\right)+\Lambda _++\varphi
   _+\right)  y_{\nu} \nu_L e_L  + 2\sqrt{2} i \Lambda _{++}  y_{\nu} e_L e_L +h.c.
\end{eqnarray}
As we can see  that  the dominant contribution to the neutrino mass is  a  constant term proportional to $y_\nu v^2/f $ from a dim-5 Weinberg operator. On the other hand, once the CP-even field $\lambda_0 = -\frac{i}{\sqrt{2}}\left(\Lambda_0 -\Lambda_0^*\right)$  obtains a tiny VEV $\sim \mathcal{O}(1)$ GeV,  which is possibly achieved  by either  bilinear 4-fermion interaction or partial compositeness,   a Majorana mass will  be generated.

\bibliographystyle{JHEP}
\bibliography{2HDCH}
\end{document}